\documentclass[trackchanges,twocolumn]{aastex701}

\usepackage{amsmath}
\newcommand{\kms}{km\,s$^{-1}$}

\newcommand{\msun}{$M_{\rm \odot}$}
\newcommand{\chandra}{{Chandra}}
\newcommand{\xmm}{{XMM-Newton}}
\newcommand{\xrism}{{XRISM}}

\graphicspath{{./}}

\begin{document}

\title{Mapping plasma properties of Cassiopeia A with XRISM/Resolve: \\ a Bayesian analysis via UltraSPEX}

\correspondingauthor{Manan Agarwal}

\author[orcid=0000-0001-6965-8642,gname='Manan',sname='Agarwal']{Manan Agarwal}
\affiliation{Anton Pannekoek Institute/GRAPPA, University of Amsterdam, Science Park 904, 1098 XH Amsterdam, The Netherlands}
\affiliation{SRON Netherlands Institute for Space Research, Niels Bohrweg 4, 2333 CA Leiden, The Netherlands}
\email[show]{m.agarwal@uva.nl}

\author[orcid=0000-0002-4708-4219,gname='Jacco',sname='Vink']{Jacco Vink}
\affiliation{Anton Pannekoek Institute/GRAPPA, University of Amsterdam, Science Park 904, 1098 XH Amsterdam, The Netherlands}
\affiliation{SRON Netherlands Institute for Space Research, Niels Bohrweg 4, 2333 CA Leiden, The Netherlands}
\affiliation{Department of Physics, Graduate School of Science, The University of Tokyo, 7-3-1 Hongo, Bunkyo-ku, Tokyo 113-0033, Japan}
\email{j.vink@uva.nl}

\author[orcid=0000-0001-9911-7038,gname='Liyi',sname='Gu']{Liyi Gu}
\affiliation{SRON Netherlands Institute for Space Research, Niels Bohrweg 4, 2333 CA Leiden, The Netherlands}
\email{l.gu@sron.nl}

\author[orcid=0000-0003-1415-5823,gname='Paul',sname='Plucinsky']{Paul P. Plucinsky}
\affiliation{Harvard-Smithsonian Center for Astrophysics, MS-3, 60 Garden Street, Cambridge, MA, 02138, USA}
\email{pplucinsky@cfa.harvard.edu}

\author[orcid=0000-0003-0890-4920,gname='Aya',sname='Bamba']{Aya Bamba}
\affiliation{Department of Physics, Graduate School of Science, The University of Tokyo, 7-3-1 Hongo, Bunkyo-ku, Tokyo 113-0033, Japan}
\affiliation{Research Center for the Early Universe, School of Science, The University of Tokyo, 7-3-1 Hongo, Bunkyo-ku, Tokyo 113-0033, Japan}
\affiliation{Trans-Scale Quantum Science Institute, The University of Tokyo, Tokyo  113-0033, Japan}
\email{bamba@phys.s.u-tokyo.ac.jp}

\author[orcid=0000-0001-9267-1693,gname='Toshiki',sname='Sato']{Toshiki Sato}
\affiliation{Department of Physics, School of Science and Technology, Meiji University, Kanagawa, 214-8571, Japan}
\email{toshiki@meiji.ac.jp}

\author[orcid=0000-0002-5359-9497,gname='Daniele',sname='Rogantini']{Daniele Rogantini}
\affiliation{Department of Astronomy and Astrophysics, The University of Chicago, Chicago, IL 60637, USA}
\email{danieler@uchicago.edu}

\author[orcid=0009-0002-4783-3395,gname='Yuken',sname='Ohshiro']{Yuken Ohshiro}
\affiliation{RIKEN Pioneering Research Institute (PRI), 2-1 Hirosawa, Wako, Saitama 351-0198, Japan}
\email{yuken.ohshiro@riken.jp}

\begin{abstract}
Mapping the physical conditions of the shocked plasma of young supernova remnants (SNR) is crucial for understanding their explosion mechanisms, ejecta structure, and large-scale asymmetries.
Using $>350$~ks of \xrism/Resolve high spectral resolution observations of Cassiopeia~A (Cas~A), the youngest known Galactic core-collapse SNR, we present the first microcalorimeter-based plasma parameter maps of any SNR. 
We tessellate Cas~A into $1\arcmin\times1\arcmin$ regions and fit the broadband spectra as thermal emission from two pure-metal ejecta components---corresponding to intermediate-mass elements (IMEs) and iron-group elements (IGEs)---plus nonthermal synchrotron radiation.
For robust inference, we introduce {\em UltraSPEX}, a Bayesian framework that couples the \texttt{SPEX} plasma code with the {\em UltraNest} nested-sampling algorithm, yielding full posterior distributions and exploration of parameter degeneracies.
Key findings include enhanced Ar/Si and Ca/Si abundance ratios near the base of the Si-rich jets, and a high Ni/Fe mass ratio ($0.08\pm0.015$) in the base of NE jet. 
IGEs ejecta exhibit systematically higher Doppler velocities and broadenings than IMEs ejecta in most regions, with maximum differences of $\sim800$~\kms\ and $\sim1200$~\kms, respectively; Ca shows distinct (faster) kinematics from other IMEs in several SE regions.
The ionization timescale and electron temperature show a robust anti-correlation, particularly for IGEs. 
This relation and measured parameter values could be explained by semi-analytical models with significant ejecta clumping (overdensities of $\sim10$ for IGEs and up to $\sim100$ for IMEs) and reduced historical reverse-shock velocities.
Nonthermal emission accounts for a substantial fraction, with at least 47\% of the 4--6~keV continuum and dominates in the western regions, where the spectrum hardens.
\end{abstract}



\section{Introduction}\label{sec:intro}
Core-collapse supernova explosions are broadly recognized to be intrinsically aspherical, as is evident from both observations \citep[e.g.,][]{wang08, lopez11,fang24} and simulations \citep{blondin03,moesta15}. The most conspicuous example of this is the young Galactic supernova remnant (SNR) Cassiopeia~A (Cas~A), $\sim350$~ years old \citep{thorstensen01} and $\sim$3.4~kpc away \citep{reed95}, for which asymmetries are observed across various wavelengths (X-rays - \citealt{hughes00a, hwang00}; optical - \citealt{lawrence95,reed95,milisavljevic13,alarie14}; infrared - \citealt{hines04, milisavljevic24}; radio - \citealt{kenny85}). The optical light echo observations of the supernova of Cas~A from different directions showed that the explosion was intrinsically asymmetric \citep{rest11} and the interaction of the forward shock with a non-uniform circumstellar medium (CSM) has further enhanced the asymmetry of the remnant \citep{vink22a,orlando22,orlando25a}.

One of the most striking highly asymmetric features is the observation of high-velocity bipolar jet-like structures emerging in nearly opposing directions --- northeast and southwest -- with velocities measured up to 14,000 \kms \citep{fesen06}, which are chemically distinct and rich in Si/S ejecta \citep{vink04a,hwang04}. 
The distribution and velocities of the heavy elements within Cas~A are also reported to be anisotropic. \cite{grefenstette15} mapped the spatial distribution of radioactive titanium-44 ($^{44}$Ti), which showed pronounced knots and large-scale offsets unrelated to other elements. The remnant also shows evidence of inverted stratification --- Fe-rich ejecta have penetrated outward through the Si-rich shell in the southeast, seen in projected radius \citep{hughes00a} as well as 3D kinematics \citep{tsuchioka22}, and in the northern region, reported in Doppler space \citep{willingale02}. 
Furthermore, the bulk motions of the ejecta are highly asymmetric, with an overall redshifted component dominating in the northwest and a prominent blueshift in the southeast \citep{holt94,hwang01}. The shock history across Cas~A is also highly variable, indicating localized differences in the interaction of the remnant with its surroundings, as evident from the reverse shock moving inward only in the western regions \cite{vink22a}. This dramatic array of asymmetries sheds light on the complexities of the SN explosion and the subsequent evolution of its remnant. Theoretical studies have managed to explain several observed asymmetries using simulations of neutrino-driven supernova explosions \citep{wongwathanarat17,orlando25b}; however, features such as jet structures are not reproduced and their origin remains unclear under self consistent simulations.

Mapping the physical properties of X-ray emission from SNRs is crucial to understanding these asymmetries and their implications for explosion mechanisms. Such studies have been conducted for the majority of the brightest and youngest SNRs --- Tycho \citep{godinaud25,uchida24}, Kepler \citep{sun19}, SN1006 \citep{li15}, W49B \citep{zhou18a}, Puppis~A \citep{mayer22} --- using data from one of \chandra, \xmm, and SRG/eROSITA.

For Cas~A, several spatially resolved X-ray parameter maps have been produced, each providing key insights into the complex structure of the remnant. \cite{willingale02} presented the first three-dimensional dynamics of the X-ray emitting plasma using \xmm\ data. \cite{yang08} mapped the element abundances using two deep \chandra\ observations ($\sim$250 ks total) to show a strong correlation between Si, S, and Ca, while distinct from Fe. \cite{hwang12} constructed detailed parameter maps of the ejecta material using data from the 1~Ms \chandra\ VLP observations \citep{hwang04}. They reported that most of the Fe ejecta in Cas~A has already been heated by the reverse shock and are associated with two populations corresponding to different burning layers. Their temperature and ionization age parameter maps remain a key reference for Cas~A studies.
\cite{lazendic06} derived the plasma properties of 17 bright X-ray knots across the remnant using the Si and S lines observed from the high-resolution \chandra\ HETGS data.
\cite{stage06} and \cite{helder08} produced nonthermal X-ray emission maps, discovering that this emission is more dominant in the west and is probably synchrotron radiation from accelerated electrons. 
\cite{patnaude11} investigated the 4.2--6 keV emission between the years 2000 and 2010, and reported a steady decline and steepening of the nonthermal X-ray emission over the entire remnant.
Most recently, \cite[][see Appendix]{vink22b} reconstructed the parameter maps by modeling Cas~A with an additional nonthermal component across the remnant for the first time. 

The goal of this study is to update the maps with precise plasma properties by leveraging significant improvements in technology and our understanding of X-ray emission in Cas~A over the years. These advancements are threefold. 
First, the advent of the X-Ray Imaging and Spectroscopy Mission \citep[\xrism,][]{tashiro25} with its microcalorimeter instrument, Resolve \citep{ishisaki25, kelley25}, has opened a window to the long-awaited spatially resolved high-resolution X-ray spectroscopy --- $\Delta E \sim4.5$ eV at 6 keV. This is a major development over the CCD and grating instruments such as those on \chandra\ and \xmm\ telescopes, particularly for the study of extended sources and diffuse emission. 
It has led to the first detection of rare odd-Z elements (P, Cl, K) in Cas~A \citep{xrism_casA_2025}.
Simulations of the high-resolution data of \xrism\ also show that it can measure ion temperatures, and identify potential features of emission mechanisms such as radiative recombination continuum \citep{greco20} or charge exchange.

Secondly, since previous mapping studies of Cas~A, nonthermal emission has been shown to be a more significant fraction of total X-ray emission. Although already more than two decades ago, \cite{willingale02} estimated that a nonthermal power-law contribution can be up to 25\%; however, since synchrotron radiation was not confirmed at the time, they did not include it in their models. 
Synchrotron emission was then first shown to be significant in the narrow outer rim regions \citep{vink03a, bamba05}. 
Subsequently, \cite{helder08} reported a more than 50\% overall nonthermal contribution in the 4.2 to 6 keV band using \chandra\ data. 
More conclusive evidence comes from NuSTAR observations, which demonstrated that the hard X-rays ($>15$ keV) from Cas~A originate from nonthermal processes \citep{grefenstette15}. This is further supported by the detections of $\gamma$-rays at GeV-TeV energies \citep{aharonian01, ahnen17, cao25}
The synchrotron nature of the nonthermal X-ray emission has recently been confirmed by detecting an overall $\sim5$\% polarization \citep{vink22b} and up to 20\% polarization degree near the outer rim \citep{mercuri25} using {IXPE}.
A nonthermal component is missing in all of the full parameter mapping studies of Cas~A, except \cite{vink22b}, which was also limited to maximum photon energies of 7~keV. \xrism/Resolve features exceptionally low instrumental background noise (due to operating in the low Earth orbit, and using an anti-coincidence detector and optimized event screening), which is crucial for detecting faint nonthermal X-ray emission up to 12~keV \citep{mochizuki25}, as opposed to \chandra, for which the background becomes significant above $\sim7-8$~keV.

Finally, there has been a marked improvement in fitting techniques and atomic databases. The traditional X-ray spectral fitting methods are prone to settling at local minima due to rugged parameter terrain with potentially many degeneracies. Recent advances in Bayesian-based parameter fitting provide more reliable and robust results \citep{buchner23}. In addition, atomic databases (corresponding to the most widely used X-ray spectral fitting softwares: \texttt{XSPEC} and \texttt{SPEX}) have been updated to include significantly more transition lines and atomic physics (see \citealt{plucinsky25}), in part owing to the high-resolution spectroscopic data provided by \xrism. To take advantage of both of these advances, we developed a new tool, {\em UltraSPEX}, which integrates \texttt{SPEX} with the Bayesian-based nested sampling algorithm, {\em UltraNest} \citep{buchner21}.

The structure of the paper is as follows: in \S~\ref{sec:data}, we describe the \xrism/Resolve observations of Cas~A and data reduction for this analysis. In \S~\ref{sec:model}, we present a detailed description of our spectral model and the different choices leading to it. The motivation, advantages and implementation details behind {\em UltraSPEX} are described in \S~\ref{sec:ultraspex}. The results and a discussion based on the parameter maps of our spectral fitting analysis are provided in \S~\ref{sec:results}. Finally, we present our conclusions in \S~\ref{sec:conclusions}.
Throughout the paper, errors are reported at a 68.3\% confidence level ($1\sigma$), unless otherwise stated.

\begin{figure}
\includegraphics[width=\columnwidth]{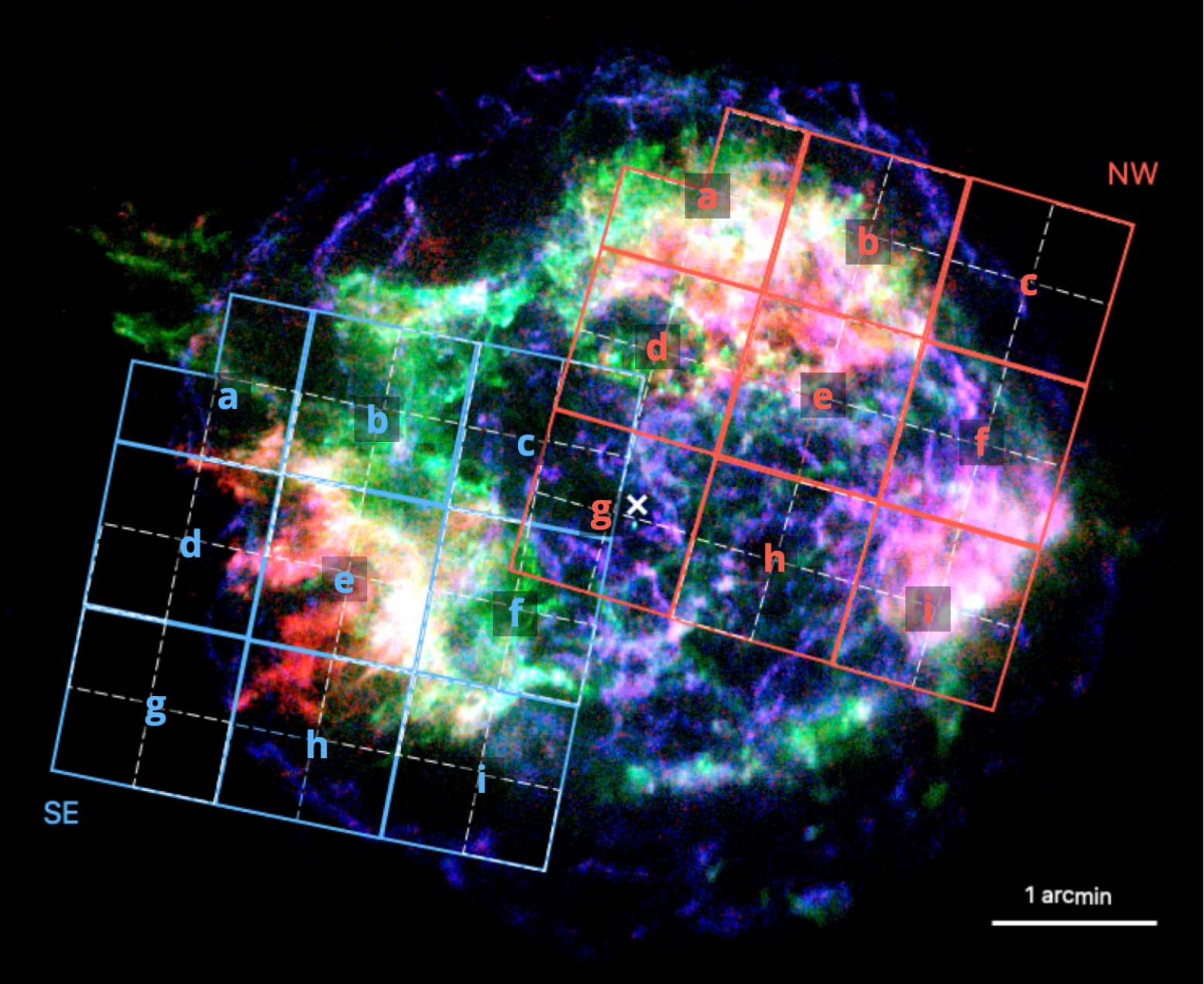}
 \caption{
The \xrism/Resolve observations of Cas~A are shown projected onto a three-color \chandra\ image (ObsID 4638)\footnote{The Chandra data can be accessed via ~\dataset[DOI: 10.25574/cdc.536]{https://doi.org/10.25574/cdc.536}} --- red: 6.50--6.75 keV (Fe K$\alpha$ complex); green: 1.76--1.94 keV (Si He$\alpha$); blue: 4.00--6.00 keV (continuum).
The SE (ObsID 000129000) and NW (ObsID 000130000) \xrism\ pointing fields of view are shown in blue and red, respectively. 
The white dashed boxes outline the \xrism/Resolve pixels and the solid boxes show the 2$\times$2 pixel binnings (``super-pixels"), labeled per observation from {\it a} to {\it i}. The white `$\times$' symbol marks the expansion center from \citet{thorstensen01}.
}
 \label{fig:pixels}
 \end{figure}

\section{Observations and Data Reduction}\label{sec:data}
Cas~A was observed twice with \xrism\ in December 2023 during the commissioning phase. The Resolve microcalorimeter onboard \xrism\ enables non-dispersive high-resolution spectroscopy and has a field of view (FOV) of $\approx3.1\arcmin\times3.1\arcmin$ with an angular resolution of $\sim 1\farcm3$ HPD (half-power diameter). The details of the two observations of Cas~A and the data reduction procedure are outlined in \cite{plucinsky25}, with only a concise summary provided here. 

The two observations target the southeast region (``SE" hereafter, ObsID 000129000) and the northwest region (``NW" hereafter, ObsID 000130000) for a Resolve exposure time (after applying standard filtering criteria) of 181 ks and 166 ks, respectively. The observation data were reprocessed and calibrated utilizing the HEASoft 6.34 software package and \xrism\ CalDB 9 (version 20240815). Only high-resolution primary grade events (`Hp') are used for the spectral analysis. An extra large (XL) redistribution matrix file (RMF) in the split-rmf format was generated by the \texttt{rslmkrmf} task using the cleaned event file. An auxiliary response file (ARF) was generated by the \texttt{xaarfgen} task using an exposure corrected Chandra image in the 0.5 to 5.0 keV band as an input sky image. The non X-ray background (NXB) spectra were extracted using the \texttt{rslnxbgen} task from a database of Resolve night-Earth data and weighted by the distribution of geomagnetic cut-off rigidity sampled during each observation.
Spectral fitting was performed using the \texttt{SPEX} software version 3.08.01 \citep{kaastra96} together with an updated atomic database, which will be released in \texttt{SPEX} version 3.08.02, we refer to this version as \texttt{SPEX} 3.08.01$^*$ \citep{plucinsky25}. The spectra and the response files (in \texttt{SPEX} file format) are optimally binned following the approach by \cite{Kaastra16} where the bin size varies depending on factors such as the spectral resolution, the number of resolution bins, and the local intensity of the spectrum, thus resulting in different optimal bin sizes for each energy range. 

We analyze the spectra in the 1.8--11.9 keV energy range from 2 $\times$ 2 Resolve pixels, corresponding to 1\arcmin $\times$ 1\arcmin\ square regions, which is commensurate with the $1\farcm3$ HPD of \xrism\ point spread function (PSF). The larger regions reduce the effects of spatial spectral mixing caused by the broad PSF, as estimated quantitatively for Cas~A in \cite{plucinsky25}. These ``super-pixels" are labeled \textit{a} to \textit{i}, as marked in Fig.~\ref{fig:pixels}.

\section{Spectral Model}\label{sec:model}
The optical spectroscopy of the light-echoes of Cas~A confirmed it as a Type IIb supernova, i.e., the progenitor had only a thin hydrogen layer left at the moment of explosion \citep{krause08,rest11}.
This is in agreement with the finding that the optical ejecta knots of Cas~A do not contain hydrogen, except for very few outermost knots \citep{fesen87}.
For the X-ray emitting ejecta of Cas~A, we cannot measure the hydrogen content, since hydrogen only produces thermal continuum radiation, which is also provided by other ions.
For that reason, there is a long tradition of modeling the thermal X-ray emission from Cas~A using a pure-metal plasma
\citep{vink96,willingale02,hwang12,vink22b}. As the greater charge of the metal ions and the higher number of electrons provided by the metals result in more efficient
thermal continuum radiation, pure-metal plasmas result in overall lower mass estimates for a given flux \citep{vink96,greco20,leahy24}. Moreover,
free-bound (i.e., radiative recombination continuum; RRC) and two-photon continuum contributions become more enhanced than free-free emission (i.e., bremsstrahlung), see \cite{greco20}.

To complicate matters, a large fraction of the X-ray continuum from Cas~A appears to be from X-ray synchrotron radiation \citep{allen97,vink03a}, perhaps even as much as 54\% to 93\% \citep{helder08}. This nonthermal continuum is associated with both the reverse shock---most notably in the (south)western part---and the forward shock. The synchrotron nature of large parts of the continuum was recently confirmed by polarization measurements using {IXPE} \citep{vink22b,mercuri25}.
\xrism/Resolve has sufficient sensitivity up to $\sim$12 keV to allow a better constraint on the nonthermal contribution.

The plasma conditions inside young SNRs are in the non-equilibrium ionization (NEI) state and show a gradient in the electron temperature ($kT_\mathrm{e}$) and ionization age ($n_{\mathrm e}t$) due to varying shock heating and projection effects. 
Moreover, the plasma properties of Cas~A vary on scales of $\sim$arcsec \citep{hwang12} and the large extraction regions in our analysis due to the \xrism\ PSF lead to further mixing of regions with different plasma conditions.
For these reasons, we adopted a plane-parallel shock (\texttt{pshock}) model \citep{borkowski01}, which \citet{bamba25} has
shown to fit \xrism\ Cas~A spectra better than a simple single-zone NEI model.
In \texttt{SPEX}, \texttt{pshock} is implemented as a linear combination of 200 NEI layers with a constant fitted $kT_\mathrm{e}$ and logarithmically increasing $n_{\mathrm e}t$, 
starting from $5\times10^{7}$ cm$^{-3}$s up to a fitted maximum $n_{\mathrm e}t$ value. 
A \texttt{pshock} model is better at capturing the expected distribution of ionization parameters, but with the caveat that the distribution may in reality be different and that besides a distribution in $n_{\mathrm e}t$ there may also be a distribution of $kT_\mathrm{e}$ contributing to the extracted spectra.

With the above considerations taken into account, we set up our X-ray spectral model as consisting of two pure-metal \texttt{pshock} components --- one targeting the intermediate-mass elements (IMEs), O$^*$, Si, S, Ar, Ca, and the other the iron-group elements (IGEs), Fe and Ni --- a nonthermal component modeled as a power-law --- a foreground absorption (\texttt{absm} in \texttt{SPEX}) plus a background model based on the available non-X-ray background (NXB) of \xrism/Resolve. This results in a model with 17 free parameters. 

The separation into IMEs and IGEs components is justified by the fact that the two groups are known to have significantly different spatial distribution, in particular in the SE, where regions with pure iron ejecta have been identified
\citep{hughes00a,hwang03}. 

In order to model the thermal components as a pure-metal plasma, we set the abundance of Si for the IMEs component and Fe for the IGEs component to a value of 100,000 times their solar abundance \citep[as per proto-Solar abundances of][]{lodders09}.
This ensures that the continuum emission for both components is dominated by IMEs and IGEs, respectively.  The other elements were allowed to vary. The Si and Fe emissivity is then completely determined by the normalization (emission measure) of the components.
Normalization is defined as the total $n_{\rm e}n_{\rm H}V$ in units of ${\rm cm}^{-3}$, where $n_{\rm e}$ and $n_{\rm H}$ are the electron and hydrogen densities, respectively. $V$ is the volume of the emitting plasma.
The distance used to calculate the normalization is 3.4 kpc.
Note that the fitted normalization values are smaller than would normally be the case, as hydrogen has become a trace element, as we assumed pure-metal plasmas.
This also affects the $n_{\rm e}/n_{\rm H}$ ratio as the electrons originate from metal ions.

Due to the problem with the Gate Valve, \xrism/Resolve has only sufficient sensitivity above 1.8 keV, thus no emission lines of O are available. Despite this, we include O in the IMEs component, as it is the dominant source of the thermal continuum emission \citep{hwang12}, given the O-rich nature of Cas~A. The inclusion of O thus adds freedom to the model to enhance the thermal continuum emission, but in reality additional thermal continuum may also originate from a small fraction of H or from He, Ne, and Mg. As these elements do not provide line emission in the \xrism/Resolve band, we cannot constrain their individual contributions. However, we expect O to be the largest source of thermal continuum, in particular since for Cas~A previous work has shown that Ne and Mg have relatively low abundances \citep{vink96,hwang12}.
In our model, the O abundance is therefore a proxy for all elements that could contribute to the thermal continuum from elements with $Z<14$,
and hence we refer to it as O$^*$.

The impact of interstellar extinction is also limited above 1.8 keV, making it difficult to constrain with \xrism/Resolve alone. In order to take extinction into account, we adopted the extinction parameter $N_{\rm H}$ from \citet{vink22b} that used \chandra\ data in the 0.6--7~keV band. The map of the $N_{\rm H}$ values used in our analysis is shown in Figure~\ref{fig:extra_maps}.

The plasma parameter mapping provided here has several 
innovations with respect to previous works,
most notably the excellent spectral resolution of \xrism/Resolve, 
along with a better grasp on the nonthermal X-ray emission above 8 keV without any use of another X-ray experiment and the use of a pure-metal spectrum model.
What we miss in this analysis is, however,
a sensitivity to the O-, Ne-, Mg-rich emission, arcsecond spatial resolution, as well as full coverage of the entire SNR.

An obvious comparison is the full plasma modeling detailed in \citet{hwang12} based on the 2004 Chandra observations. Their model was based on an \texttt{XSPEC vpshock} model with a pure-metal plasma. 
Only for regions with strong nonthermal emission an X-ray synchrotron component was added, and in a few other regions with strong Fe-K emission a single NEI component was used. In contrast, the \xrism/Resolve spectra are extracted with larger grid sizes of 1\arcmin $\times$ 1\arcmin\ regions, and all spectra were treated with the same model, i.e., all included a nonthermal component.

A caveat in the model used here, similar to previous studies, is that we do not have a separate model for the thermal emission from the shocked circumstellar medium (CSM).  Although the CSM component is likely subdominant for the X-ray spectrum of Cas~A as the continuum is dominated by nonthermal emission and the ejecta continuum, and the line emission is also dominated by the ejecta. However, a recent study of the so-called ``green monster" identified by JWST \citep{milisavljevic24} and also analyzed in X-rays \citep{vink24a}, showed that the thermal CSM component appears to dominate  the line emission from the central and southwestern regions of Cas~A. This has some relevance for part of Cas~A covered by Resolve for the NW observation.

\section{UltraSPEX: Bayesian-based spectral fitting tool}\label{sec:ultraspex}

X-ray spectral fitting has been conventionally performed manually aided by advanced gradient descent-based algorithms like Levenberg-Marquardt \citep{levenberg1944, marquardt1963}, where the workflow typically involves iteratively adjusting parameters by hand to minimize a fit statistic (such as C-statistic). Although this is computationally efficient, it can be subjective, susceptible to converge to local minima, difficult to automate to numerous spectra, and estimating parameter uncertainties is still computationally slow. As these drawbacks are widely known today, the Bayesian approach to spectral fitting is growing in popularity. Although the first application to X-ray spectra was more than two decades ago by \citet{vanDyk01}, it is only in the last few years that the Bayesian framework has truly gained momentum and is being extensively applied to wide-ranging X-ray studies --- AGNs \citep{buchner14,rogantini22}, X-ray binaries \citep{kosec23}, pulsars \citep{choudhury24}, interstellar medium \citep{rogantini21} and SNRs (Tycho - \citealp{ellien23,godinaud25}, Cas~A - \citealp{vink24a}). 
This rapid adoption is driven by the increased availability of high-performance computational resources needed to run these algorithms and the need for more complex models required to fit high-resolution spectroscopy data, such as from \xrism.

In the Bayesian framework,
the parameter space is explored using the likelihood distribution, but the goal shifts from finding likelihood maxima to evaluating the posterior probability distribution of the model over the data. We refer to recent reviews by \citet{ashton22} and \citet{buchner23} for technical details on the Bayesian statistics.
For X-ray spectral analysis, Bayesian inference can be implemented using different sampling algorithms, with Markov Chain Monte Carlo (MCMC) and Nested Sampling \citep[NS,][]{skilling04}.
MCMC can be the more efficient of the two and generates parameter uncertainties and correlations \citep{albert25}. However, it has difficulties in exploring a high-dimensional, multimodal, and degenerate parameter landscape and is highly sensitive to initial parameter guesses, leading to incomplete exploration of plausible regions of parameter space and underestimates of the parameter uncertainties.
NS is a powerful Monte Carlo technique that is well suited for complex problems. Unlike MCMC, NS starts from the entire parameter space and evolves a collection of ``live points" to map all minima, including multiple modes. This approach is insensitive to the initialization point and ensures extensive coverage of the parameter space.
NS is also efficient in calculating Bayesian evidence, which MCMC methods do not directly provide. 

In order to implement this Bayesian framework, we developed a Python tool which integrates the NS algorithm {\em UltraNest} \citep[v4.3.2][]{buchner19, buchner21}, with \texttt{PYSPEX}, which is a python interface to \texttt{SPEX} (v3.08.01$^*$). We refer to this NS Python tool as {\em UltraSPEX}, which is publicly available via 
GitHub \footnote{\url{https://github.com/MananAgarwal/UltraSPEX}} and 
Zenodo\footnote{\url{https://doi.org/10.5281/zenodo.16921445}} repositories. 
{\em UltraSPEX} is largely inspired by Bayesian X-ray Analysis \citep[BXA][]{buchner14} which integrates {\em UltraNest} with \texttt{XSPEC} and \texttt{Sherpa}.
Due to the higher customisability of \texttt{SPEX} models which can be exploited to improve model computation times, \texttt{SPEX} is well suited for Bayesian analysis. In addition, the extensive atomic database of \texttt{SPEX} (\texttt{SPEXACT}) and the ability to optimally bin \citep{Kaastra16} the spectral response files (from $\sim$ 0.8 GB to $\sim$ 20 MB for \xrism\ XL RMF) make it ideal for high-resolution X-ray spectroscopy data of \xrism/Resolve.
Given these features of \texttt{SPEX}, a similar integration with an NS algorithm, {\em MultiNest} \citep{feroz09}, was developed by \citet{rogantini21} called \textsc{BaySpex}. {\em UltraNest} and {\em MultiNest} are similar but differ in their sampling strategies and robustness, with {\em UltraNest} generally being more robust and less prone to biases at the cost of being more computationally expensive \citep{buchner16, hoogkamer25}. UltraNest offers better scalability with dimensionality, parallelization capabilities, dynamic adjustment of the number of live points for improved parameter space exploration, and more reliable uncertainty estimation for complex or high-dimensional problems. 

\begin{figure*}
\centerline{
  \includegraphics[width=0.49\textwidth]{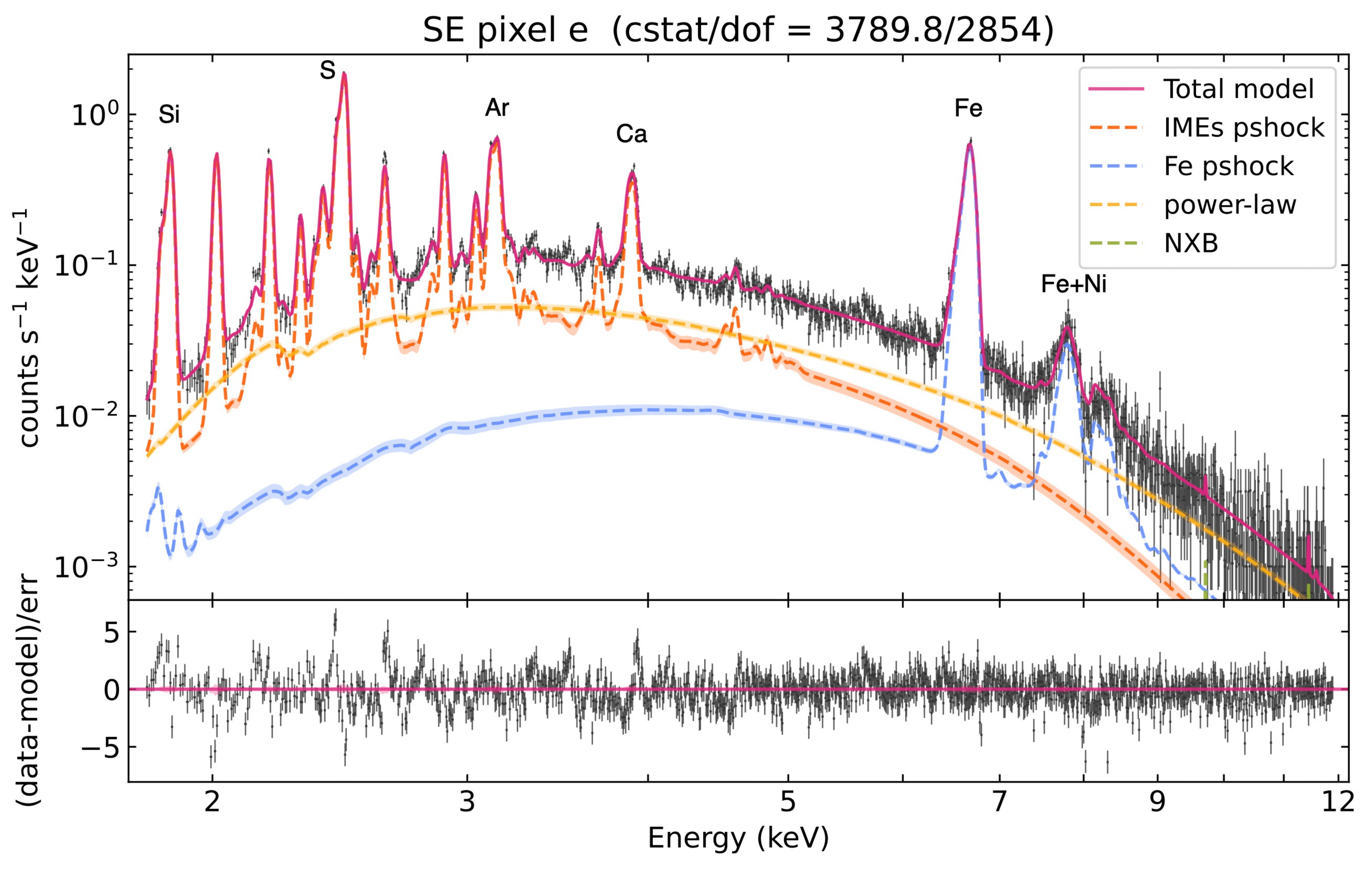}
  \hspace{2mm}
  \includegraphics[width=0.49\textwidth]{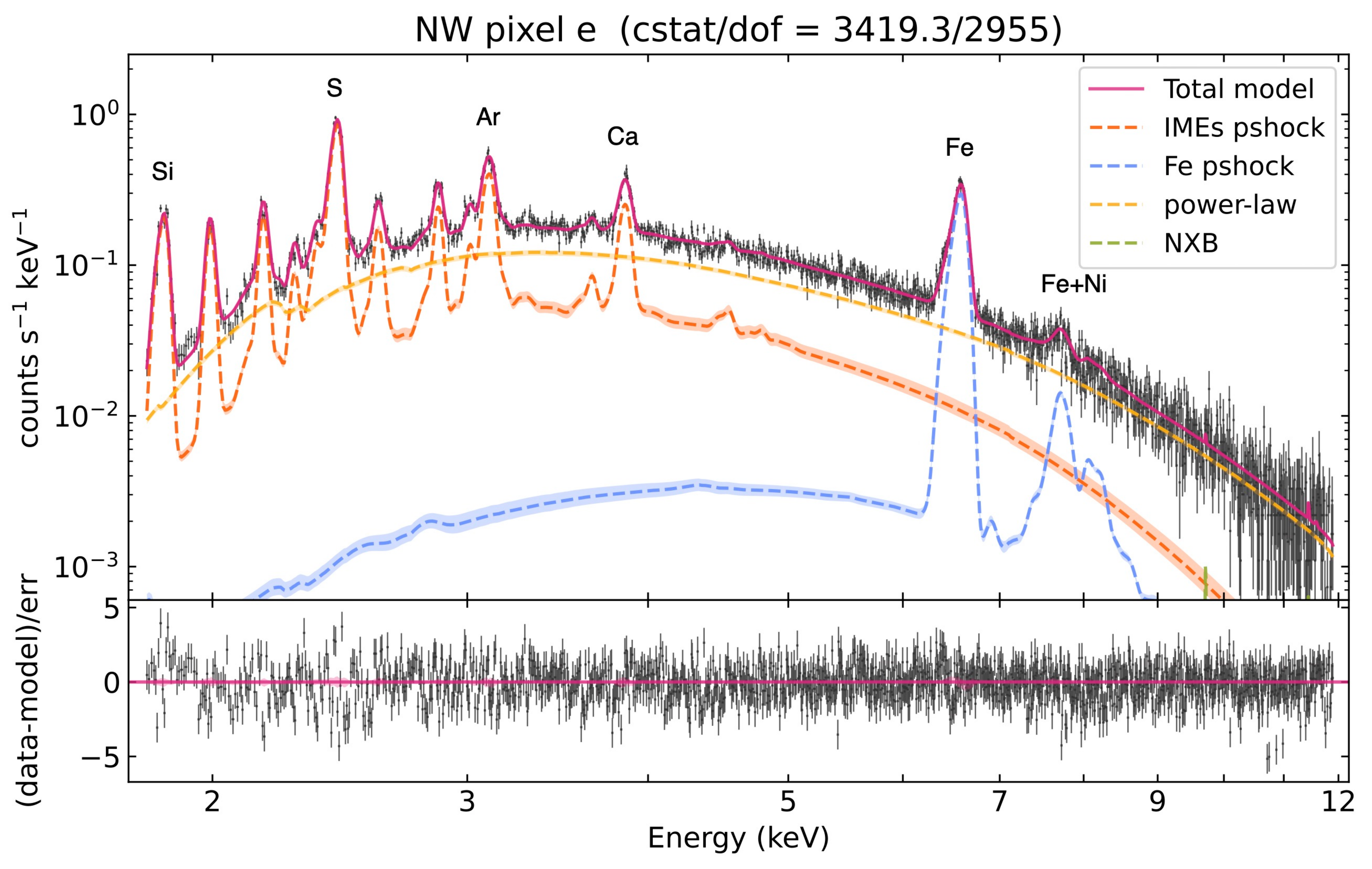}
}
\caption{
\xrism/Resolve spectra in the 1.8--11.9 keV energy range from the central super-pixel {\em e} ($2\times2$ pixel) for the SE (left) and NW (right) pointing, with their corresponding fitted spectral model using {\em UltraSPEX}. 
The total model is shown in a solid red line and the model components are shown in dashed lines with the corresponding 1 sigma uncertainties marked in shaded bands. The orange, blue, yellow and green dashed line corresponds to the pure-metal IMEs \texttt{pshock} component, pure-metal Fe-group \texttt{pshock} component, the power-law component and the NXB component, respectively.
The data are represented by the black data points and have been rebinned to minimum signal to noise ratio of 5 for display purposes only. The lower panels display the residuals between the data and the fitted model.
}
\label{fig:pixel_e_spectra}
\end{figure*}

\subsection{Optimizing UltraSPEX for \xrism\ data analysis}

We successfully tested {\em UltraSPEX} on \xrism\ data by performing fits to a simple model composed of Gaussians in \citet{vink25}, where a single spectral fit took less than 2 hours when parallelized over 4 cores on an Apple M2 Max processor. 
Scaling up to a more complex physical model, as described in section \ref{sec:model}, required a number of optimizations, some of which are specific to our model definition and a priori knowledge about Cas~A characteristics. 

In \texttt{SPEX}, the model computation time scales with the square of the number of spectral bins and with the square of the number of ions in the model. To speed up the computation, we utilized optimally binned spectra and response files from 2 $\times$ 2 Resolve pixels \citep[also employed in][]{vink25}
and we use the \texttt{SPEX} command ``\texttt{ions ignore}" to remove the rare elements (e.g., P, Cl, K, Ti, Cr, Mn), which contribute little to the total thermal flux from our model.
The ``\texttt{ions ignore}" command accelerates model calculations by omitting line emission from selected ions, in contrast to setting the corresponding elemental abundances to zero, which does not affect the computation time.  
Skipping rare elements from the model also helps reduce the dimensionality of the fitting parameter space. We use the same \texttt{SPEX} command to also ignore all the ions whose emission lines are not detected in the \xrism/Resolve passband with the Gate Valve closed.

To further speed up the calculations, we apply a multiplicative Gaussian component to the \texttt{pshock} model to account for total line broadening (i.e., both thermal and kinematic Doppler broadening), instead of using the intrinsic \texttt{SPEX} thermal Doppler broadening, which accounted for nearly 50\% of the model computation costs. Although using the intrinsic \texttt{SPEX} thermal broadening is more accurate as the broadening depends on the element mass, this effect is negligible ($<1$\% effect on broadening values), since a separate broadening is applied to the IMEs and IGEs components in our model---which are both confined to relatively small energy bands---and since the broadening in Cas~A is dominated by the velocity broadening (order of 1000 \kms).
These optimizations brought down the average model computation time by a factor of 15 (from $\sim$ 20 sec to $\sim$ 1.4 sec using the Apple M2 Max processor).

Using this \texttt{SPEX} setup, we performed spectral fits with the Reactive Nested Sampler using a step sampler in {\em UltraNest} with 400 live points and a log-evidence accuracy of 0.5.
Each spectral fit evaluated $\sim$ 900,000 models and took on average $\sim$ 8 days and used $\sim$ 25 GB of memory when parallelized over 16 cores on a computing cluster with AMD EPYC 7401 processor.
The priors (parameter ranges) used and the median values of the posterior distribution for all the 17 free parameters are provided in the Appendices~\ref{app:all_spectra} and \ref{app:corner_plots}.

\begin{figure*}
\centerline{
  \includegraphics[width=0.33\textwidth]{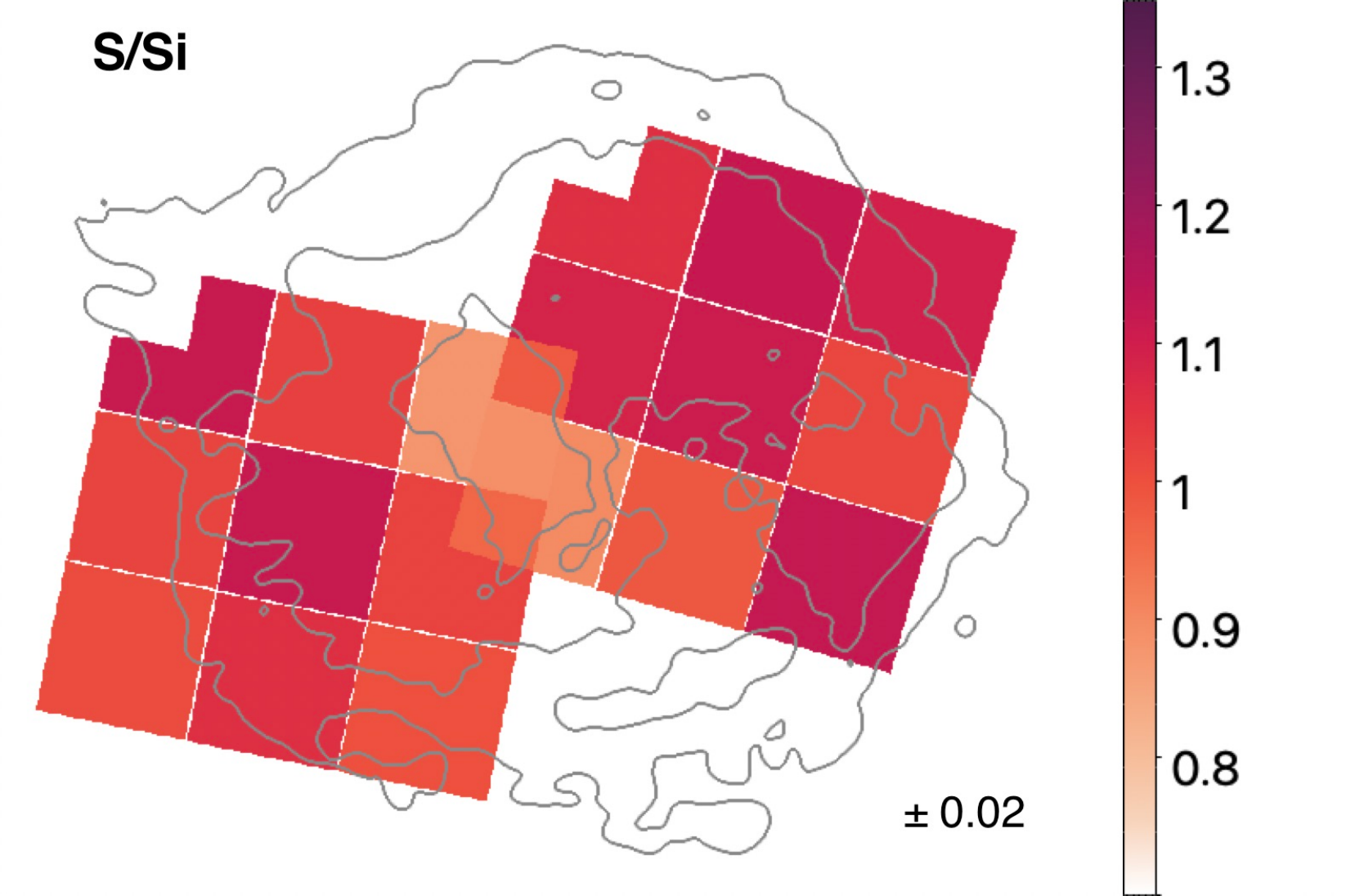}
   \hspace{2mm}
  \includegraphics[width=0.33\textwidth]{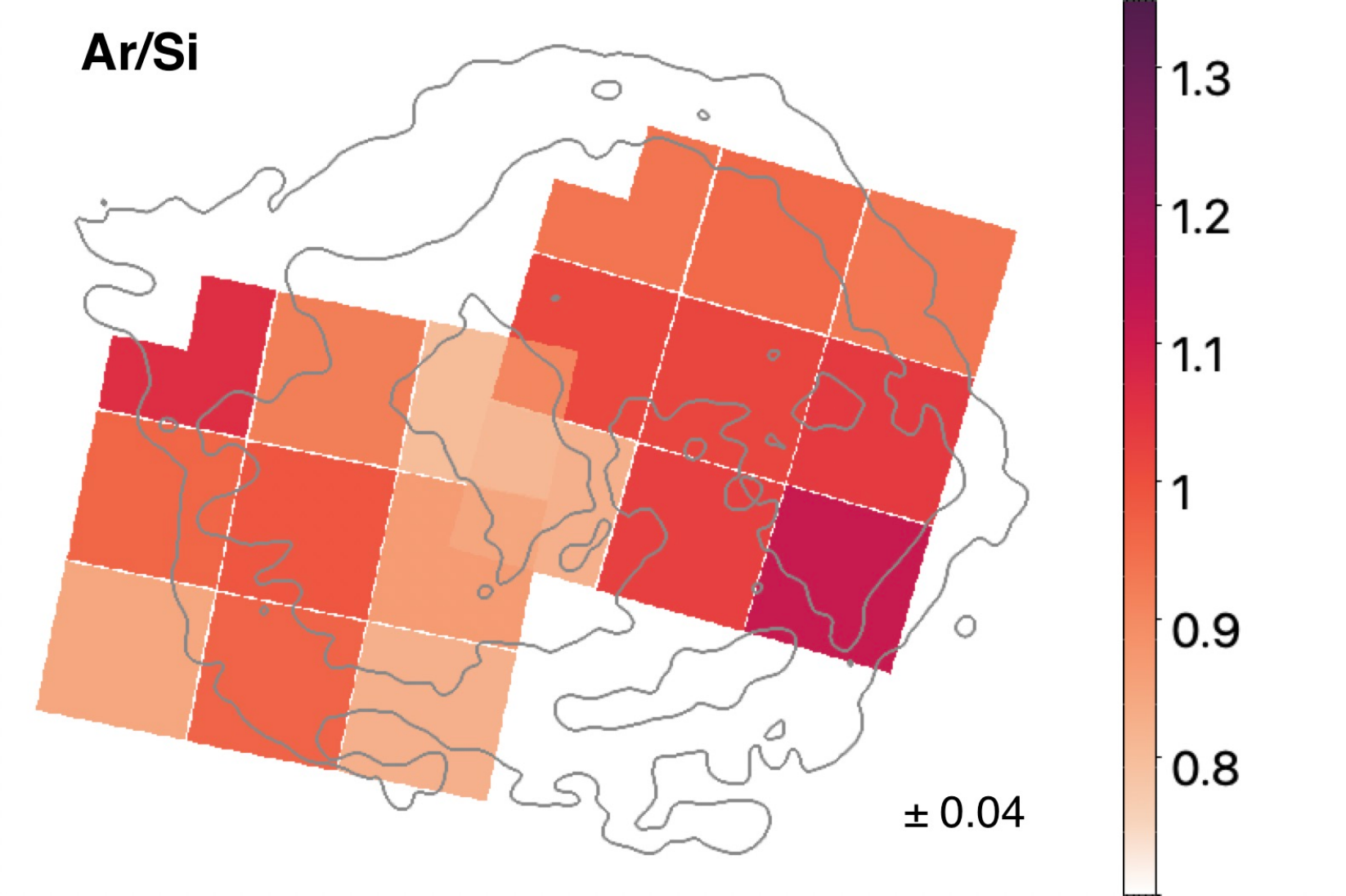}
   \hspace{2mm}
  \includegraphics[width=0.33\textwidth]{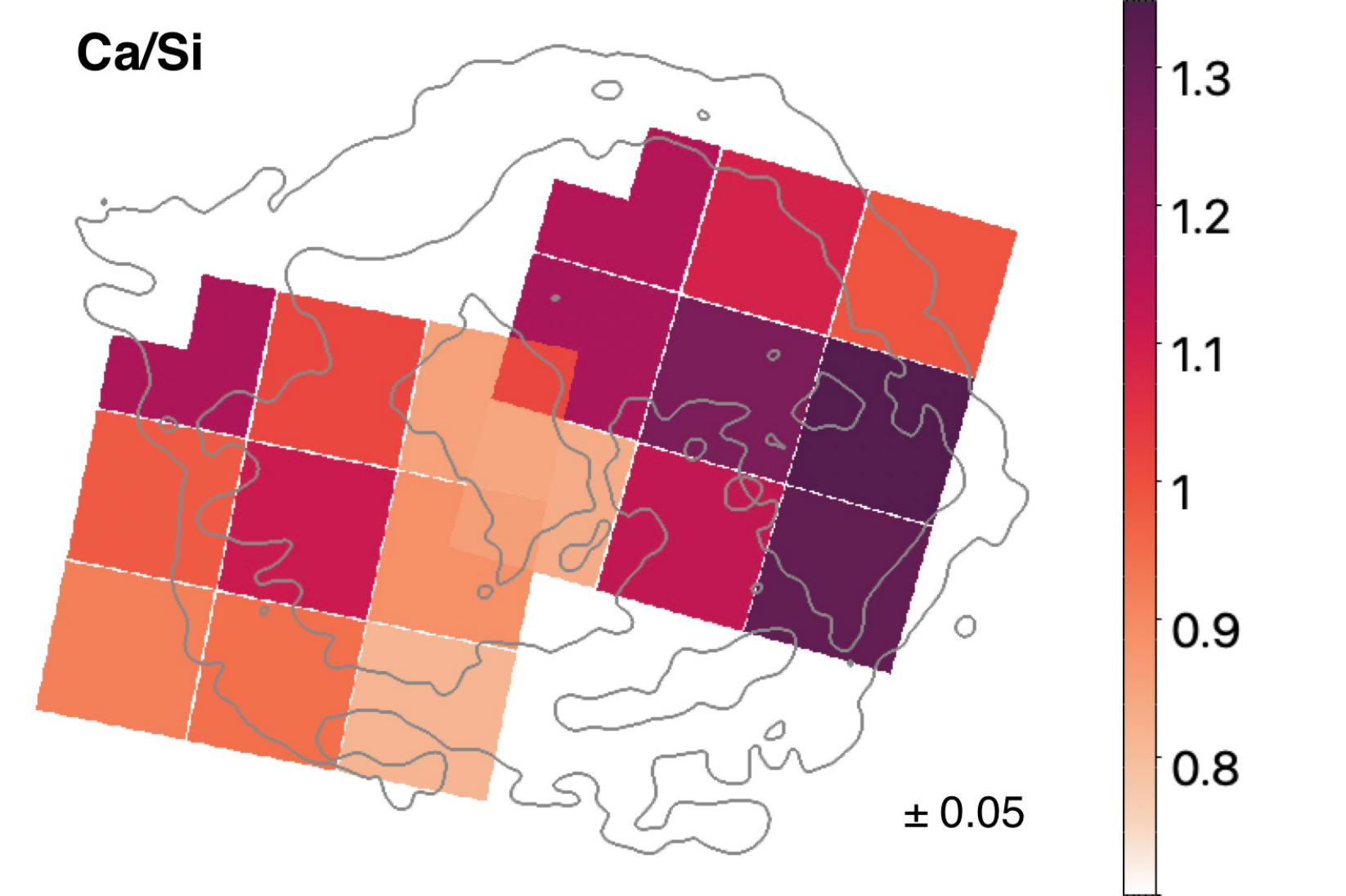}
}
\caption{
The panels show the fitted abundance ratios to solar abundance ratios \citep[as per][]{lodders09} for S/Si, Ar/Si, and Ca/Si (Si was fixed to 1e5 times solar value).
The average $1\sigma$ error is indicated on the bottom right of each subplot. In all the figures with parameter maps, for the overlapping pixel regions near the center, an average parameter value is shown. The contours in grey show the outline of Cas~A as derived from the \chandra\ observation from 2004 (ObsID 4638) using a broadband image.
}
\label{fig:abundance_maps_IMEs}
\end{figure*}

\begin{figure}
\hspace{0.1\columnwidth}
\includegraphics[width=0.8\columnwidth]{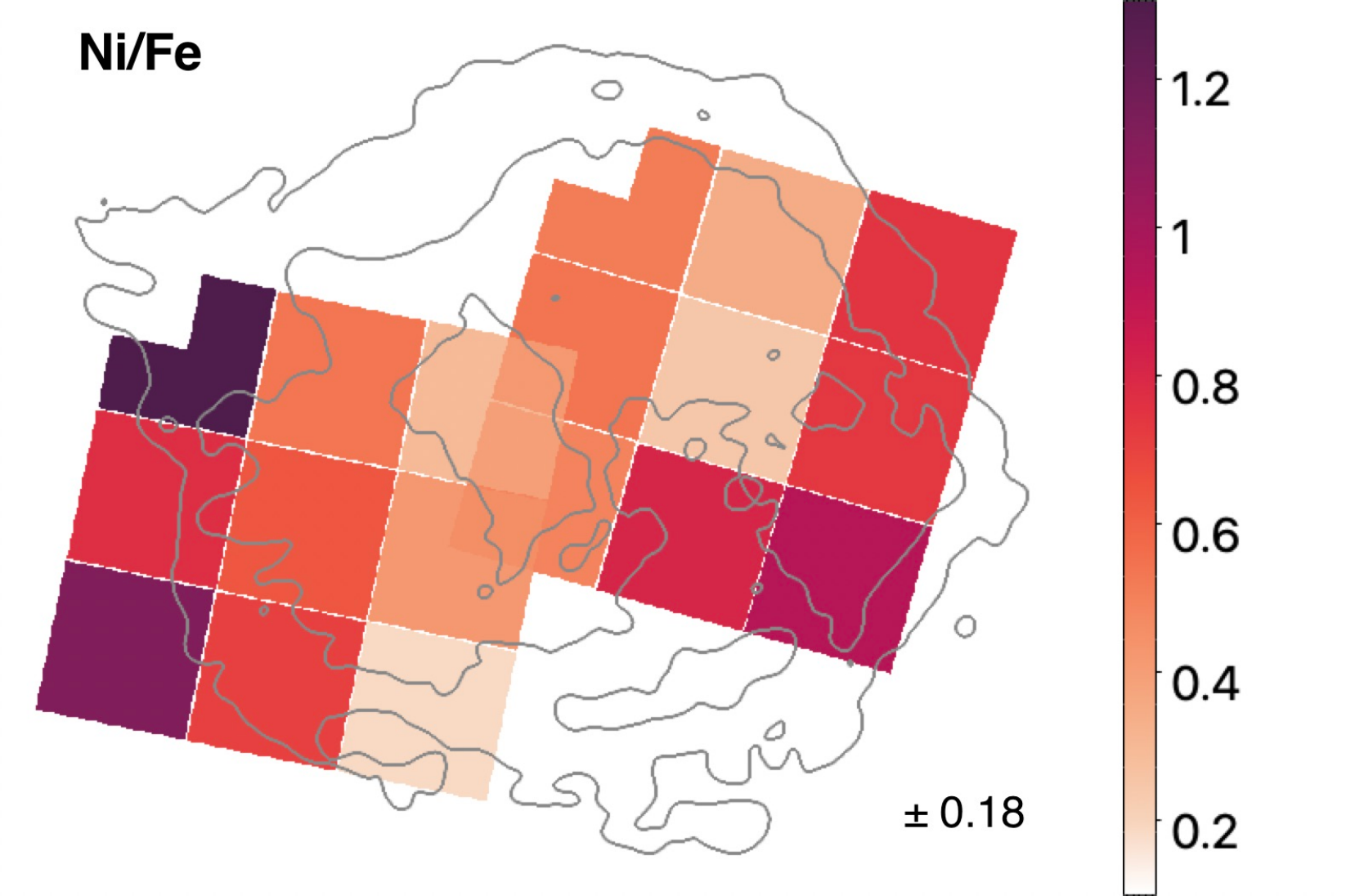}
\caption{The fitted abundance ratio of Ni to Fe compared to the solar abundance ratio (Ni/Fe)$_{\odot}$. The average one sigma error for the fitted value is indicated on the bottom right.
}
\label{fig:abundance_maps_IGEs}
\end{figure}

\section{Results and Discussion}\label{sec:results}

The fitted model consisting of two pure-metal NEI (\texttt{pshock}) components, plus a power-law continuum, reproduces the overall spectrum of each super-pixel remarkably well.
Figure~\ref{fig:pixel_e_spectra} shows two example model fits along with the constituent components for the central super-pixels of the SE and NW pointing. 
In the fitted spectrum of SE pixel {\em e}, we find some residuals for line profiles of Si Ly$\alpha$ ($\sim$2.0 keV), S He$\alpha$ ($\sim$2.4 keV), S Ly$\alpha$ ($\sim$2.6 keV), and Ca He$\alpha$ ($\sim$3.9 keV), which are indicative of the need for a more complex Doppler-velocity components to model the IMEs than a single \texttt{pshock} component as shown in \cite{vink25} and \cite{suzuki25} for \xrism/Resolve spectra. Such a model is beyond the scope of this work due to computational constraints of the Bayesian framework,
and we hope to incorporate this in a follow-up study.

\begin{figure*}
\centerline{
  \includegraphics[width=0.33\textwidth]{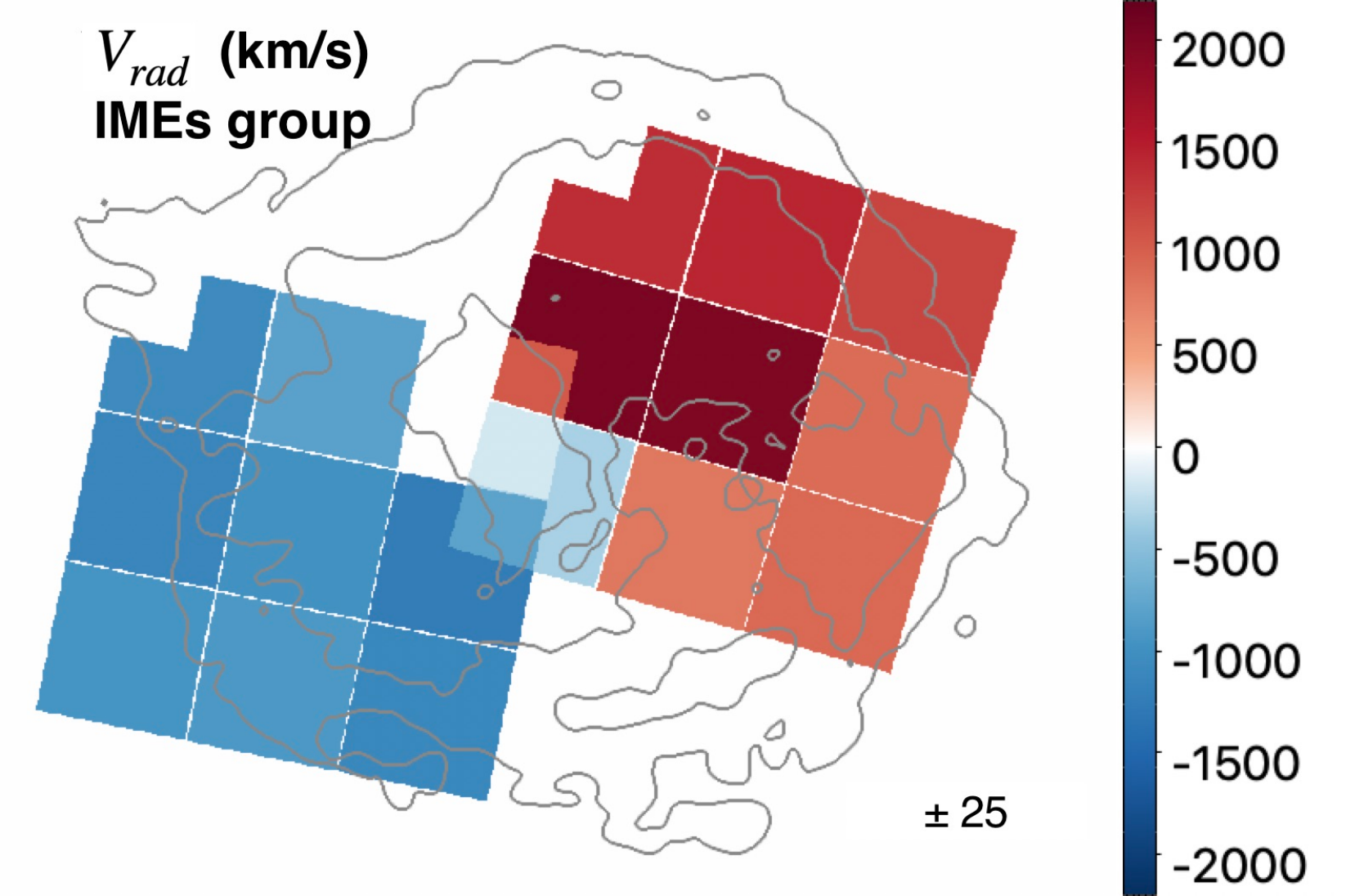}
   \hspace{2mm}
  \includegraphics[width=0.33\textwidth]{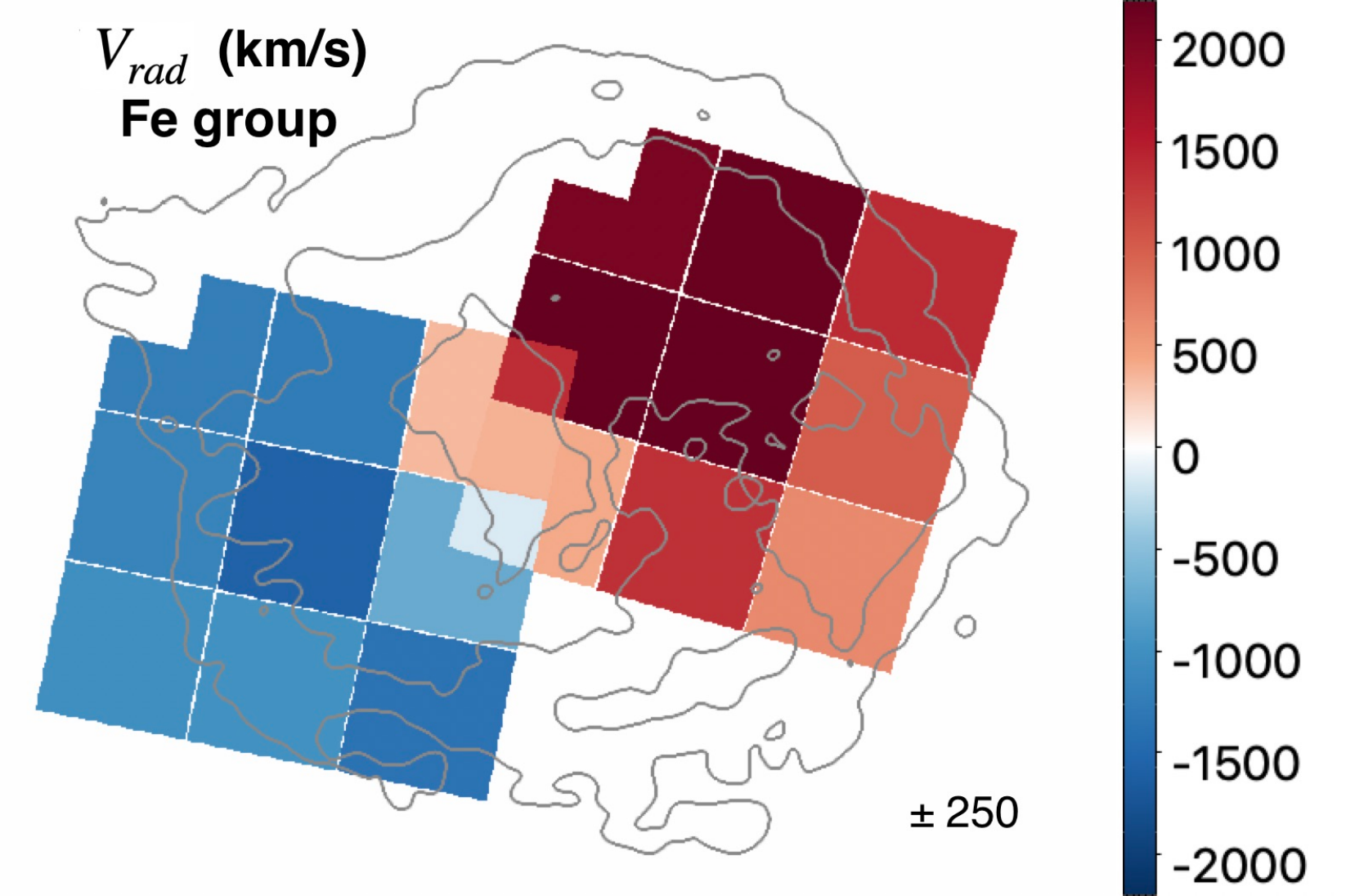}
   \hspace{2mm}
  \includegraphics[width=0.33\textwidth]{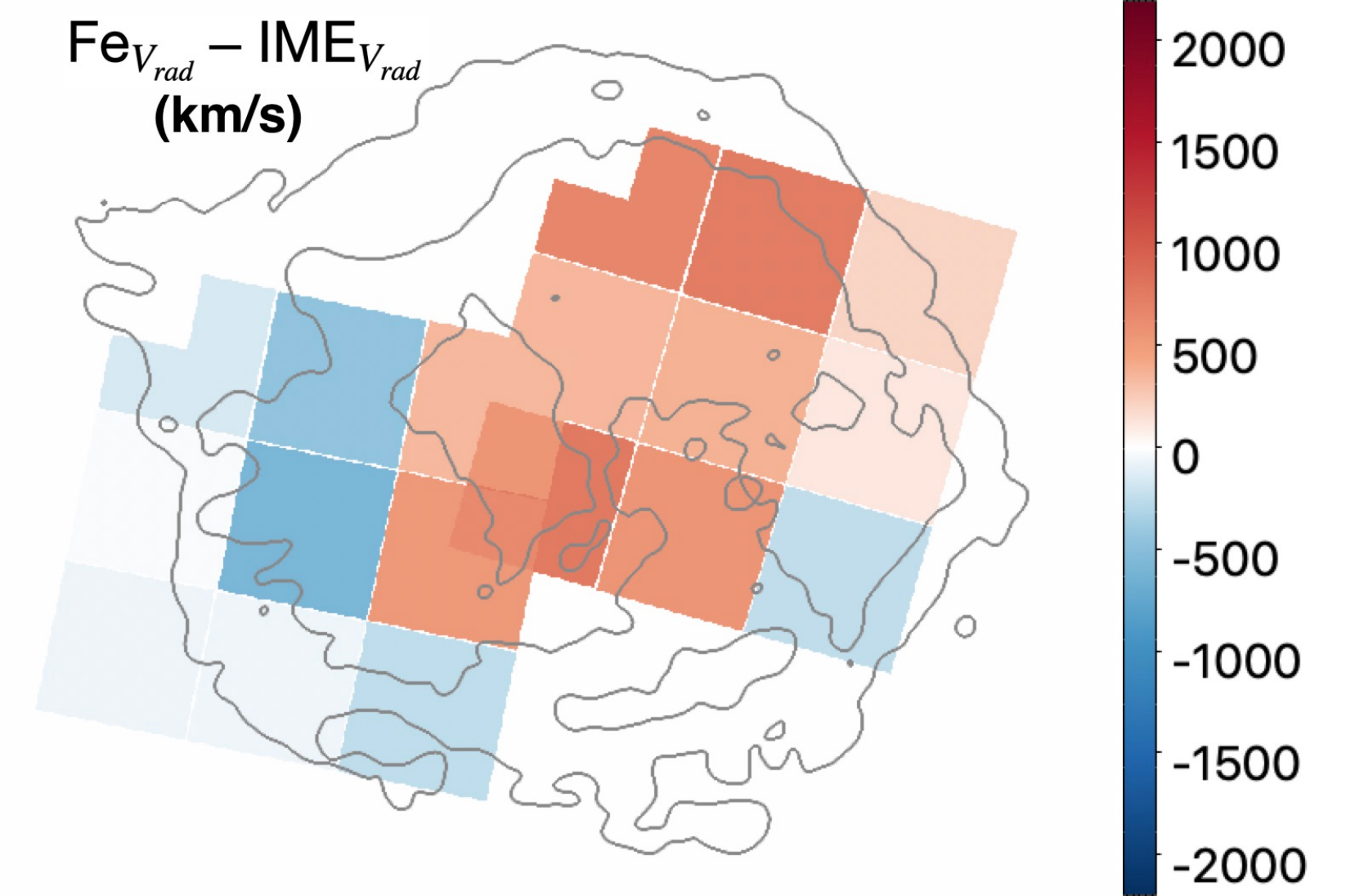}
}
\vspace{3mm}
\centerline{
  \includegraphics[width=0.33\textwidth]{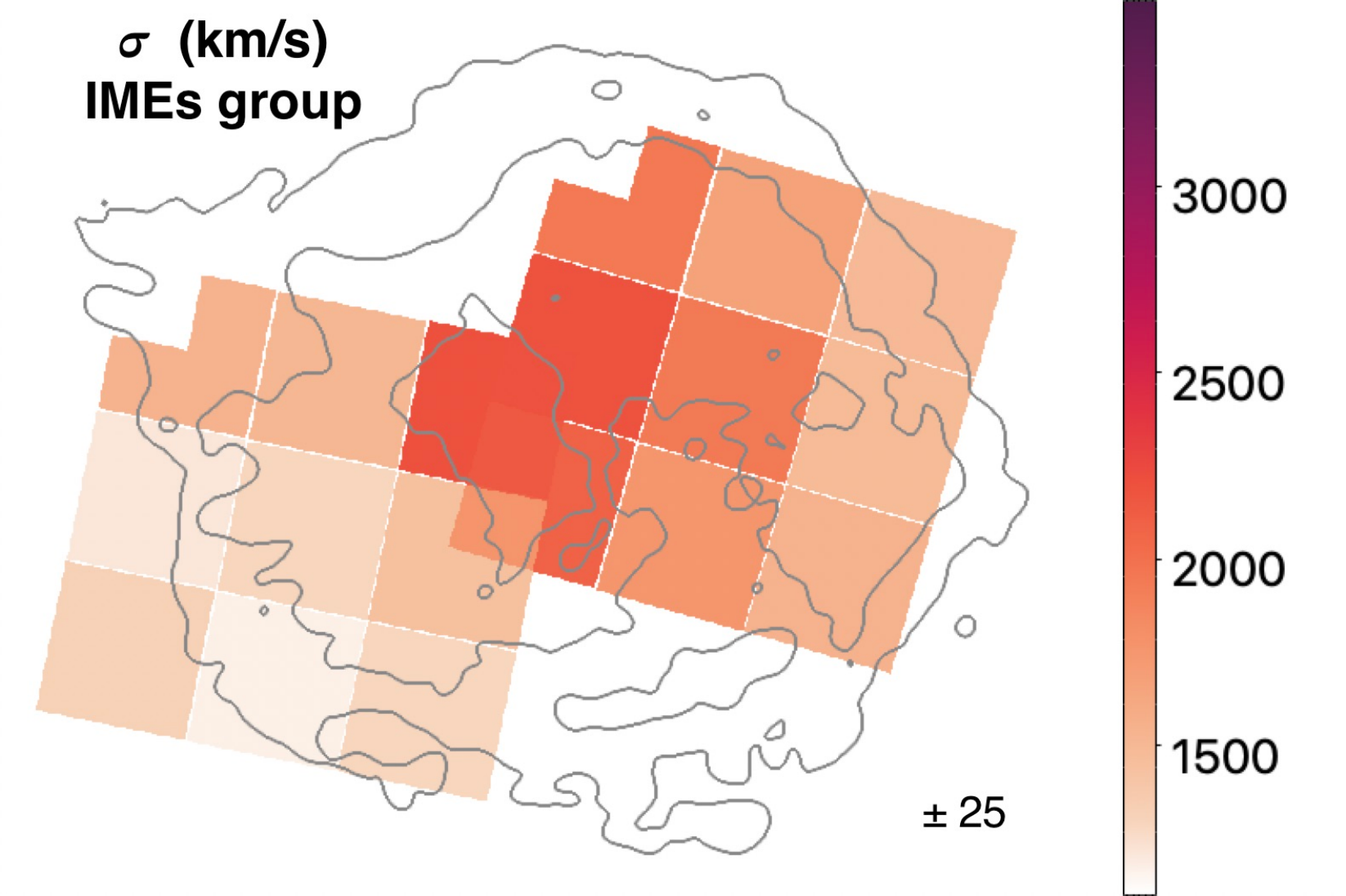}
   \hspace{2mm}
  \includegraphics[width=0.33\textwidth]{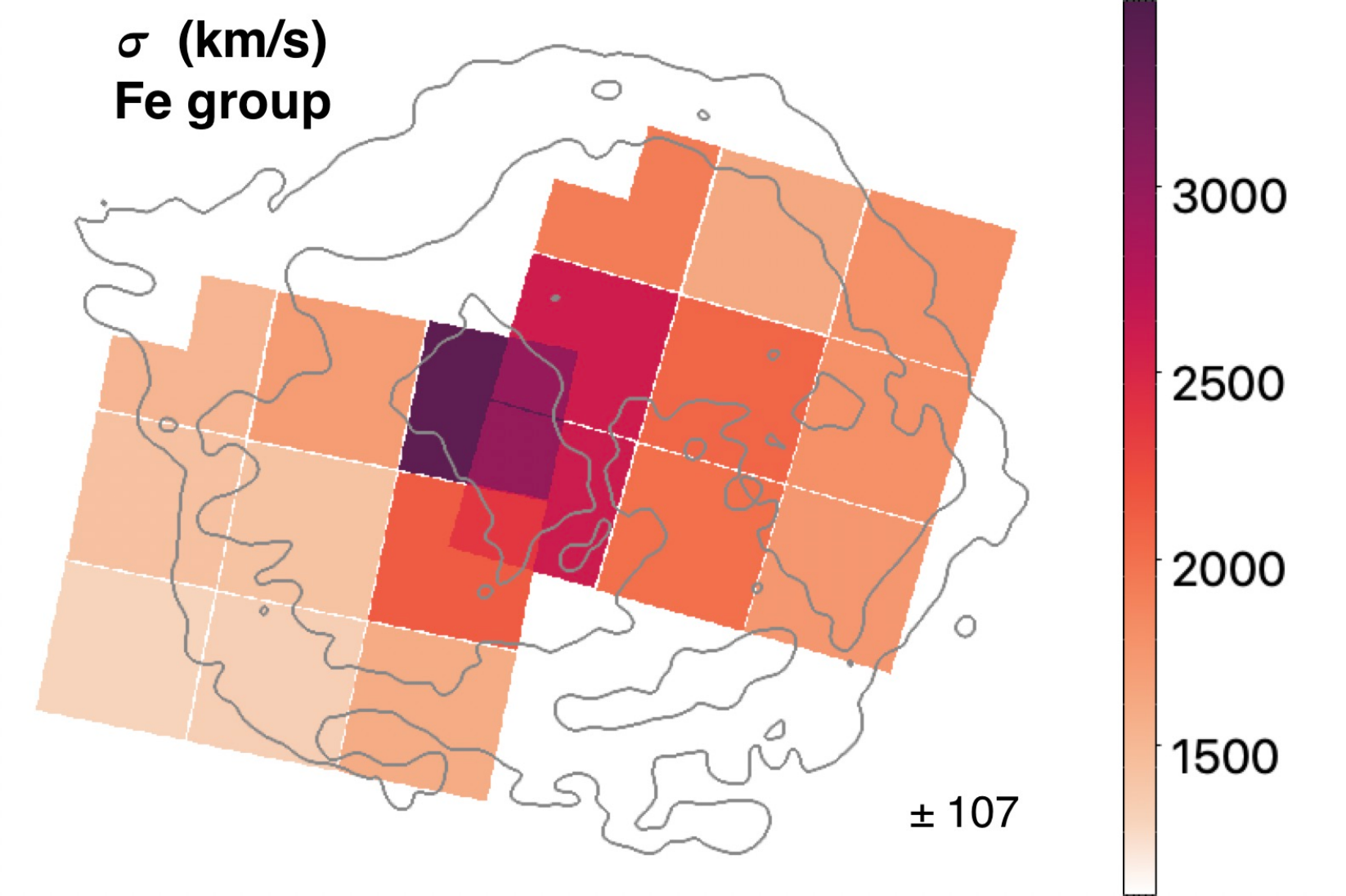}
   \hspace{2mm}
  \includegraphics[width=0.33\textwidth]{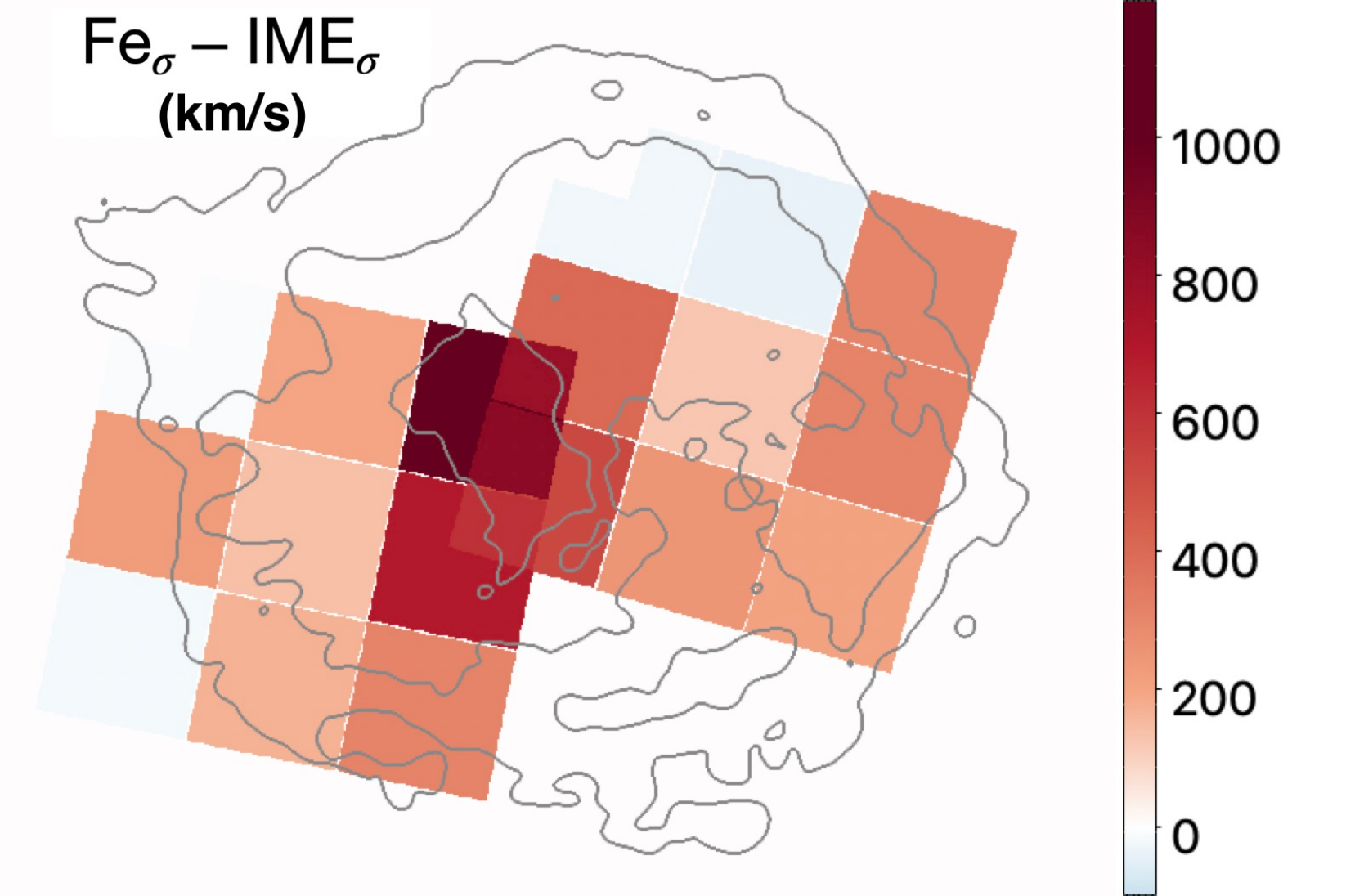}
}
\caption{
Top row: The fitted redshift maps for the IMEs (top left) and IGEs (top center) \texttt{pshock} components are shown. The map of difference in velocites (IGEs--IMEs) between the two is shown in the top right plot.
Bottom row: The maps of fitted velocity dispersion measurements for the IMEs, IGEs and their difference (IGEs--IMEs) are presented left to right.
All values are reported in units of \kms and the average 1 sigma errors are marked at the bottom right of each plots.
}
\label{fig:velocity_maps}
\end{figure*}

We also see smaller residuals near energies 2.2 keV, 2.8 keV, and 3.5 keV, which can be explained by the presence of rare elements (P, Cl, K) that were not included in our current model but are detected in the \xrism\ data \citep{plucinsky25, xrism_casA_2025}. 
The fitted spectrum of the NW pixel {\em e} does not show any sharp peaks in the residuals.
The spectrum for all other super-pixels with the fitted models and residuals is given in Figures~\ref{fig:SE_spectra} and \ref{fig:NW_spectra} in the Appendix. 
A map of the C-statistic over degrees of freedom (cstat/dof) is also given in the Appendix (Figure~\ref{fig:extra_maps}), which shows the overall good quality of the spectral fits.
Our current model is a useful approximation to the observed high-resolution spectra, which can constrain the average chemical and physical plasma properties across the remnant.
The fitted model parameters for each super-pixel are listed in the Tables~\ref{tab:SE_median_par} and \ref{tab:NW_median_par}, and are represented as a montage of maps in Figures~\ref{fig:abundance_maps_IMEs},~\ref{fig:abundance_maps_IGEs},~\ref{fig:plasma_parameter_maps},~\ref{fig:velocity_maps} and \ref{fig:powerlaw_maps}. 
These model parameter values are the median values of the parameter posterior distribution (Appendix~\ref{app:corner_plots}) obtained from the fits using {\em UltraSPEX}.

\subsection{Elemental abundance ratios}

Our spectral model has the abundances of Si and Fe fixed to 100,000 times their solar values \citep[as per][]{lodders09} to model pure-metal plasmas. Thus, the model fits only constrain the element abundance ratios -- with respect to Si for the IMEs component and with respect to Fe for the Fe-group component -- instead of the absolute element abundances. We report the fitted abundance ratio with respect to their solar abundance ratio; for example, the S to Si ratio is $(S/Si)/(S/Si)_{\odot}$. 
The abundance ratio maps for IMEs (S/Si, Ar/Si and Ca/Si) and Ni/Fe are shown in Figure~\ref{fig:abundance_maps_IMEs} and Figure~\ref{fig:abundance_maps_IGEs}, respectively.
The map of O$^*$/Si is given in the Appendix (Figure~\ref{fig:extra_maps}) which shows that the abundance ratio is well constrained with most regions within $\pm$0.5 times solar ratio, despite the absence of O spectral lines in the fitted energy range, and thus indicating the robustness of the fitted results.

The S/Si abundance ratio map shows only a small variation over the face of the remnant, with values ranging from 0.88 to 1.12. This indicates that S traces Si, which is expected as they are produced in the same nucleosynthesis layer and have a similar spatial distribution as seen using \chandra\ data \citep{hwang00,willingale02, yang08}. The Ar/Si and Ca/Si maps show more variation with a higher than solar abundance ratio in the northeast and southwest, that is, along the directions of the Si-rich jet-like structures in Cas~A \citep{hwang04, vink04a}. This is most prominent for Ca in the southwest, which is also the region with the most nonthermal contribution. To test if this is an artifact of spectral fitting with possible dependence on the power-law component, we performed the spectral fits again for all the super-pixels with power-law index fixed to 3.2 and still got high abundance ratios for Ar/Si and Ca/Si in the same regions.
In addition, we observe a discrepancy in the velocity measurements of Ca with respect to the IMEs (Appendix~\ref{app:ca_velocities}), which also indicates that Ca and Si/S do not directly trace each other and our model comprising all IMEs in a single \texttt{pshock} component is a simplification of the true picture.

High ratios of Ar/Si and Ca/Si at the base of the jets suggest
it consists of 
incomplete Si-burning material \citep{ikeda22}, which is consistent with the fact that the jets are rich in Si and poor in Ne/Mg \citep{vink04a,hwang04} and show high Cr/Fe ratio
\citep{sapienza25}. That said, the limited spatial resolution prevents us from being definitive about attributing these enhancements to the jets, indicating that further work is necessary.

The abundance map of Ni/Fe shows the ratio to be either below or consistent with the solar ratio everywhere except for the base of the NE jet (SE pixel {\em a}) where the abundance ratio is $1.31^{+0.25}_{-0.26}$. 
This corresponds to a mass ratio of Ni/Fe $=0.08^{+0.015}_{-0.016}$.

Note that there is an inconsistency reported in plasma modeling of Ni between the latest atomic databases -- \texttt{XSPEC AtomDB v3.1.0} \citep{arnaud96} and \texttt{SPEX v3.08.01$^*$}, with \texttt{AtomDB} fits resulting in higher abundances for Ni, see \cite{plucinsky25} for details. In our analysis with \texttt{SPEX}, despite its tendency to obtain lower Ni abundances, we still measure high mass ratios for Ni/Fe.

This high ratio of Ni/Fe near the NE jet indicates an enrichment by products of the complete Si burning region near the explosion center. In \cite{ikeda22}, the elemental abundances in the jet region suggest that the jet was formed in the incomplete Si-burning regime. However, our results may indicate mixing of material from the innermost complete Si-burning region into the jet \citep[see also][]{sapienza25}. Including a possible connection with the high-entropy plumes discussed in \cite{sato21}, further detailed investigations of the elemental composition in the jet region with \xrism\ are anticipated, requiring complete coverage of Cas~A.
However, in the meantime, an analysis of the jet region using the Xtend data of the existing \xrism\ observations is being analyzed (Kurashima et al., in prep). An analysis of the other Fe-peak elements (Ti, Cr, Mn, Ni) using Resolve data is presented in \cite{sato26}.


\subsection{Doppler shifts and Doppler broadening}\label{subsec:doppler_shift_and_broadening}

\begin{figure}
  \includegraphics[width=\columnwidth]{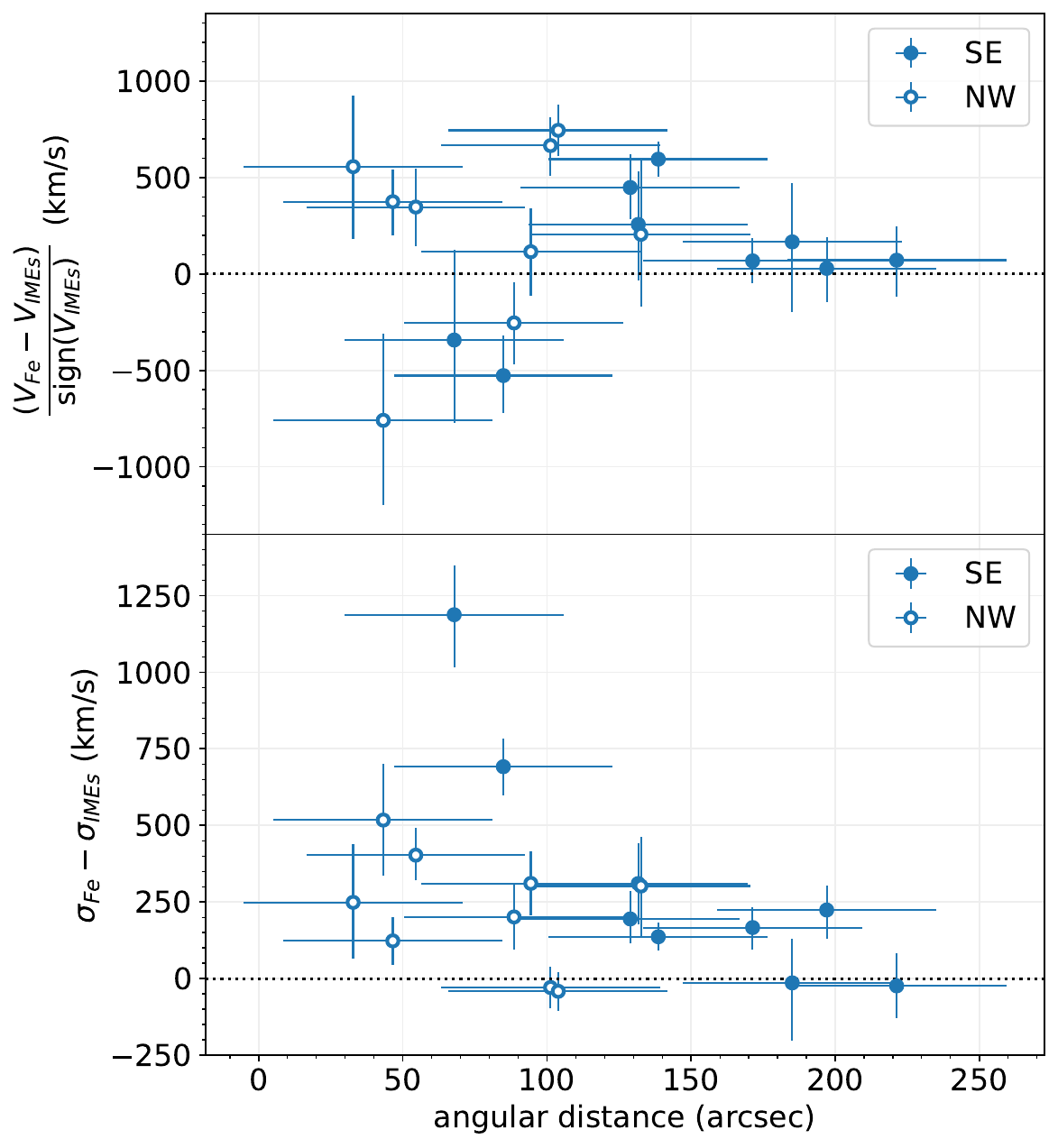}
\caption{Top panel: Scatter plot of difference in Fe-group Doppler velocity and IMEs Doppler velocity versus the projected angular distance.
Here, sign$(V_{\rm IMEs})$ denotes the sign of the Doppler velocity of the IMEs, taken to be positive for redshifted and negative for blueshifted velocities.
Bottom panel: Scatter plot of difference in Fe-group Doppler broadening and IMEs Doppler broadening versus the projected angular distance.
The dotted line marks the line of no difference between the Fe-group and IMEs. The error bars on the angular distance have a size of $38\arcsec$, which is about half the angular resolution of \xrism.
}
\label{fig:delta_z_delta_sig_vs_angdist}
\end{figure}

\subsubsection{Measurements}\label{subsubsec:doppler_measurements}

The asymmetric bulk Doppler velocity between the SE and NW regions has been evident since the early X-ray observations of Cas~A \citep{markert83, holt94,hwang01b,willingale02}. This was further confirmed using \xrism/Resolve data \citep{vink25,bamba25,suzuki25} which eliminated systematic effects such as blended lines or ionization variations as potential causes of the asymmetry.

The asymmetry of Doppler velocities reported here is consistent with other studies:
Figure \ref{fig:velocity_maps} (top row) shows that the IMEs component velocities range between -1250 \kms\ and 2000 \kms\ and Fe-group velocities are in the range of -1700 \kms\ to 2400 \kms. 
We measure higher absolute values of the radial velocities in the NW for both IMEs and IGEs.

In most regions, the Fe-group is observed to be faster than the IMEs, i.e., more blueshifted in the blueshift regions and more redshifted in the redshift regions, except for two super-pixels near the explosion center in the SE direction --- SE pixel {\it f} and NW pixel {\it g}, and a super-pixel near the SW Si-rich jet --- NW pixel {\it i}, where Fe-group is slower as shown in Figure~\ref{fig:velocity_maps} (top right) and Figure~\ref{fig:delta_z_delta_sig_vs_angdist} (top panel). 
Moreover, the figures show that the difference in velocity decreases with increasing angular distance from the explosion center.
The higher Doppler velocity measured for the Fe-group suggests that there could be more regions where Fe ejecta have protruded past the IMEs, 
similar to Fe knots in the SE regions of Cas~A \citep{hughes00a}. A three-dimensional velocity map of the Fe ejecta will be presented in a forthcoming paper (Bamba et al., in preparation).

We measure Doppler broadening for the IMEs and Fe-group components as presented in the bottom row of Figure~\ref{fig:velocity_maps}. The broadening for the Fe-group is measured to be as high as $\sim$ 3700 \kms, compared to a maximum broadening of $\sim$ 2200 \kms\ for the IMEs component. 
Despite these high Doppler broadening values in the central regions, we do not find hints of double peaked line profiles (indicating the two sides of the reverse shocked ejecta shell) in the fitted residuals, as also reported by \cite{vink25} using a more detailed line profile fitting.
Similarly to Doppler velocity, the differences in Doppler broadenings between IMEs and IGEs are higher near the center.
The lower panel of Figure~\ref{fig:delta_z_delta_sig_vs_angdist} shows that the Fe-group is generally broader than the IMEs, except for a few regions (super-pixels in blue on the bottom right plot of Figure~\ref{fig:velocity_maps}) where it is consistent with the same Doppler broadening. 

These differences in the kinematic properties of IMEs and IGEs are suggested by \citet[][hereafter BA25]{bamba25} but were not conclusive due to the high uncertainties in their analysis at per Resolve pixel level compared to our Bayesian fits at a $2\times2$ pixel resolution. Additionally, we fit a single broadband physical mode as opposed to BA25 where they compare two independent narrow-band fits to the Fe-K complex (fit with a \texttt{pshock} model) and the He-like Si/S lines (fit with Gaussians).

BA25 reported blueshifted narrow Fe-K lines with Doppler broadening below 850 \kms\ for individual Resolve pixels near the eastern periphery of the remnant. In our analysis, we also find relatively reduced broadening for the Fe-group component in the southeastern pixels. However, the lowest value we measure is 1300 \kms\ for the SE pixel {\em g}, suggesting that the narrow features identified by BA25 likely correspond to small-scale structures not captured at our spatial resolution.

For IMEs, detailed line-profile fitting revealed the presence of a narrow line component for the NW pixel {\em h} \citep{vink25}.
This characteristic was reported for the Si/S Ly$\alpha$ and He$\beta$ transitions. The spectral model in our current analysis cannot reproduce this detail, as all IME lines are fitted with a single line broadening component. Inspecting the spectral fit residuals of the Si/S line emission does reveal narrow features not captured by the current model.
The IMEs component in our model assumes the same Doppler shift and Doppler broadening for all the IMEs. The detailed analysis of Si and S by \cite{vink25} and \cite{suzuki25} using data from \xrism\ showed that the true picture is more complex with different Doppler velocities for H- and He-like lines.

In addition, while investigating the best-fits, we remarked a few super-pixels ({\em d}, {\em e}, and {\em h}) in the SE with sine wave-shaped residuals for Ca He$\alpha$, indicating a higher blueshifted velocity for Ca compared to the rest of the IMEs. To quantify this velocity shift, we fit the redshift of the IMEs component only to the Ca He$\alpha$ line in the 3.8--4 keV energy range. The greatest improvement in the C-statistic is measured for the SE pixel {\em e} with $\Delta$C-stat $= 93$ for the redshift $-1445^{+45}_{-32}$ \kms. The highest velocity difference was measured for the SE pixel {\em d} with Ca at $-1656^{+103}_{-88}$ \kms\ compared to the overall IMEs velocity of $-1125^{+5}_{-17}$ \kms. More details on this analysis, along with the fitted spectra and the Doppler velocity shifts for Ca He$\alpha$ for these three SE super-pixels are given in the Appendix~\ref{app:ca_velocities}.
A similar residual is not apparent for Ar He$\alpha$; it should be noted that this line is contaminated with some prominent lines of H- and He-like S, which could blend out the difference in redshift.

The differences in Doppler velocities, at least for a few regions in the SE, as well as the non-uniform Ca/Si abundance ratios (Fig.~\ref{fig:abundance_maps_IMEs}) suggest that Ca may originate,
at least partially, from another layer of the supernova than the bulk of the IMEs. 
Si is mainly produced during (explosive) O-burning, whereas Ca is primarily produced  during incomplete Si-burning \citep[e.g.,][]{curtis19}. 
The higher Doppler velocity of Ca in the SE also suggests that on average Ca overtook
the Si-rich ejecta in the SE region.

\subsubsection{Thermal Doppler broadening}\label{subsubsec:thermal_doppler_broadening}

The measured line broadening ($\sigma_{\rm IMEs}$ and $\sigma_{\rm IGEs}$) is a combination of Doppler velocity broadening and Doppler thermal broadening, with the (kinematic) velocity broadening dominating throughout the remnant. However, towards the periphery of the remnant, the line-of-sight motions are minimal, and thus the contribution of thermal Doppler broadening could become significant. 
Although the outermost regions in Fig.~\ref{fig:velocity_maps} show high Doppler shifts for both IMEs and IGEs, as seen in Fig.~\ref{fig:delta_z_delta_sig_vs_angdist} the redshifts of IMEs and IGEs are consistent with each other at large angular distances, however, their Doppler broadening are significantly different (IGEs-IMEs $\sim$200--300 \kms), for NW super-pixel {\em c} and SE super-pixels {\em d} and {\em h}.

Attributing this difference solely to thermal Doppler broadening may provide a hint at differences in the shock velocity with which IMEs and IGEs have been shocked in these regions, providing a view into their different
reverse shock histories.
To do so, we assume that the Doppler velocity broadening is the same for IMEs and IGEs in these regions and attribute the difference in total broadening to thermal broadening \citep[$\sigma_{v,i,th}^2=kT_i/m_i$, e.g.][]{vink25}. 
There are some caveats,
since the IMEs ang IGEs have different ejecta distribution, which can result in different Doppler velocity broadening even with the same Doppler shifts, but we are focusing only on outermost regions where the effect of bulk velocity broadening is expected to be small.

Under this assumption and taking Fe and Si as representative elements of the IGEs and IMEs, respectively, we get 
\begin{equation}
    \sigma_{\rm IGEs}^2 - \sigma_{\rm IMEs}^2 = \frac{kT_{\rm Fe}}{m_{\rm Fe}}- \frac{kT_{\rm Si}}{m_{\rm Si}}
    \approx \frac{1}{m_{\rm Si}} \left(\frac{kT_{\rm Fe}}{2}- kT_{\rm Si}\right)
    \label{eq:delta_sigma_IGEs-IMEs},
\end{equation}
which sets a lower limit on the difference in ion temperatures ($kT_{\rm Fe} - kT_{Si}$).
We use here the fact that $m_{\rm Fe}\approx 2m_{\rm Si}$.
Evaluating Eq.~\ref{eq:delta_sigma_IGEs-IMEs}, we find that the ion temperatures of Fe are more that of Si by $300\pm180$ keV and $150\pm60$ keV, in the NW super-pixel {\em c} and SE super-pixels {\em d} \& {\em h}, respectively. 

A difference in thermal broadening between IMEs and IGEs,
under the assumption of non-equilibration of temperatures, implies different shock velocities, 
with the temperature for each species $i$ when heated by a high-Mach-number shock is given by $kT_{\rm i} = (3/16) m_{\rm i} V_{\rm S,i}^2 $, where $V_{\rm S,i}$ is the shock velocity with which specie $i$ is shocked. Expressing Eq.~\ref{eq:delta_sigma_IGEs-IMEs} in terms of shock velocities,
\begin{equation}
    \sigma_{\rm IGEs}^2 - \sigma_{\rm IMEs}^2 = \frac{3}{16} \left(V_{\rm S,IGEs}^2-V_{\rm S,IMEs}^2\right)
    \label{eq:delta_sigma_IGEs-IMEs_wrt_shock_vel},
\end{equation}
shows that the shock velocity of IGEs is higher than that of IMEs with $V_{\rm S,IGEs}^2 = (2300\pm700)^2+V_{\rm S,IMEs}^2$~\kms\ and $V_{\rm S,IGEs}^2 = (1600\pm400)^2+V_{\rm S,IMEs}^2$~\kms\ for the NW and SE super-pixels, respectively. 
It also sets the lower limit on the shock velocity for IGEs. Since Eq.~\ref{eq:delta_sigma_IGEs-IMEs_wrt_shock_vel} is applicable right after the shock but the ions have already started to equilibrate, the true difference in shock velocity could be even slightly higher. 

The difference in thermal broadening, therefore, suggests that
IGEs were heated by faster shocks than IMEs. 
In the canonical semi-analytical models for SNRs evolving
in a steady wind, see Appendix~\ref{app:ionization_age}, the reverse shock velocity in the frame of the ejecta is constant about $\sim$40~yr after the explosion.
In this scenario, the IGEs could have been shocked very early on. But it could also be that Cas~A had a more complicated reverse shock history. Another suggestion is that clumping could have affected the shock velocity of the IMEs. Shocks moving into a clump reduce their shock velocity. In Section~\ref{subsubsec:net_vs_kT-implications_of_anticorrelation} we provide a further discussion on the effects of clumping in the context of the measured $n_{\rm e}t$ values.

\subsection{Ionisation timescales and electron temperatures}\label{subsec:net_vs_kT}

\begin{figure*}
\vspace{3mm}
\centerline{
  \includegraphics[width=0.33\textwidth]{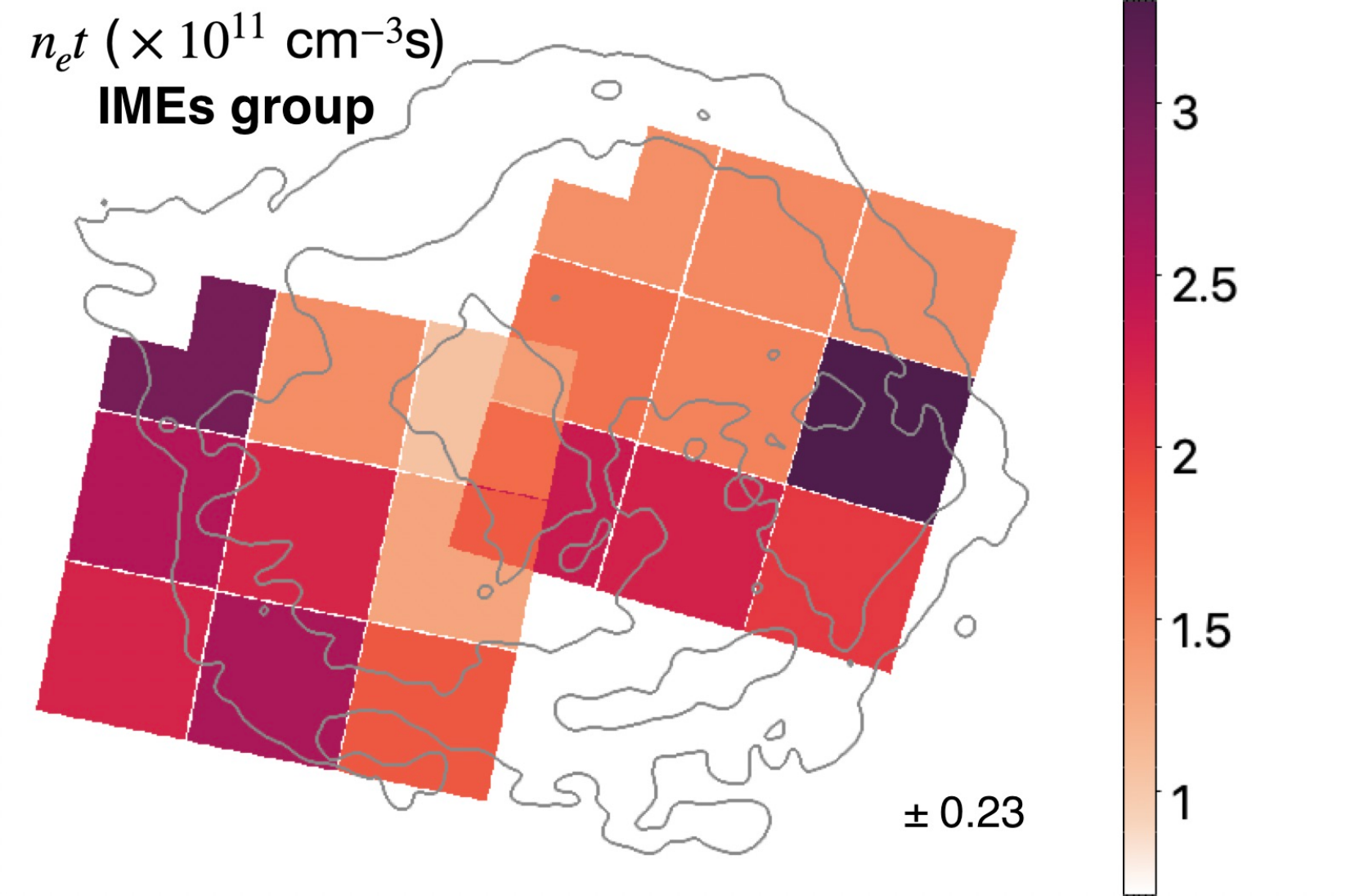}
   \hspace{2mm}
  \includegraphics[width=0.33\textwidth]{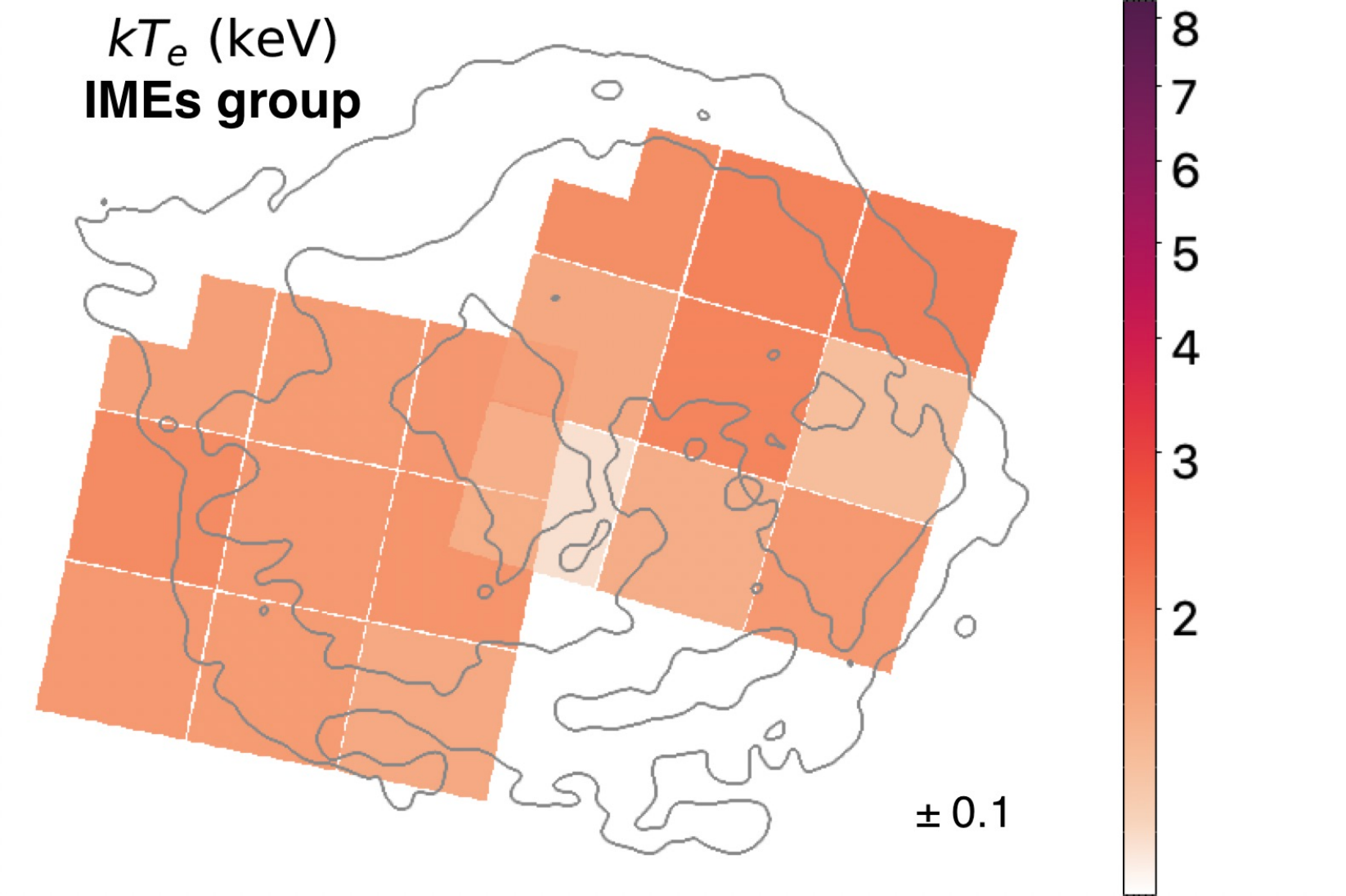}
   \hspace{2mm}
  \includegraphics[width=0.33\textwidth]{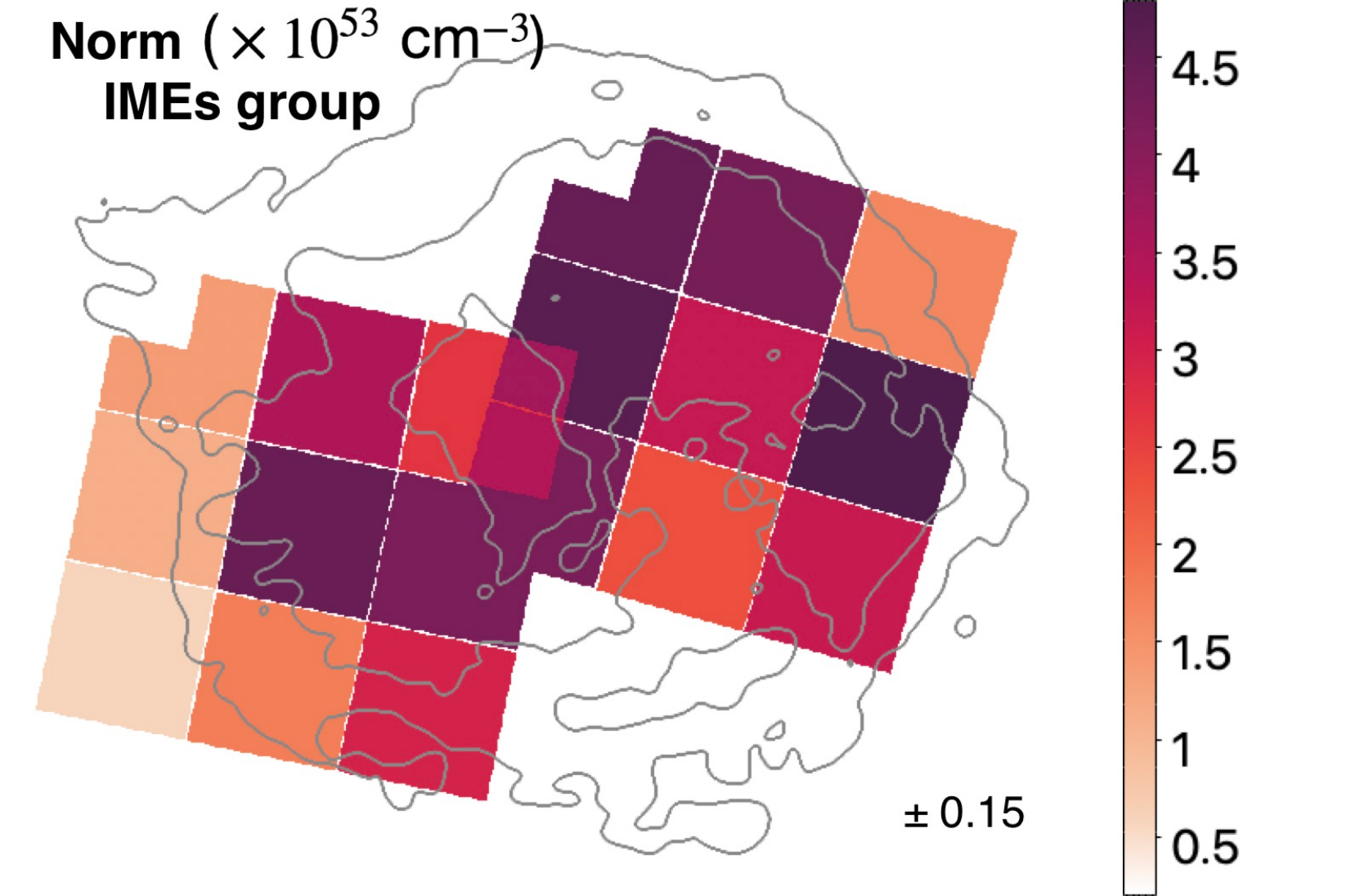}
}
\vspace{3mm}
\centerline{
  \includegraphics[width=0.33\textwidth]{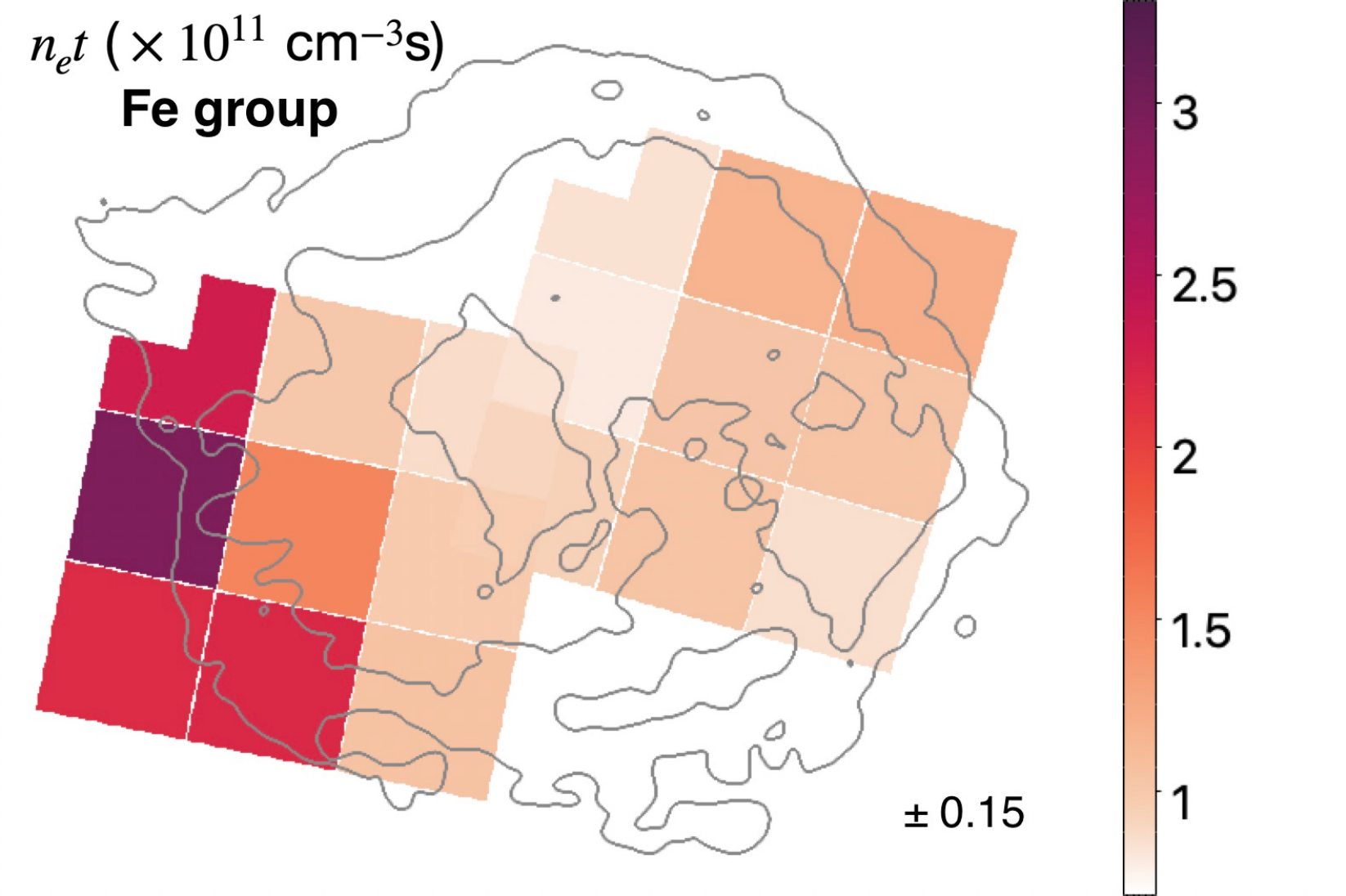}
   \hspace{2mm}
  \includegraphics[width=0.33\textwidth]{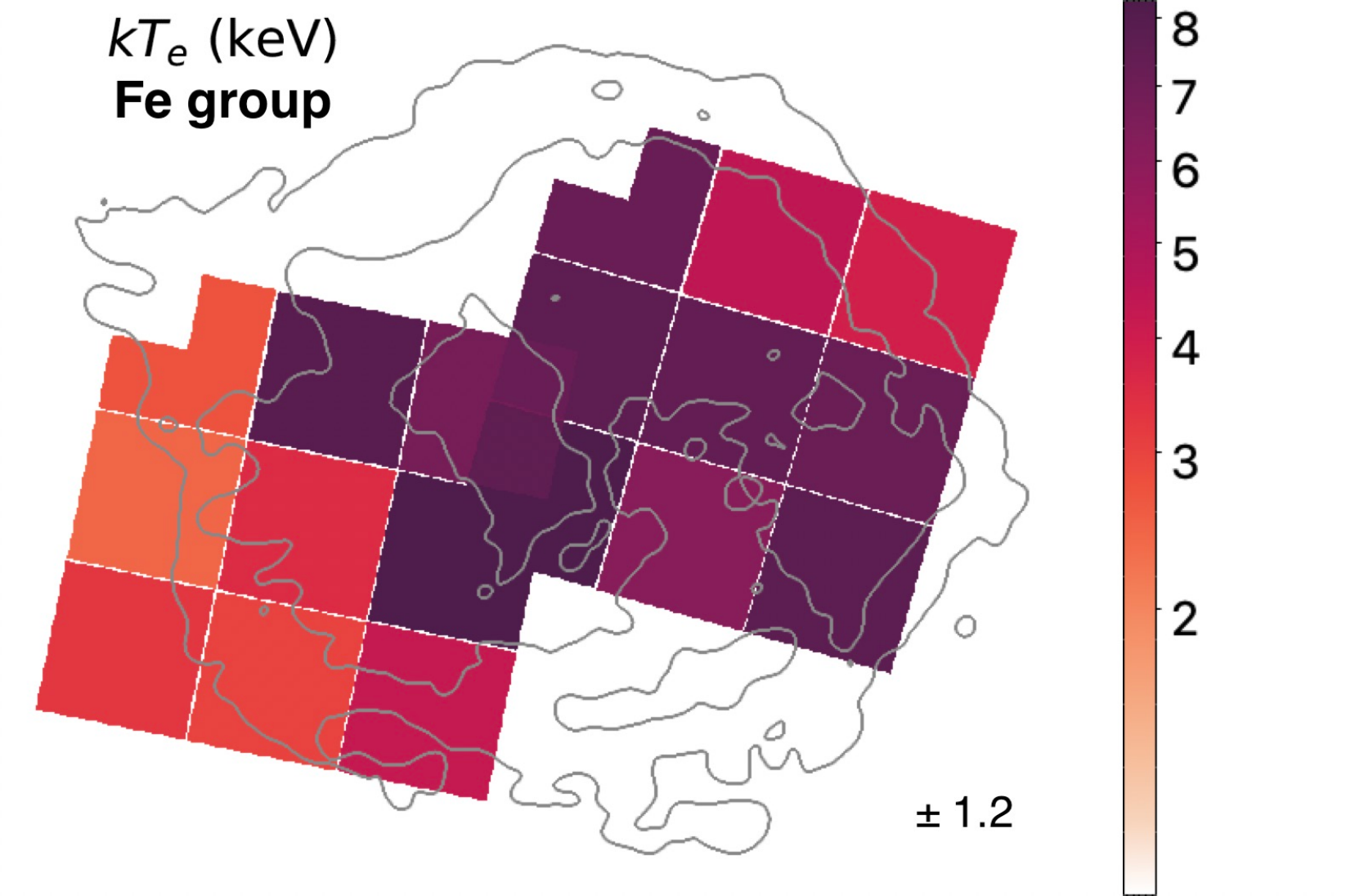}
   \hspace{2mm}
  \includegraphics[width=0.33\textwidth]{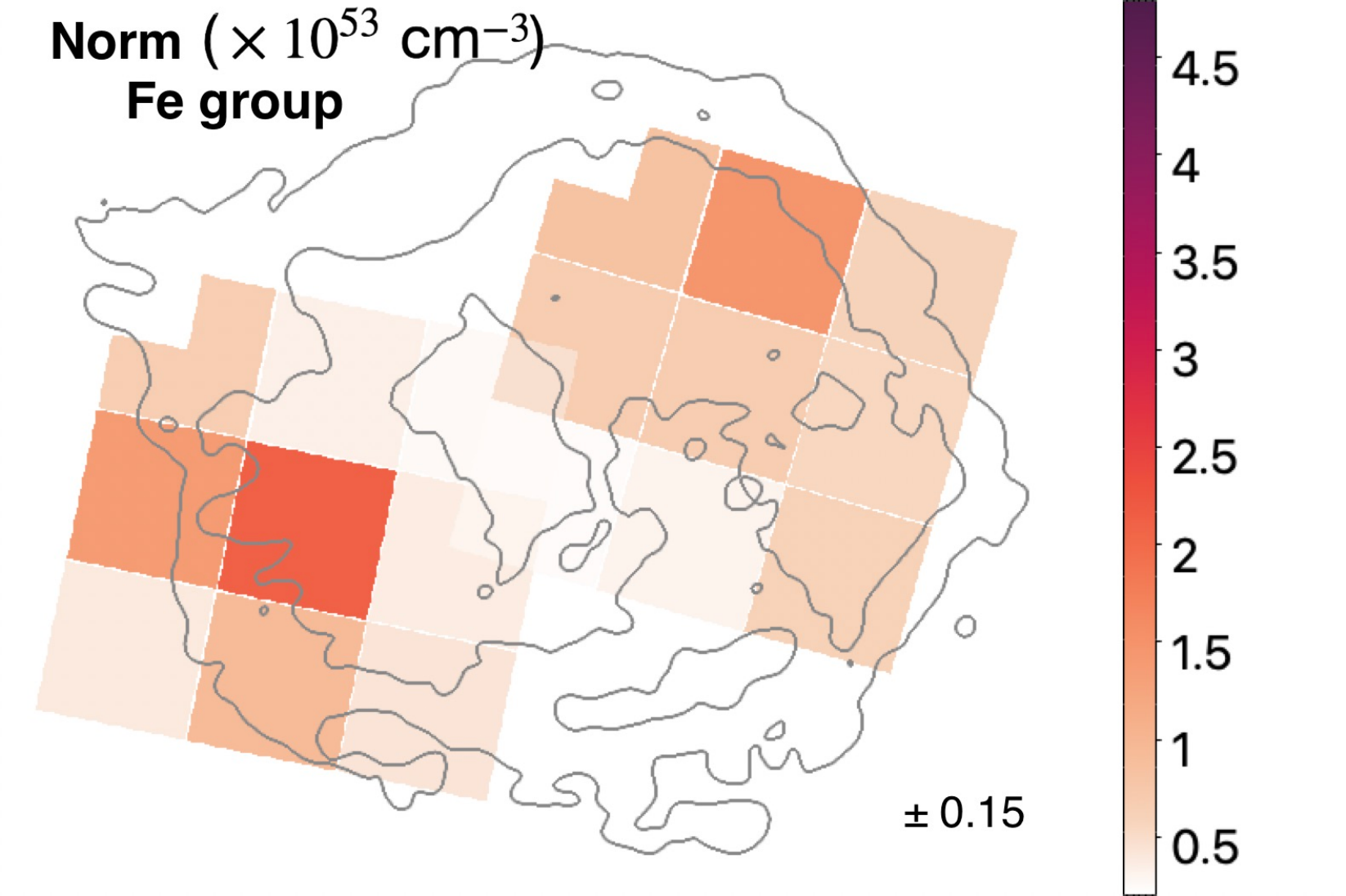}
}
\caption{
Top row: left to right are the maps of fitted model parameters -- ionisation timescale ($n_{\mathrm{e}}t$), electron temperature ($kT_\mathrm{e}$), and emission measure (normalization), respectively, for the IMEs component modeled as pure-metal plasma.
The average one sigma error is indicated on the bottom right of each subplot.
Bottom row: Same as the top row plots but for Fe-group component.
}
\label{fig:plasma_parameter_maps}
\end{figure*}

\begin{figure}
  \includegraphics[width=\columnwidth]{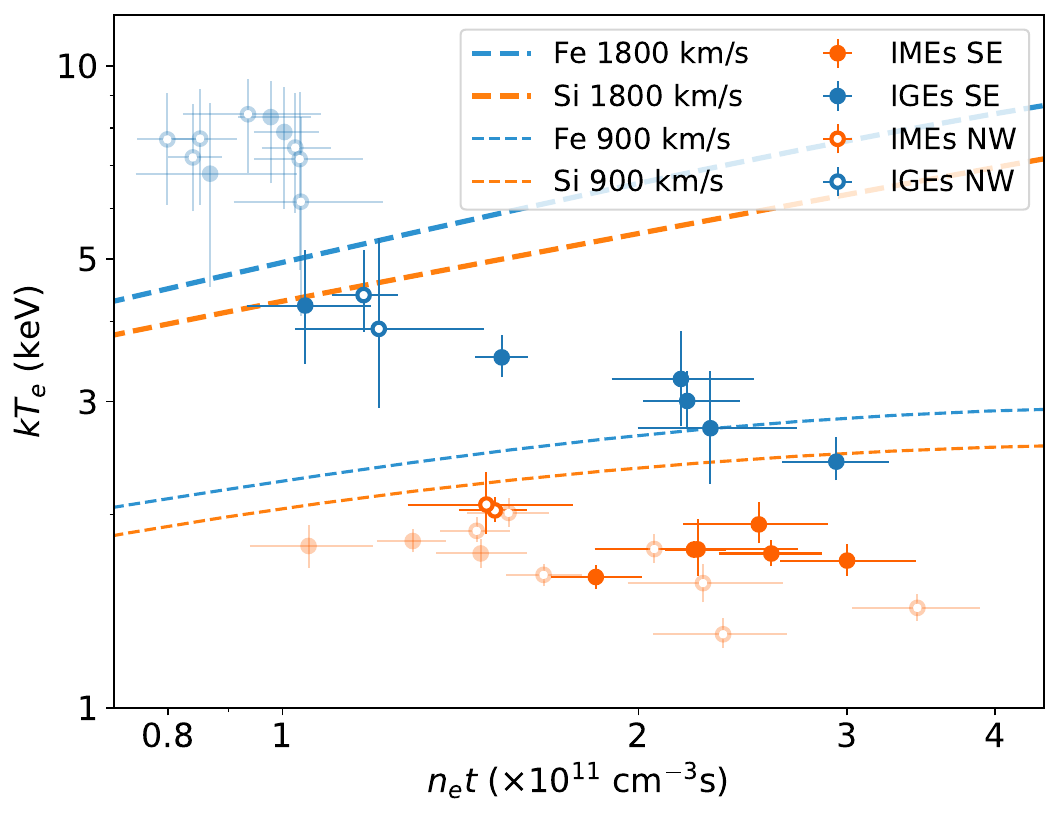}
\caption{
The distribution of the fitted $n_{\rm e}t$ and $kT_{\rm e}$ for the IMEs (orange markers) and IGEs (blue markers). The superpixels in the SE and NW are marked with solid and open circles, respectively. The low opacity points are the regions where the $kT_{\rm e}$ parameter of IGEs component is biased towards higher values due to high synchrotron emission (as detailed in Section~\ref{subsubsec:net_vs_kT-measurements}).
The thermal equilibration of shock ejecta is traced using the numerical models by \citet{ohshiro24} for pure Fe ejecta (blue dashed line) and pure Si ejecta (orange dashed line). The evolution tracks are shown for the expected reverse shock velocity --- 1800 \kms\ (thick line) --- and half of that --- 900 \kms\ (thin line).
}
\label{fig:net_vs_kt}
\end{figure}

\subsubsection{Measurements}\label{subsubsec:net_vs_kT-measurements}
Our spectral modeling reveals several noteworthy trends in the ionisation timescales ($n_{\mathrm{e}}t$) and electron temperatures ($kT_\mathrm{e}$) of the ejecta.
Figure~\ref{fig:plasma_parameter_maps} presents the maps of $n_{\mathrm{e}}t$, $kT_\mathrm{e}$, and the emission measure (EM; normalization) for the IMEs (top row) and the IGEs (bottom row).
The $n_{\mathrm e}t$ for the IMEs ranges between $1\times10^{11}$--$3.4\times10^{11}$ cm$^{-3}$s and shows a clear spatial dependence: highest values at the base of the two opposing Si-rich jet-like structures of Cas~A, and lowest near the southeastern inner region (SE pixel {\em c}). Interestingly, this same super-pixel has been identified as exhibiting strong Li-like emission lines of Si and S with \xrism\ \citep{vink25}, consistent with our finding that the plasma has not yet reached higher stages of ionization here. 
The $n_{\mathrm e}t$ for the IGEs spans between $8\times10^{10}$--$3\times10^{11}$ cm$^{-3}$s.
A comparison between the ejecta-rich regions further shows that the $n_{\mathrm{e}}t$ values are lower in the northern regions than in the southeast, both for IMEs and IGEs. A similar N-SE asymmetry has been suggested in previous analyzes for Si/S \citep{lazendic06,suzuki25}, indicating that these variations may be a robust property of the remnant rather than modeling artifacts and applies to all elements.

The distribution of $kT_\mathrm{e}$ for the IMEs component covers a narrow range between 1.3--2.1 keV and does not show a large variation over the face of the remnant. The values are well constrained with an average 1-sigma uncertainty of $\pm0.1$ keV.
The $kT_\mathrm{e}$ corresponding to the Fe-group component is higher than that of IMEs and ranges between 2.4--8.4 keV with an average uncertainty of $\pm1.2$ keV. The fitted $kT_\mathrm{e}$ values in the western and central regions exceed 5 keV. 

The posterior distribution of IGEs $kT_\mathrm{e}$ for some super-pixels (example SE pixel {\em f} and NW pixel {\em g}) spans a wide range and reaches the upper limit of 10 keV set as a prior, indicating that the values are not well constrained. This is also clear from the corner plots of these super-pixels (example Appendix Fig.\ref{fig:corner_NW_pix_e}), where we find that above 5 keV, the $kT_\mathrm{e}$ value has little dependence on any other parameter except for the normalization of the IGEs \texttt{pshock} component.
\cite{helder08} and \cite{vink22b} also reported higher electron temperatures, often reaching the upper limit, for Fe when modeling Cas~A spectra with pure-Fe components. This can be expected since a pure-metal Fe-group component primarily fits the Fe K$\alpha$ complex, making the fits biased toward higher $kT_\mathrm{e}$. 
Furthermore, in these regions, a power-law alone fits well the continuum emission (discussed in Section~\ref{subsec:results_nonthermal_emission}) which biases the Fe-group component to lower the thermal continuum (Bremsstrahlung, two-photon emission and radiative recombination continuum) contributed by the IGEs. To this end, the model tends to favor lower normalization and higher $kT_\mathrm{e}$, which makes the emission lines peak well over the corresponding continuum emission, as can be seen in the fitted Fe-group component of NW pixel {\em e} in Figure~\ref{fig:pixel_e_spectra}. 
The Fe-L lines which could help break this parameter degeneracy are inaccessible due to \xrism\ Gate Valve closed.
However, another possibility remains that our model choice of pure-metal ejecta is not well suited for these regions and rather the model requires an additional forward shocked CSM component. 
For further interpretation, we decide to be prudent and exclude these regions from our discussion; they are marked with a reduced opacity in Figure~\ref{fig:net_vs_kt} which shows the distribution of $kT_\mathrm{e}$ with respect to $n_{\mathrm{e}}t$ (discussed in detail in the following subsection).
We measure a lower $n_{\mathrm{e}}t$ and higher $kT_\mathrm{e}$ for the Fe-group than the IMEs for the corresponding super-pixel, except for the SE pixel {\em d}.

Parameter maps derived from \chandra\ data by \citet{vink22b} show mean values of $n_{\mathrm{e}}t \approx 2.2 \times 10^{11}$ cm$^{-3}$s and $kT_\mathrm{e}$ $\approx 1.4$ keV for a Si-rich NEI component. Similar values for $n_{\mathrm{e}}t$ and $kT_\mathrm{e}$ in ejecta-dominated regions were also reported by \citet{hwang12}. Our measurements of the IME component are consistent with these $n_{\mathrm{e}}t$ estimates, although the $kT_\mathrm{e}$ values we obtain tend to be higher. 
For the Fe-rich NEI component, \citet{vink22b} found mean values of $n_{\mathrm{e}}t \approx 1.1 \times 10^{11}$ cm$^{-3}$s and $kT_\mathrm{e}$ $\approx 3.0$ keV, with many regions reaching the imposed upper limit of 4 keV. In contrast, \citet{hwang12} modeled Fe-rich regions using a pure-Fe NEI component with fixed parameters of $n_{\mathrm{e}}t = 8 \times 10^{11}$ cm$^{-3}$s and $kT_\mathrm{e}$$ = 1.95$ keV. 
This is much higher $n_{\mathrm{e}}t$ than the best fitted values in our analysis.

\subsubsection{Implications of the anti-correlation of $n_{\mathrm{e}}t$ and $kT_\mathrm{e}$}\label{subsubsec:net_vs_kT-implications_of_anticorrelation}

Our analysis reveals an anti-correlation between $n_{\mathrm{e}}t$ and $kT_\mathrm{e}$, as shown in the log-log scatter plot of Figure~\ref{fig:net_vs_kt}. This relation is more pronounced for the Fe-group component.
While an anti-correlation between $n_{\mathrm{e}}t$ and $kT_\mathrm{e}$ (and between normalization and $kT_\mathrm{e}$) is well known as a result of parameter degeneracies in individual spectral fits and is indeed visible in our corner plots (Appendix~\ref{app:corner_plots}), the fact that this trend emerges systematically across different regions of the remnant is intriguing. In particular, as standard NEI model evolution predicts the opposite trend for a collisionally heating plasma, where electron temperatures are expected to increase with higher ionization ages \citep{katsuda13, orlando16, vink25, xrism_n132d_2024}.

Similar large-scale anti-correlations have been previously reported in X-ray mapping studies of SNRs. \citet{godinaud25} found such an anti-correlation for IMEs in Tycho SNR with \chandra\ spectra modeled as pure ejecta plasma, but not for the Fe component, in contrast to our analysis of Cas~A where the trend is observed for the Fe-dominated plasma. \citet{sun19} also identified this anti-correlation for Kepler SNR, using a single NEI component with near Solar abundances fitted to \chandra\ data. Both studies attributed the anti-correlation to astrophysical differences, with various regions of the ``banana-shaped" distribution corresponding to plasma shocked under different conditions. \citet{godinaud25} suggested that different regions of the SNR occupied a different location in the $n_{\mathrm{e}}t$-$kT_\mathrm{e}$ plot (see their Figure 11). Similarly, \citet{sun19} proposed that their $n_{\mathrm{e}}t$-$kT_\mathrm{e}$ distribution shows two branches, one corresponding to newly shock-heated CSM and another for cooler ejecta material. In contrast, \citet{li15} studied SN~1006 using \xmm\ and observed a similar trend, but argued that the apparent anti-correlation arises primarily from parameter degeneracy rather than intrinsic plasma properties.

For Cas~A, parameter mapping performed by \citet{hwang12} using a single \texttt{pshock} model (without a power-law component) did not reveal such an anti-correlation. Instead, they reported a gradual narrowing of the distribution of $n_{\mathrm{e}}t$ values with increasing $kT_\mathrm{e}$ (see their Figure 2). 
\cite{hwang12} included an Fe-component in some regions, however, the $n_{\mathrm{e}}t$ parameter was fixed to 8$\times10^{11}$ cm$^{-3}$s, which was generally higher than the fitted $n_{\mathrm{e}}t$ for the IMEs component, this is contrary to our findings.
\cite{lazendic06} also produced the distribution of $n_{\mathrm{e}}t$ vs $kT_\mathrm{e}$ using the \chandra\ HETGS data of Si lines in 17 small isolated regions. They measured $kT_\mathrm{e}$ of $\sim$1 keV and $n_{\mathrm{e}}t$ of a few $10^{11}$ cm$^{-3}$s for most regions and reported no significant variation over the remnant. Although they do not claim any correlation between $n_{\mathrm{e}}t$ vs $kT_\mathrm{e}$, referring to their Figure B10, we find a slight negative slope. They do, however, suggest that the red-shifted regions were shocked more recently and with lower $n_{\mathrm{e}}$.

The persistence of the anti-correlation in our study with well-constrained parameters, by using high-resolution spectroscopy with \xrism\ and Bayesian inference methods that diminish the effects of degeneracies, indicates that the effect cannot be fully explained by fitting uncertainties alone. Rather, it points to underlying physical differences across the remnant. The fact that the anti-correlation is strongest in the Fe-rich ejecta suggests a different shock history for IGEs compared to IMEs, and a link to the remnant’s complex asymmetric explosion.

To interpret the measured values, we compare with the numerical NEI models 
which trace the evolution of post shock plasma temperatures of each species as they gradually equilibrate via Coulomb interactions. Immediately behind the collisionless shock (assuming no ``collisionless electron heating"), the temperature of each species is proportional to its mass ($m_i$) and square of the shock velocity ($V_{sh}$) --- $kT_i \propto m_i V_{sh}^2$. 
We adopt the \texttt{IONTENP} model by \cite{ohshiro24} which incorporates the evolution of $kT_\mathrm{e}$ as opposed to a constant $kT_\mathrm{e}$ assumed in simple NEI models.
Using this model, we investigate the expected $n_{\mathrm{e}}t$ and $kT_\mathrm{e}$ values for the pure Fe and pure Si ejecta shocked at a velocity of 1800 \kms, which is the expected value of the reverse shock from analytical models as described in the Appendix~\ref{app:ionization_age}. As shown in Figure~\ref{fig:net_vs_kt}, and we find that our fitted data points (dark blue and dark orange --- regions where $kT_\mathrm{e}$ is well constrained for the Fe-component and are associated with the reverse shock heated ejecta) are consistently below the evolution curve and show the opposite trend. Furthermore, the IMEs points are even below 900 \kms\ (i.e. half of the expected value).
The most straightforward justification for this would be that the reverse shock velocity (in the frame of the ejecta), $v_{rs,ej}$, was lower in the past when the stellar ejecta was shock heated. An acceleration of the $v_{\rm rs,ej}$ has indeed been noted in some magnetohydrodynamics simulations \citep{kirchschlager24}, though, $v_{\rm rs,ej} <$ 900 \kms\ are unlikely. Furthermore, if we extend this argument to explain the difference between IMEs and IGEs, it implies that IMEs were shocked at a lower $v_{\rm rs,ej}$ than IGEs. However, in Cas~A Fe has overtaken IMEs \citep[][Section~\ref{subsec:doppler_shift_and_broadening}]{hwang00} and therefore should be shock heated first and when $v_{\rm rs,ej}$ was lower.  Due to these two reasons, we argue that the slower $v_{\rm rs,ej}$ does not fully explain the observations. 
But it is perhaps a factor, since a varying reverse shock velocity has been observed over different regions of Cas~A \citep{vink22a,wu24,fesen25} and this has also been proposed by \cite{suzuki25} to explain the lower $n_{\mathrm{e}}t$ values in the N than SE for Si/S.

The other way of approaching these deviations from the models would be to understand why the $n_{\mathrm{e}}t$ values are higher than expected and/or $kT_\mathrm{e}$ values lower.
The \texttt{pshock} model includes a range of $n_{\mathrm{e}}t$ and the fitted value is the maximum of this range, so the measurements are biased towards higher $n_{\mathrm{e}}t$. However, this will be a systematic offset that will not range the trend to positive correlation and also the $n_{\mathrm{e}}t$ values from previous studies using simple NEI models are in range of a few $10^{11}$ cm$^{-3}$s, consistent with our measurements.
In the Appendix~\ref{app:ionization_age}, we argue that the estimated values of $n_{\mathrm{e}}t$ --- using the explosion time for $t$ and a uniform $n_{\mathrm{e}}$ from the total ejecta mass, $M_{ej}$ --- are an order of magnitude lower than the fitted values. This can be alleviated by clumping, as clumps have a higher density than average density --- increasing $n_{\rm e}t$ --- and the EM scales with $n_{\rm e}^2$, biasing the emission toward the denser clumps.
Thus, we argue that clumping of the stellar ejecta
is required to explain the measurements. In the Appendix~\ref{app:ionization_age}, we estimate an overdensity of $\sim$10 for IGEs and as high as $\sim$100 for IMEs, in line with estimates of \cite{lazendic06} for Si-rich knots. The high $n_{\mathrm{e}}$ for IMEs also supports the lower $n_{\mathrm{e}}t$ values for Fe compared to IMEs (except for the SE super-pixel {\em d}), despite Fe having been shocked earlier. 

The clumping of the ejecta will also give rise to a slowing
down of the shock velocity inside the knots, scaling with the density ($\rho$) contrast ($\rho/\bar{\rho}$)
as $v_{\rm s} \approx \bar{v}_{\rm s}\sqrt{\bar{\rho}/\rho}$
--- with the barred symbols indicating the intraclump quantities ---
which also results in the shock becoming more oblique \citep{klein94}.
Inside the clump, the reverse shock is therefore less efficient at heating and can explain the lower than expected $kT_\mathrm{e}$ values. Furthermore, the evolution models might also be overestimating the $kT_\mathrm{e}$, as effects such as adiabatic cooling due to expansion, radiative losses, and energy losses to particle acceleration are not included.

\begin{figure*}
\centerline{
  \includegraphics[width=0.33\textwidth]{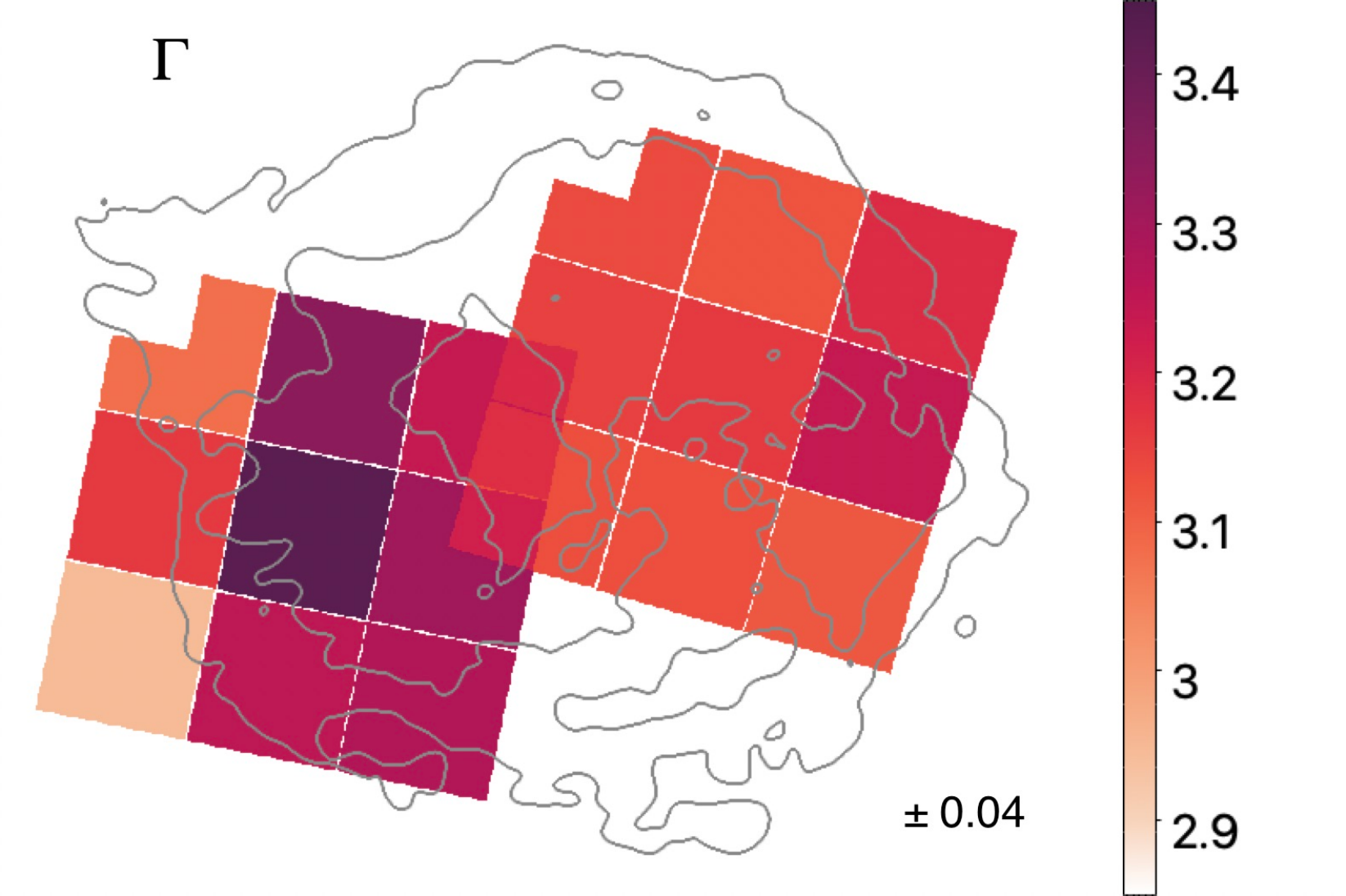}
    \hspace{2mm}
  \includegraphics[width=0.33\textwidth]{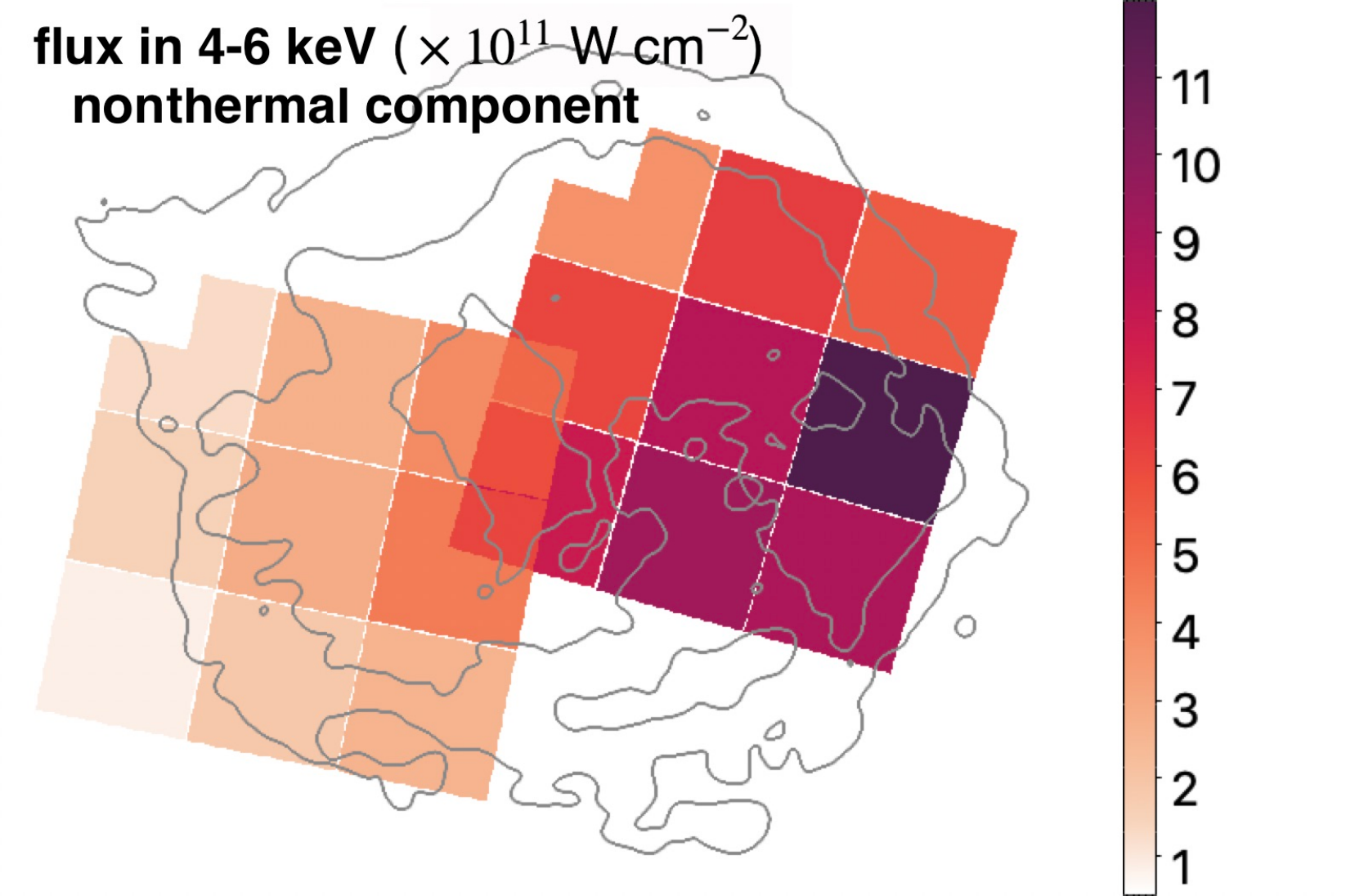}
    \hspace{2mm}
  \includegraphics[width=0.33\textwidth]{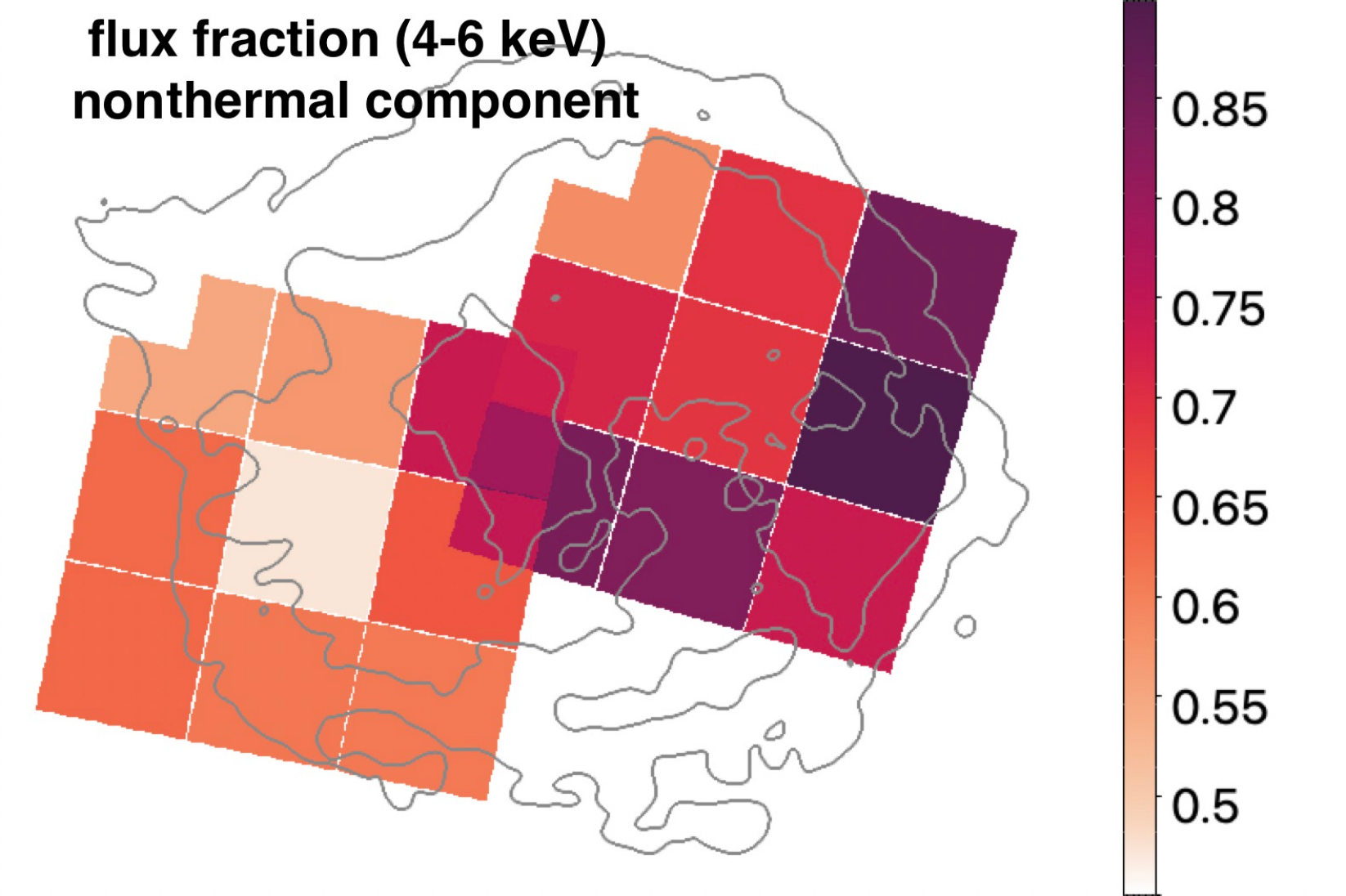}
}
\caption{
Left: Map of fitted power-law index ($\Gamma$). Darker color indicates a softer/steeper spectrum. The average one sigma error is $\pm 0.04$. 
Center: The energy flux map of the fitted power-law component in the 4 to 6 keV energy band. 
Right: Map of nonthermal-only (power-law) flux to the total flux ratio in the 4--6 keV band. The flux fractions maps for each component in the entire fitted energy range of 1.8--11.9 keV are presented in the Appendix (Figure~\ref{fig:flux_fraction_maps})
}
\label{fig:powerlaw_maps}
\end{figure*}

These arguments explain why the observed $kT_\mathrm{e}$ values are lower and the $n_{\rm e}t$ values larger than the predictions of semi-analytical models. However, more detailed modeling is needed to quantify how the observed anti-correlation between $kT_{\rm e}$ and $n_{\rm e}t$ is related to the history of the reverse shock velocity, and the variations in ejecta density due to clumping.

The \texttt{pshock} NEI model was employed because it better describes the range of ionization states in the observed plasma --- most apparent from the low-energy tail of the Fe K complex \citep{bamba25}.
The model assumes that these ionization states reflect different
time scales since the plasma was shocked, assuming a plane
parallel shock model.
However, it is possible that the variation in $n_{\rm e}t$ as incorporated in the \texttt{pshock} model has an important component due to a range in densities ($n_{\rm e}$) in the shocked ejecta, 
in addition to a range in time scales ($t$).


\subsection{Nonthermal emission}
\label{subsec:results_nonthermal_emission}

The \xrism/Resolve spectrum, benefiting from low instrumental background and energy coverage up to $\sim$12 keV, allows better constraints on the synchrotron emission from Cas~A. The extension to higher energies is particularly valuable for disentangling the nonthermal and thermal contributions. Figure~\ref{fig:powerlaw_maps} presents maps of the fitted power-law index ($\Gamma$), the energy flux in the 4--6~keV band, and the fractional contribution of the power-law component to the total flux in the same energy range. Consistent with the east–west asymmetry reported by \citet{helder08}, we find enhanced synchrotron emission in the western regions, characterized by lower $\Gamma$ values (i.e., harder spectra). The central pixel in SE (pixel {\em e}) shows the steepest spectrum ($\Gamma = 3.43$) and the lowest nonthermal contribution --- 47\% of the flux in 4--6~keV band and 17\% in the total fitted energy range of 1.8--11.9~keV (Appendix, Fig.~\ref{fig:flux_fraction_maps}).

Across the regions covered by Resolve, the fitted photon index ranges from 2.94 to 3.43, with a typical $1\sigma$ uncertainty of $\pm$0.04. These values are comparable to those derived from fits to {{NuSTAR}} spectra (PSF $\sim$60$''$ half-power diameter) data above 15 keV. \cite{grefenstette15} reported $\Gamma = 3.06 \pm 0.06$ ($2\sigma$ error) for the outer filaments and $\Gamma = 3.35 \pm 0.06$ for central knots and \cite{greco23} measured $\Gamma$ in the range 3.02--3.46. 
The synchrotron fraction in the 4–6 keV continuum spans a wide range, from a minimum of 47\% (SE pixel {\em e}) to 90\% (NW pixel {\em f}), with a flux-weighted mean contribution of 84.4\%.

These results of large synchrotron fractions reinforce previous findings of \cite{helder08} and \cite{vink22b} using \chandra\ data.
\cite{helder08} analyzed the \chandra\ spectra in the 4.2--6 keV range to estimate an overall 54\% nonthermal contribution to all continuum emission and \cite{vink22b} reported even higher contributions ranging from 38\% to nearly 100\%. Additional evidence comes from broadband hard X-ray studies. Fits to Cas~A spectra above $\sim$15 keV when extrapolated to lower energy (4--6~keV band) reproduce the observed continuum emission without requiring an additional thermal contribution ({{BeppoSAX}} PDS, {{INTERGAL}} ISGRI - \citealt{helder08}, {{NuSTAR}} - \citealt{grefenstette15, greco23}). In some cases, the extrapolated models even exceed the observed continuum in the 4–6~keV range \citep[Fig.~5 of][]{grefenstette15}. As discussed in the previous section, the high synchrotron emission provides a natural explanation for why the Fe-group plasma component fits are biased towards minimizing the thermal continuum levels, yielding higher electron temperatures and lower normalizations in plasma models.

Our results confirm that synchrotron radiation is a dominant contributor to continuum emission across Cas~A, with power-law fractions exceeding at least $\sim$45\% in the 4--6~keV band. Therefore, any spectral modeling of Cas~A should explicitly include a nonthermal power-law component, particularly when analyzing regions comparable to or larger than the \xrism\ PSF.

\subsection{Radiative Recombination Continuum}

RRC is produced when a free electron recombines with an ion, emitting a photon with energy equal to the sum of the kinetic energy of the electron and the binding energy of the bound state into which it falls. This process yields an emission with a sharp edge at the energy corresponding to the binding energy of the ion's recombined state, followed by an exponential drop at higher energies that is proportional to the electron temperature.
In an X-ray spectrum within the 0.3 to 12 keV range, the RRC features from K-shell and L-shell recombinations of several IMEs and IGEs are expected. 
RRC emission has only been detected in mixed-morphology SNRs with recombining (overionized) plasma such as W49B \citep{ozawa09, xrism_w49b_2025} and IC443 \citep{yamaguchi09}.
\cite{greco20} suggested that the RRC features should also be observed in pure-metal ejecta plasma. They proposed, based on simulated high-resolution X-ray spectra (assuming \xrism\ Gate Valve open), a Fe-rich clump in the southeast of Cas~A as a promising region for detecting RRC features of Fe L. Since we model the Cas~A spectra as emission from pure-metal ejecta, we carefully inspect all the spectra for RRC characteristics, especially the SE pixel {\em e}, which has the least non-thermal continuum and is close to the region recommended by \cite{greco20}.

We do not detect any distinct RRC features in Cas~A in the regions covered by \xrism\ so far. 
The RRC features associated with pure metal plasmas are part of the model we fitted. However, there are several reasons why these features are not distinctly visible in the observed spectra.
First of all, X-ray synchrotron radiation and thermal continuum from oxygen are the most dominant continuum radiation components, and these wash out the RRC features in the 2-10 keV band from the K-shell recombinations of elements from Si to Fe,
with their continua at least one order of magnitude below the synchrotron and oxygen continua. Fig.~\ref{fig:thermal_continuum_plots} in the Appendix shows the thermal continuum contribution of each element in our fitted model for SE super-pixel {\em e}. Other reasons that impact the visibility of the RRC features in the spectra are the high velocity broadening, which smoothes out the sharp emission edges, and the fact that the RRCs nearly coincide with transitions from high {\em n} transitions of the Rydberg series. These high {\em n} transitions are now clearly visible in the \xrism\ spectra \citep{plucinsky25}. Furthermore, the Fe-K RRC is nearly absent in our models, as Fe is not sufficiently highly ionized from the fitted plasma properties (see Appendix~\ref{app:thermal_continuum}).

Finally, the super-pixel regions we used are big and always include substantial synchrotron continuum and blend emission from plasma with different velocity components leading to larger line broadening. In the future, with newAthena \citep{newAthena25}, one can isolate bright regions with less X-ray synchrotron contributions, which would help to identify RRC features. Moreover, its broader X-ray spectral coverage should enable the detection of L-shell RRCs of Fe, and K-shell RRCs of oxygen.

\section{Conclusions}\label{sec:conclusions}    

We present the first comprehensive plasma parameter mapping of an SNR with \xrism/Resolve microcalorimeter --- a spatially resolved high-resolution X-ray spectroscopy instrument. 
We used $>350$ ks \xrism/Resolve observations of Cas~A to perform broadband spectral fits in the 1.8--11.9 keV range to model the thermal and nonthermal X-ray emission across the remnant. 
We adopted an absorbed model with two \texttt{pshock} components to separately model the IMEs and IGEs as pure-metal ejecta, plus a power-law corresponding to the synchrotron emission. 
A caveat of the model is the absence of a separate CSM component and a single \texttt{pshock} for the IMEs, which is inadequate to perfectly capture complex line profiles \citep{vink25}.
Despite this, the spectral model fits the observed data well across the different regions of the SNR.

We developed a Bayesian-based spectral fitting tool, {\em UltraSPEX}, which facilitates the analysis of the high-resolution X-ray spectra by integrating \texttt{SPEX} with the {\em UltraNest} nesting sampling algorithm.
This has unique advantages in modeling the \xrism\ data by combining the extensive atomic database and the optimized handling of large RMFs by \texttt{SPEX} with the robust Bayesian-based parameter space exploration of {\em UltraNest}. 
This provides a robust, scalable, spectral fitting method that allows for a detailed modeling of the plasma properties, their corresponding uncertainties (posterior distributions), and an understanding of potential parameter correlations and degeneracies.

We apply this to construct maps of the kinematic and plasma properties of the X-ray emitting plasma in Cas~A at a resolution of $1'\times1'$ (i.e., $2\times2$ Resolve pixels). The key results of this work are emphasized below:


\begin{itemize}
    \item The elemental abundance maps of Ar/Si and Ca/Si have an enhanced abundance ratio near the base of the Si-rich jets of Cas~A, indicative of a likely enrichment from the incomplete Si-burning layer. S and Si trace each other over the remnant.
    \item A high mass ratio for Ni/Fe of $0.08 \pm 0.015$ (corresponding to higher than solar abundance ratio) is measured near the base of the NE jet structure. This implies mixing of material from the innermost complete Si-burning region into the jet.
    \item The IGEs have both higher Doppler velocity and Doppler broadening than the IMEs in most regions. The maximum difference between IGEs and IMEs in Doppler velocity is $\sim800$ \kms\ and in Doppler broadening it is $\sim1200$ \kms. The kinematic differences are highest near the center and decrease radially outward from the explosion center.
    \item Ca is measured to have a higher Doppler velocity of $\sim-1450$ \kms\ compared to the overall IMEs velocity of $\sim-1000$ \kms\ in three super-pixel regions in the SE.
    \item The measured ionization timescales and electron temperatures show an anti-correlation; it is more prominent for IGEs than IMEs. This trend has previously been observed for Type Ia SNRs with different telescopes but is poorly explained. We reveal it for Cas~A using high-resolution spectroscopy and Bayesian inference, showing that it is a robust feature.
    \item The fitted values of $n_{\mathrm{e}}t$ are higher and $kT_{\rm e}$ are lower than expected according to semi-analytical models with a reverse shock velocity (with respect to ejecta) of 1800 \kms. We suggest a combination of a lower shock velocity in the past and a higher density ($n_{\rm e}$) due to clumping to explain the measurements. We estimate overdensities of $\sim10$ for IGEs and as high as $\sim100$ for IMEs. 
    \item An additional hint for the importance of clumping is provided by the larger broadening of IGEs lines at the edge of the SNR, assuming the difference in line broadening between IGEs and IMEs is dominated by thermal broadening.
    \item The thermal and nonthermal emission is well disentangled, enabled by the \xrism/Resolve spectrum, which features a low background and reaches high energies ($\sim12$ keV) where nonthermal emission becomes dominant. Spatial mapping of the power-law index and its flux contribution reveals that synchrotron emission accounts for at least 47\% of the 4--6 keV continuum flux across Cas~A, with the western regions exhibiting a stronger nonthermal component and a harder spectrum.
    \item No RRC features were detected in the current \xrism/Resolve observations of Cas~A.
\end{itemize}

The results presented here highlight the power of \xrism\ in mapping the X-ray properties of SNRs and other extended sources. 
The analysis discussed here utilizes data from two \xrism\ observations of Cas~A, which cover only about 60\% of the SNR. 
We need complete coverage of Cas~A with \xrism\ to produce comprehensive maps of plasma parameters throughout the remnant.
Additionally, future analyzes would benefit from a more complex model, while incorporating data from other instruments --- \xrism/Xtend, \chandra, \xmm, NuSTAR --- to access a broader range of spectral energies.



\begin{acknowledgments}
We like to thank the anonymous referee for insightful comments, especially regarding the thermal Doppler broadening, which helped to improve the paper.
The authors are grateful to Jelle de Plaa for help in implementing the Bayesian interface for SPEX.
We thank Leïla Godinaud and Fabio Acero for fruitful discussions on Bayesian-based spectral fitting for SNRs. We also thank Ryo Yamazaki and Jan-Willem den Herder for their constructive suggestions. 
MA thanks the University of Tokyo for an extended visit during which a part of this work was conducted. 
The research on this project by MA and JV was (partially) funded by NWO under grant number 184.034.002.
PP acknowledges support from the National Aeronautics and Space Administration (NASA) XRISM grants 80NSSC18K0988 and 80NSSC23K1656, and the Smithsonian Institution and the Chandra X-ray Center through NASA contract NAS8-03060.
This work was partly supported by the Japan Society for the Promotion of Science (JSPS) Grants-in-Aid for Scientific Research (KAKENHI) Grant Numbers, JP23K25907 (AB), JP23K13128 (TS).
\end{acknowledgments}





%
\facility{XRISM}



\appendix
    \restartappendixnumbering



\section{Fitted spectra}\label{app:all_spectra}
\begin{figure*}
\centerline{
  \includegraphics[width=0.49\textwidth]{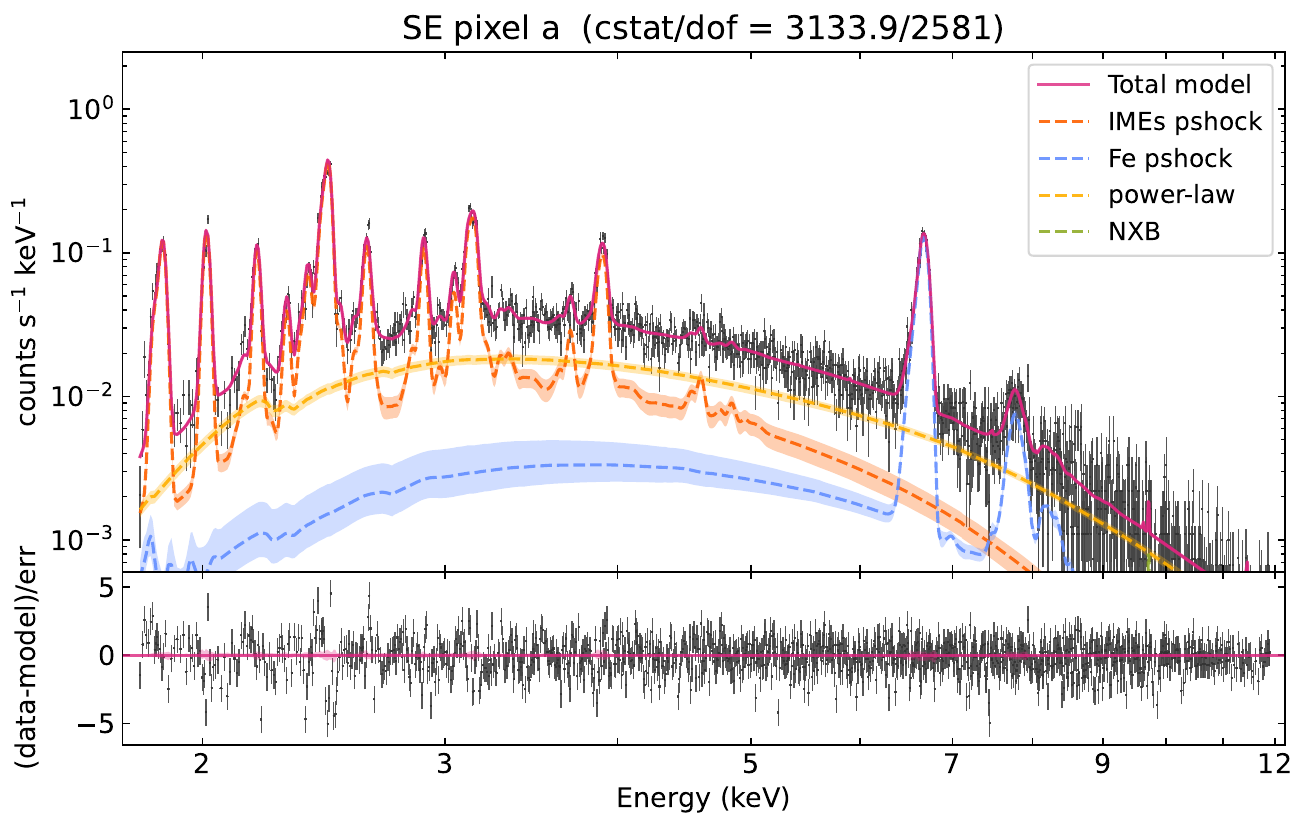}
  \hspace{2mm}
  \includegraphics[width=0.49\textwidth]{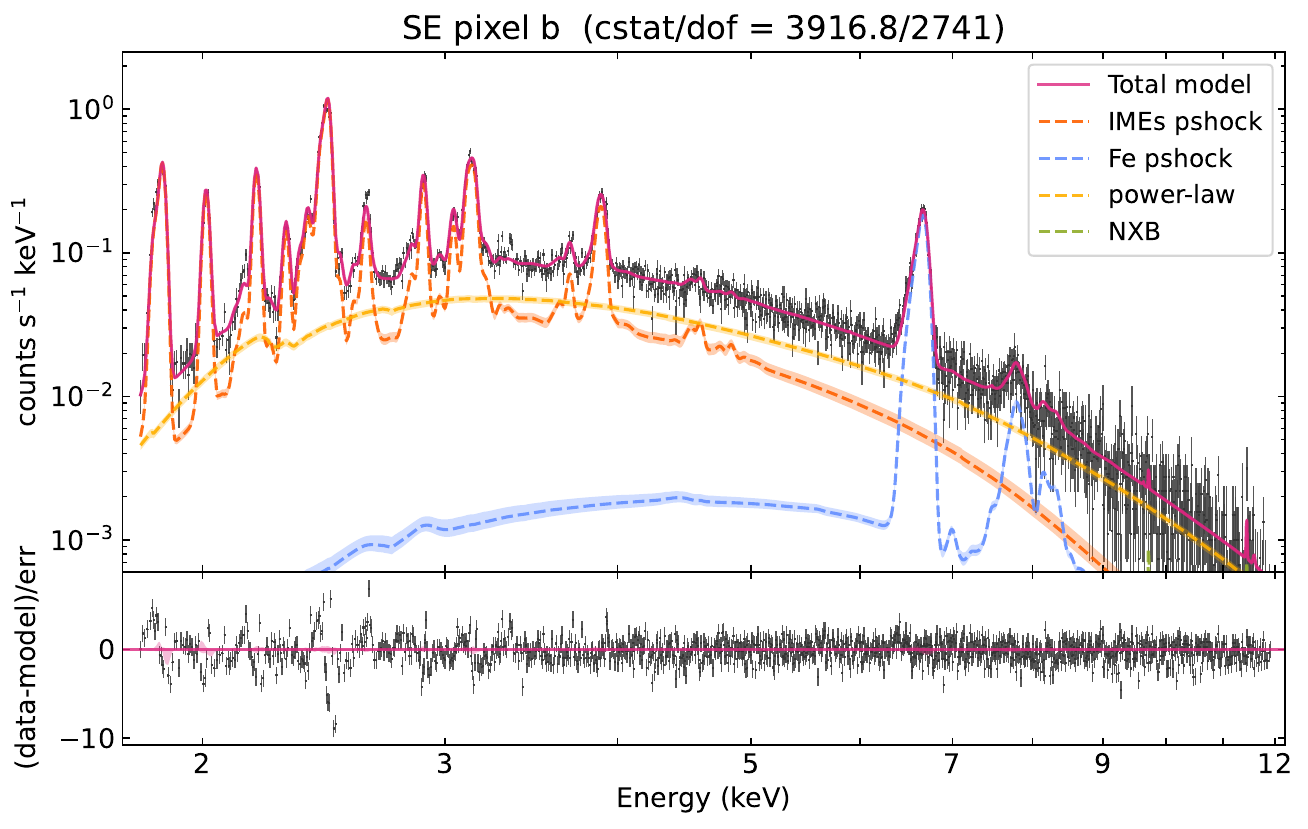}
}
\vspace{1mm}
\centerline{
  \includegraphics[width=0.49\textwidth]{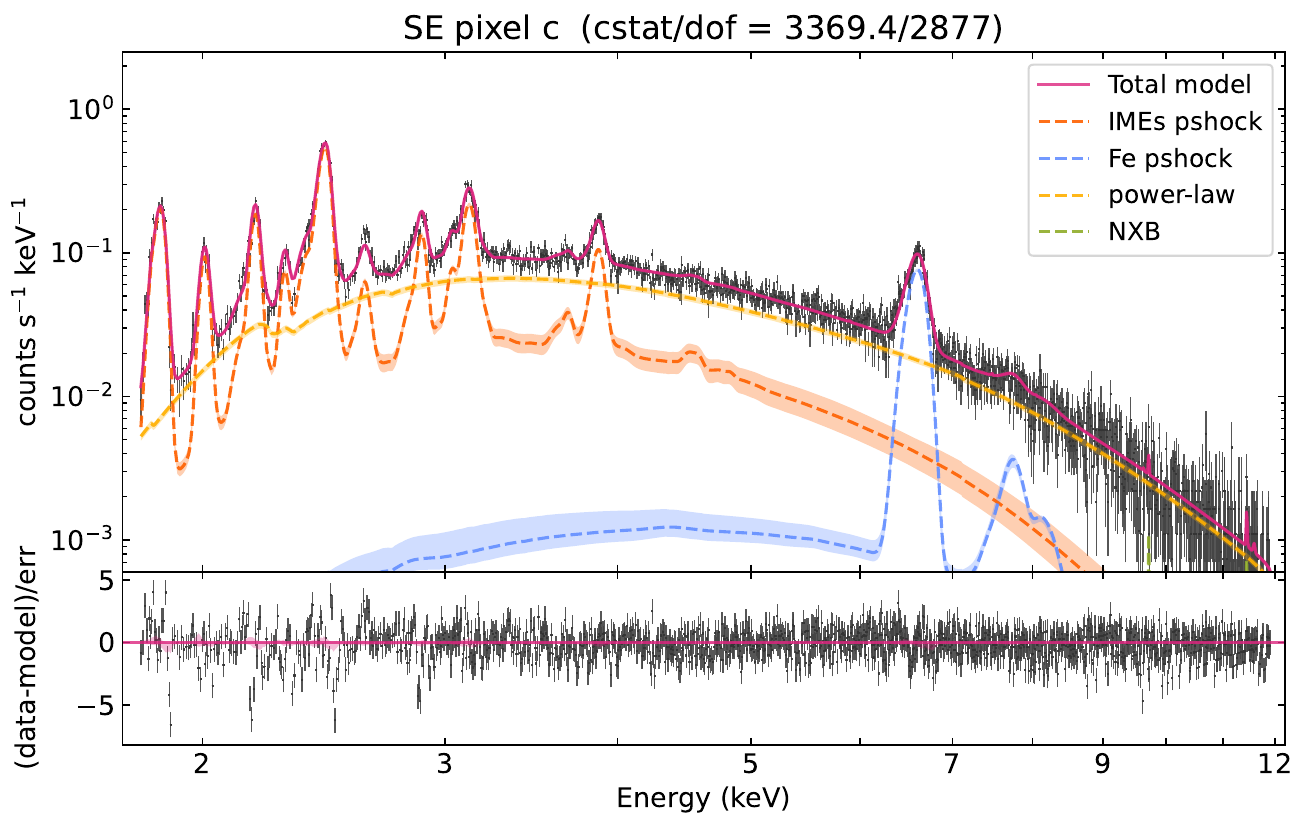}
  \hspace{2mm}
  \includegraphics[width=0.49\textwidth]{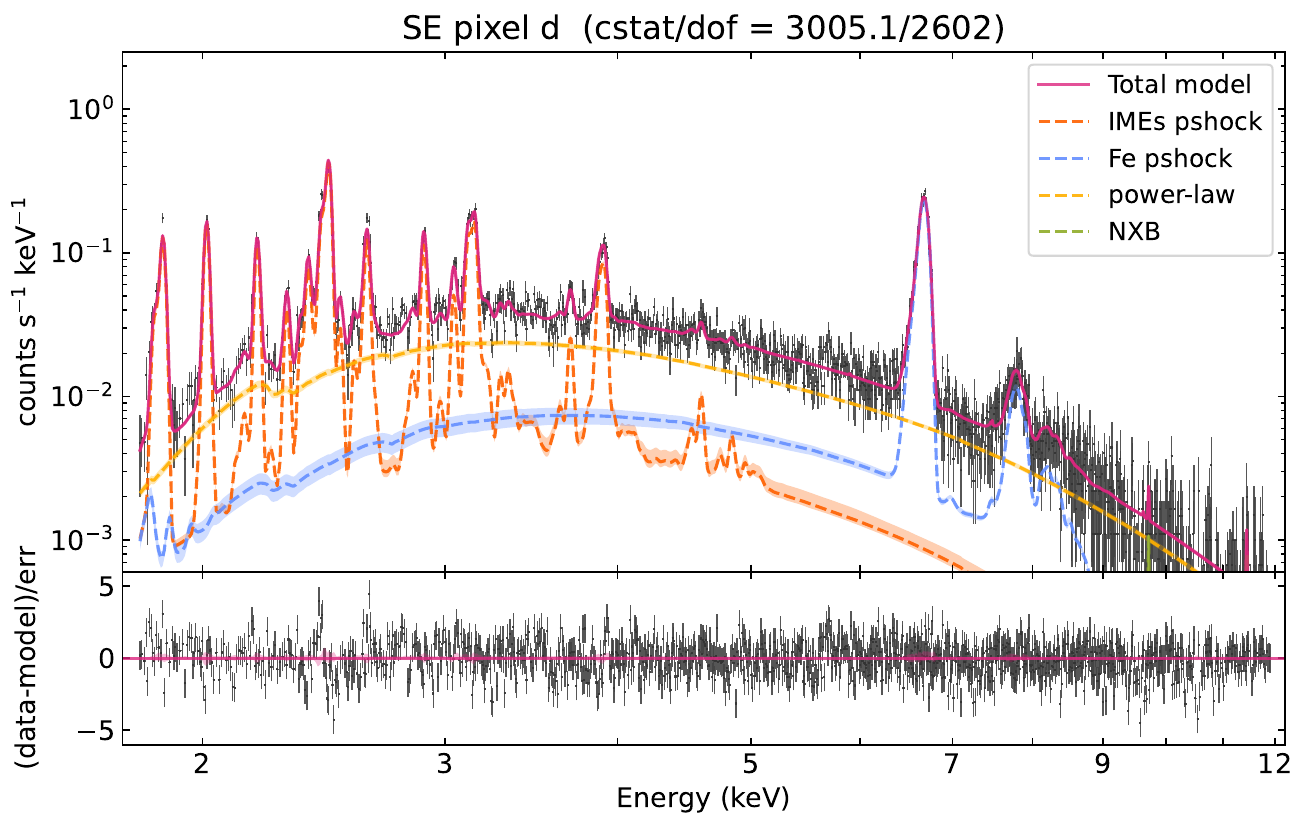}
}
\vspace{1mm}
\centerline{
  \includegraphics[width=0.49\textwidth]{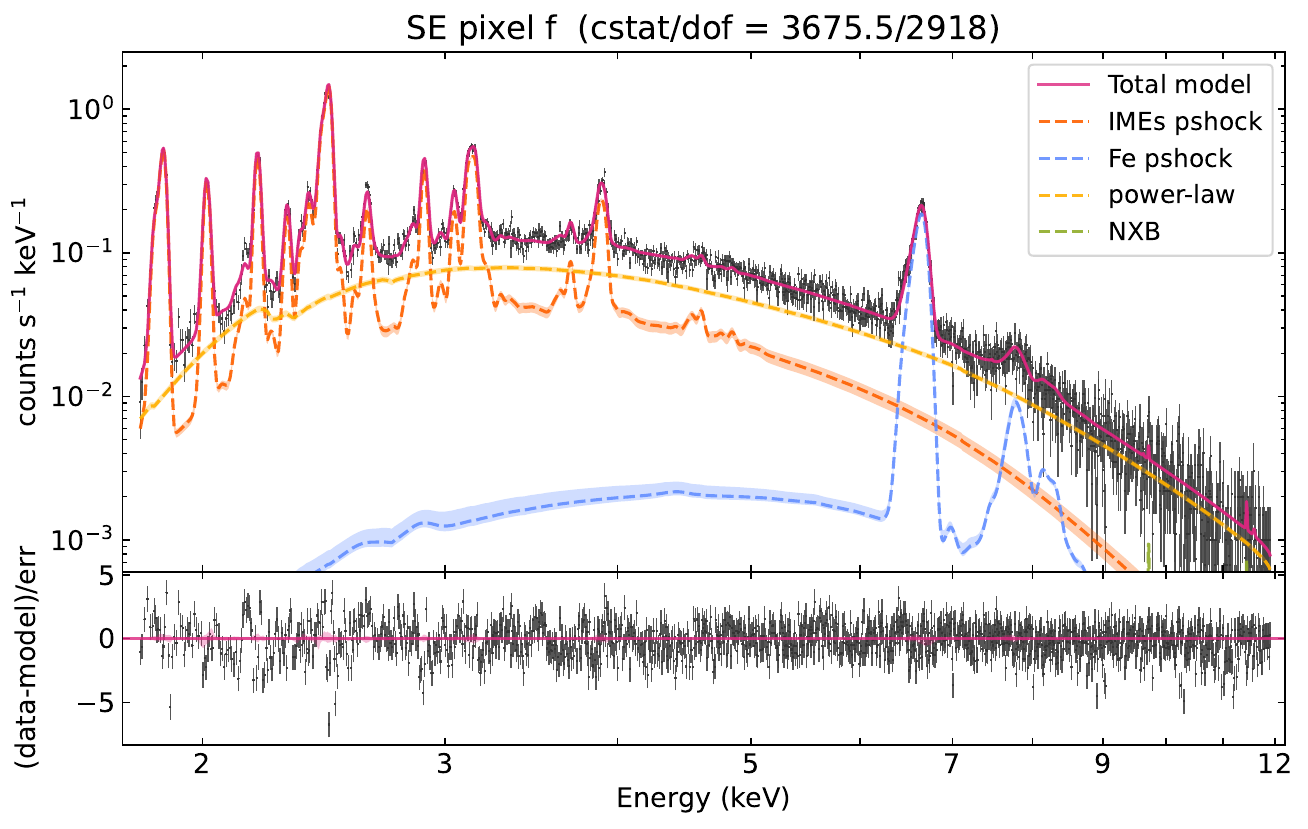}
  \hspace{2mm}
  \includegraphics[width=0.49\textwidth]{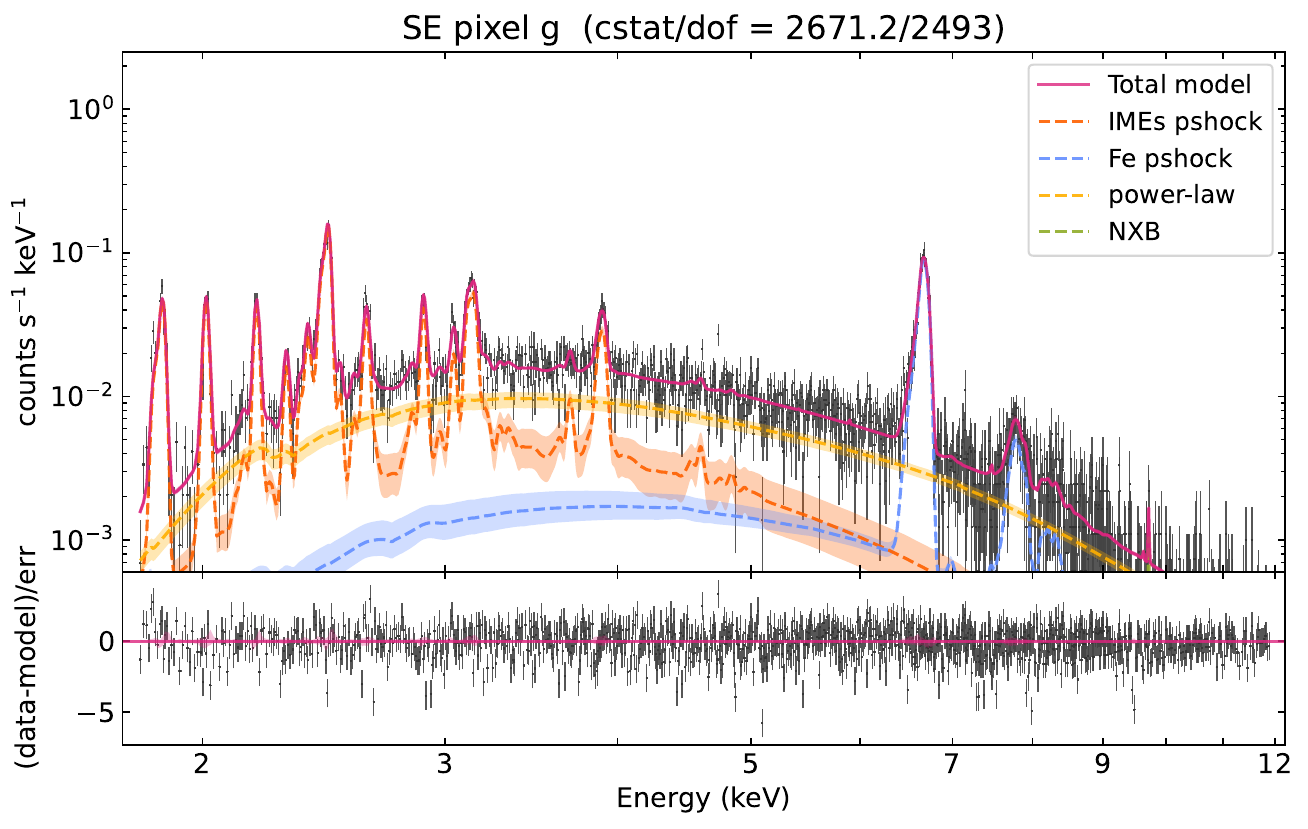}
}
\vspace{1mm}
\centerline{
  \includegraphics[width=0.49\textwidth]{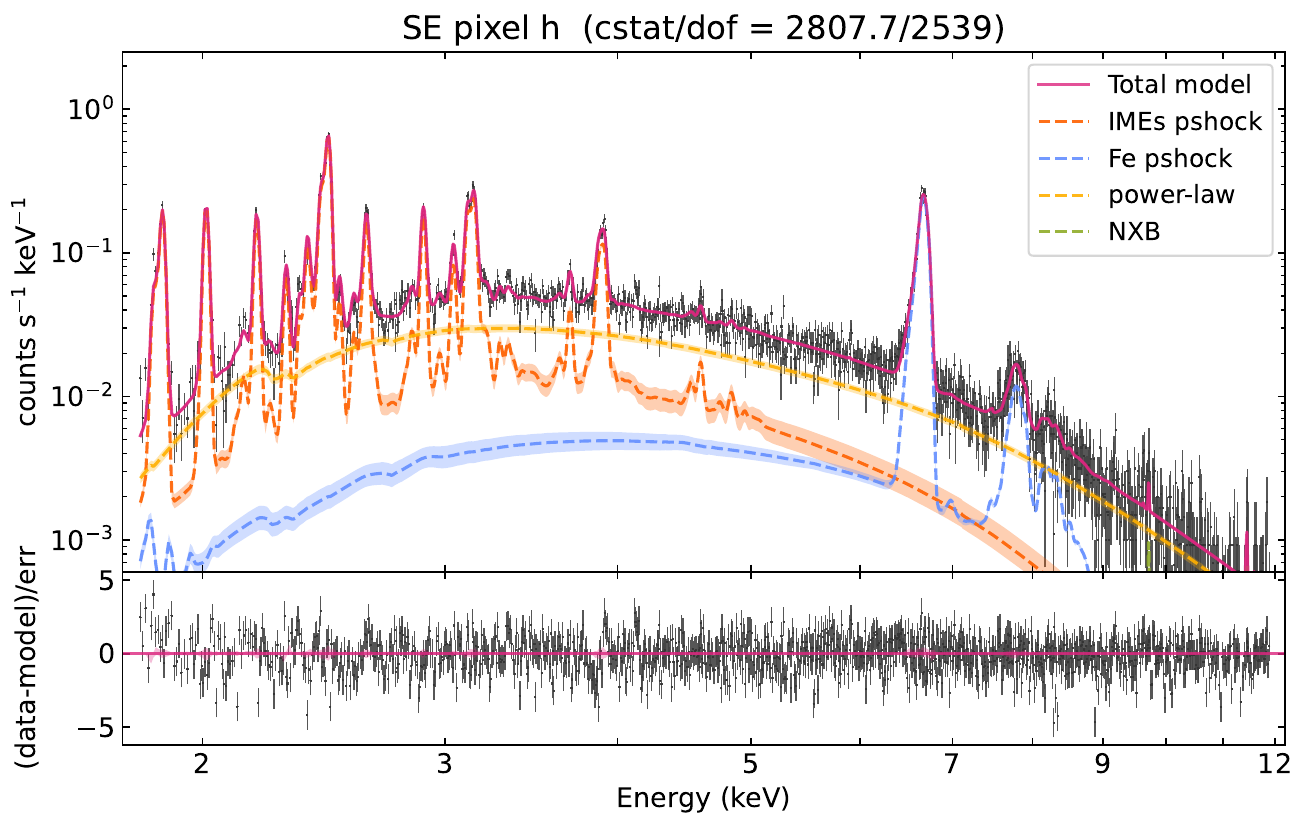}
  \hspace{2mm}
  \includegraphics[width=0.49\textwidth]{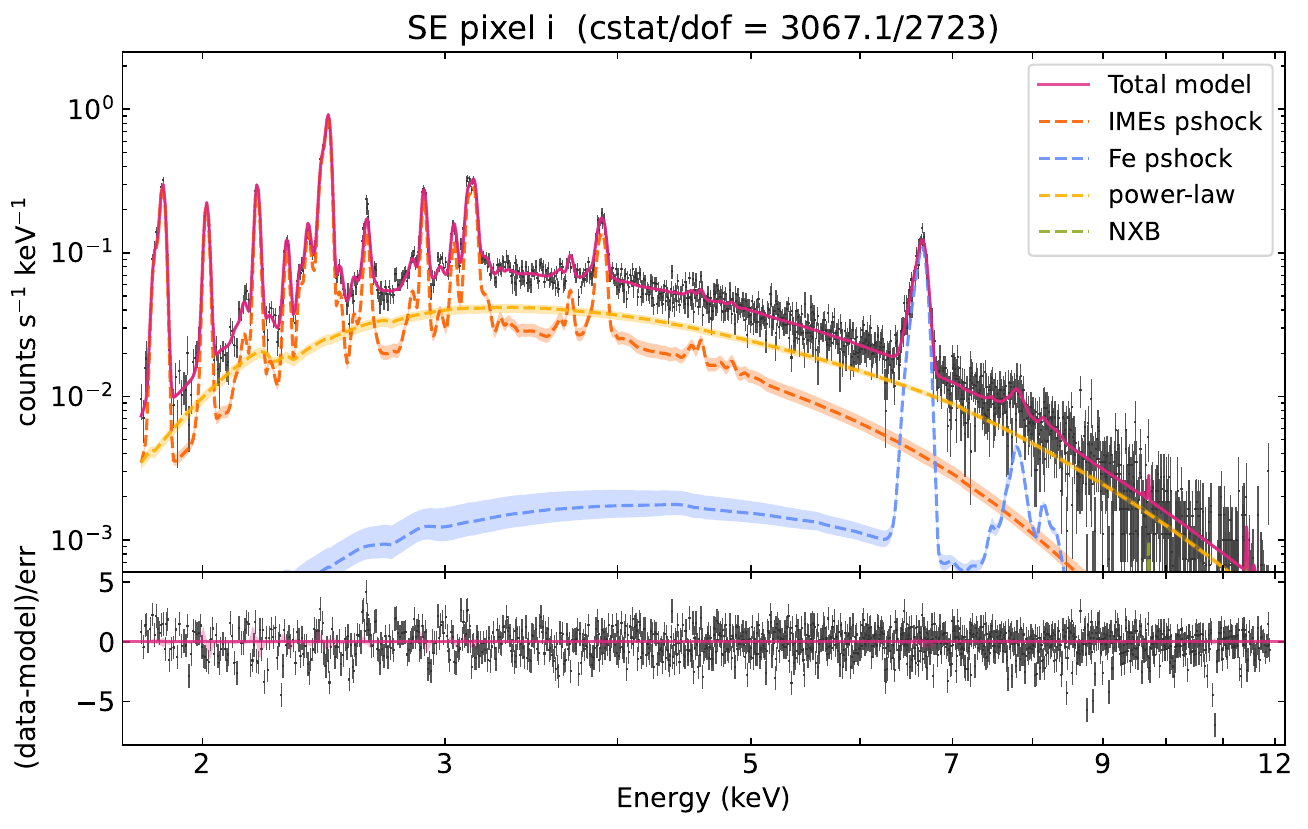}
}
\caption{Fitted spectra for all the super-pixels in the SE pointing, except pixel {\em e}. 
The total model is shown in a solid red line and the model components are shown in dashed lines with the corresponding $1\sigma$ uncertainties marked in shaded bands.
The data are represented by the black data points and have been rebinned to have minimum signal to noise ratio of 5 for display purposes only.
}
\label{fig:SE_spectra}
\end{figure*}

\begin{figure*}
\centerline{
  \includegraphics[width=0.49\textwidth]{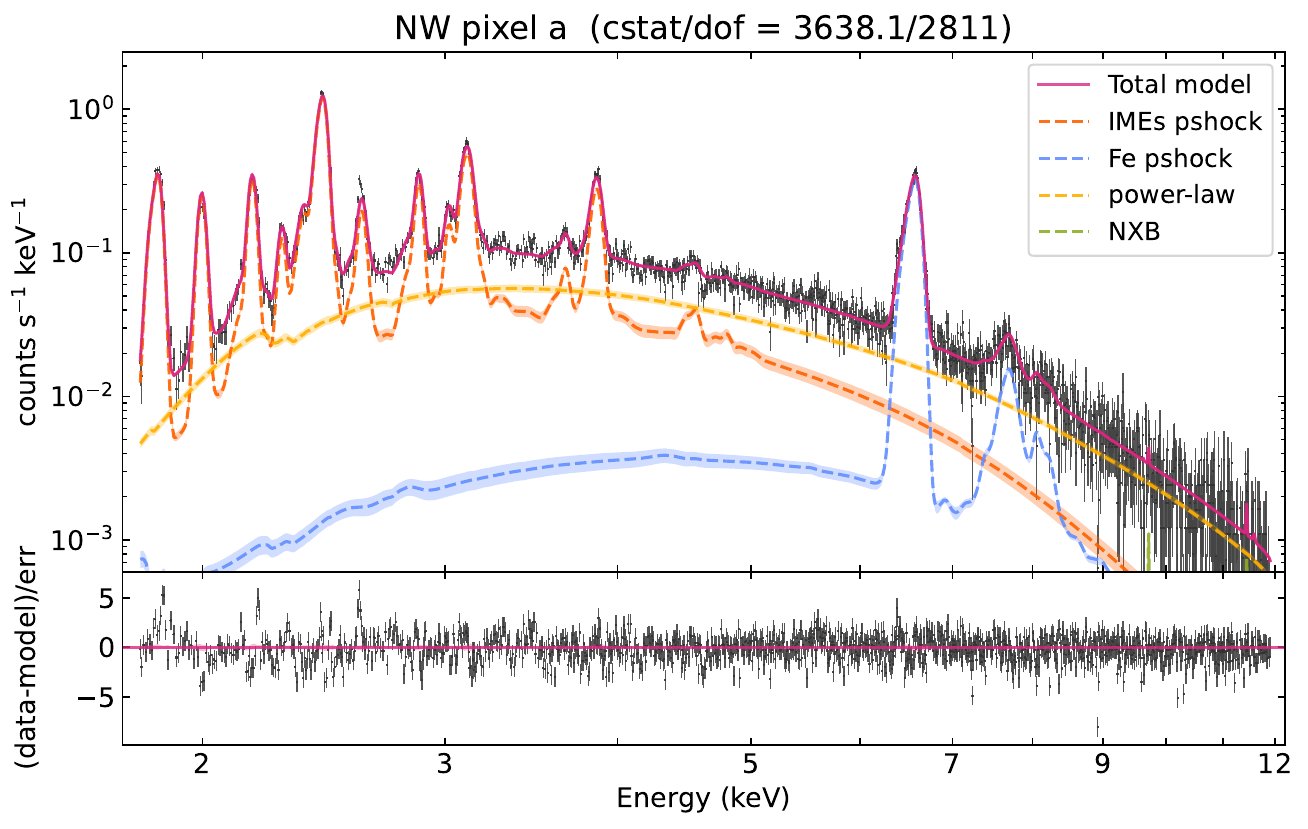}
  \hspace{2mm}
  \includegraphics[width=0.49\textwidth]{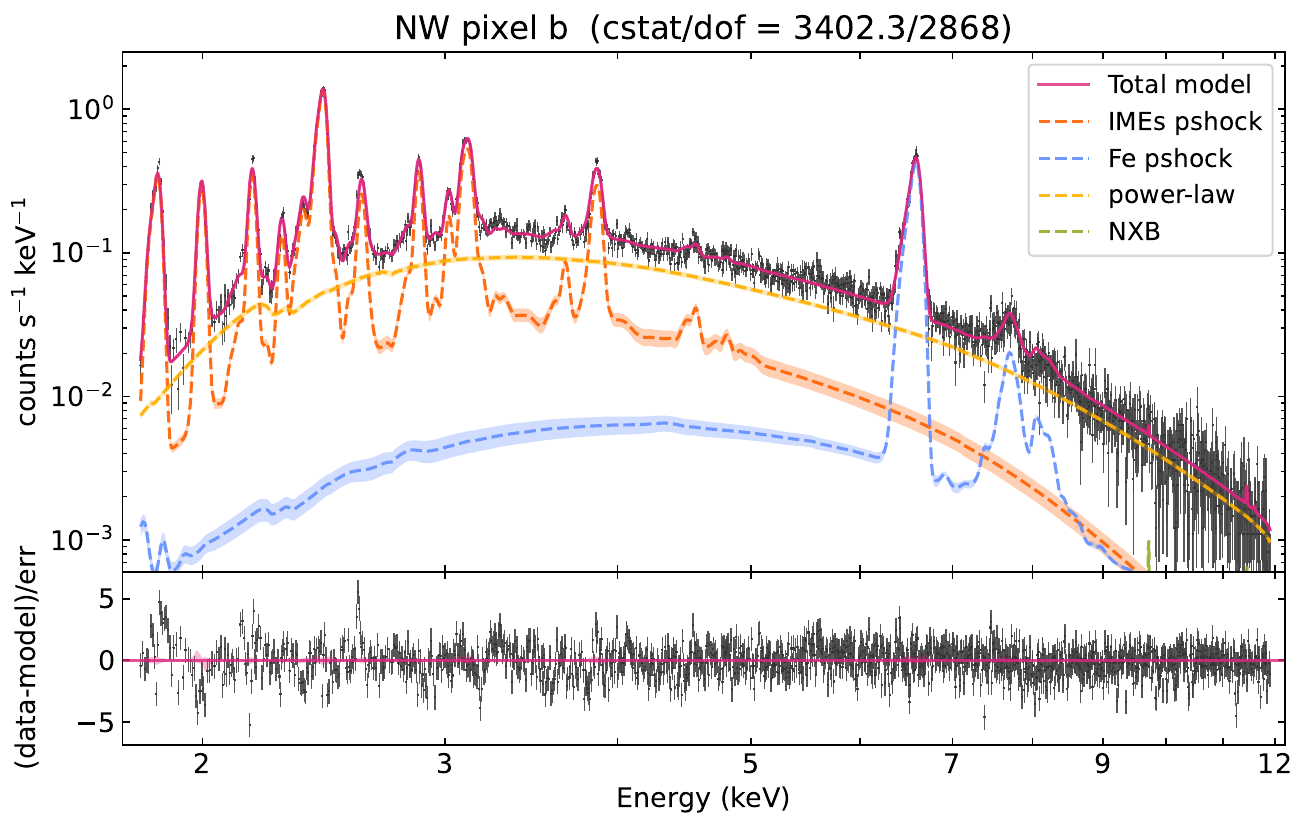}
}
\vspace{1mm}
\centerline{
  \includegraphics[width=0.49\textwidth]{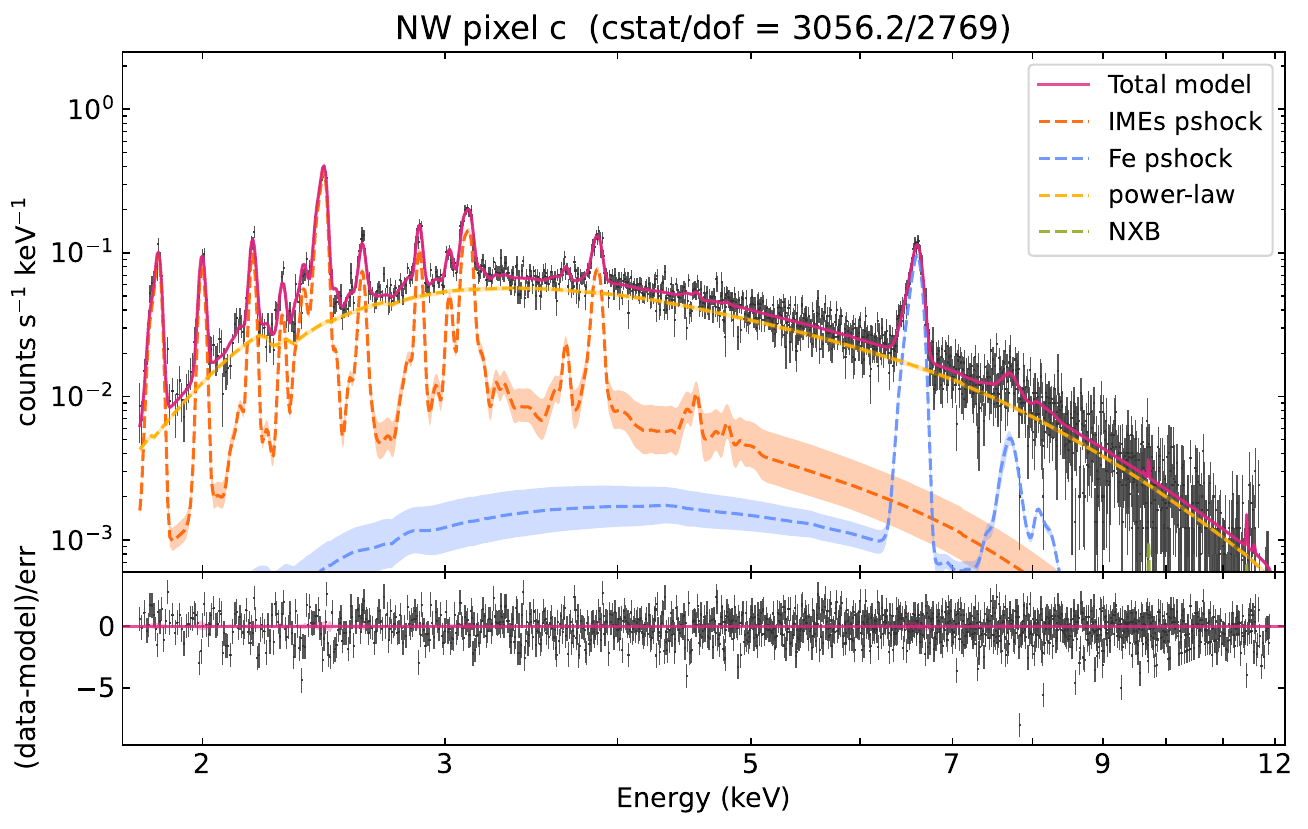}
  \hspace{2mm}
  \includegraphics[width=0.49\textwidth]{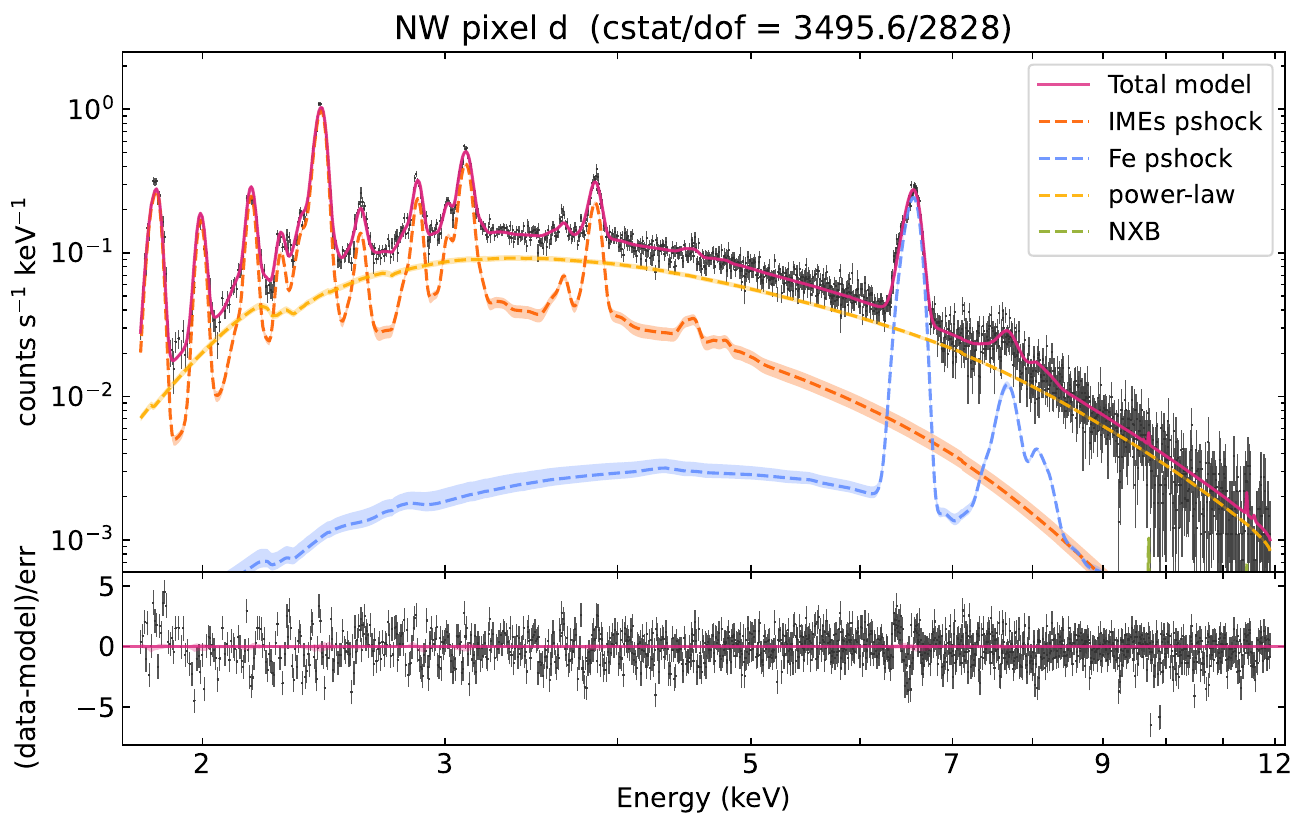}
}
\vspace{1mm}
\centerline{
  \includegraphics[width=0.49\textwidth]{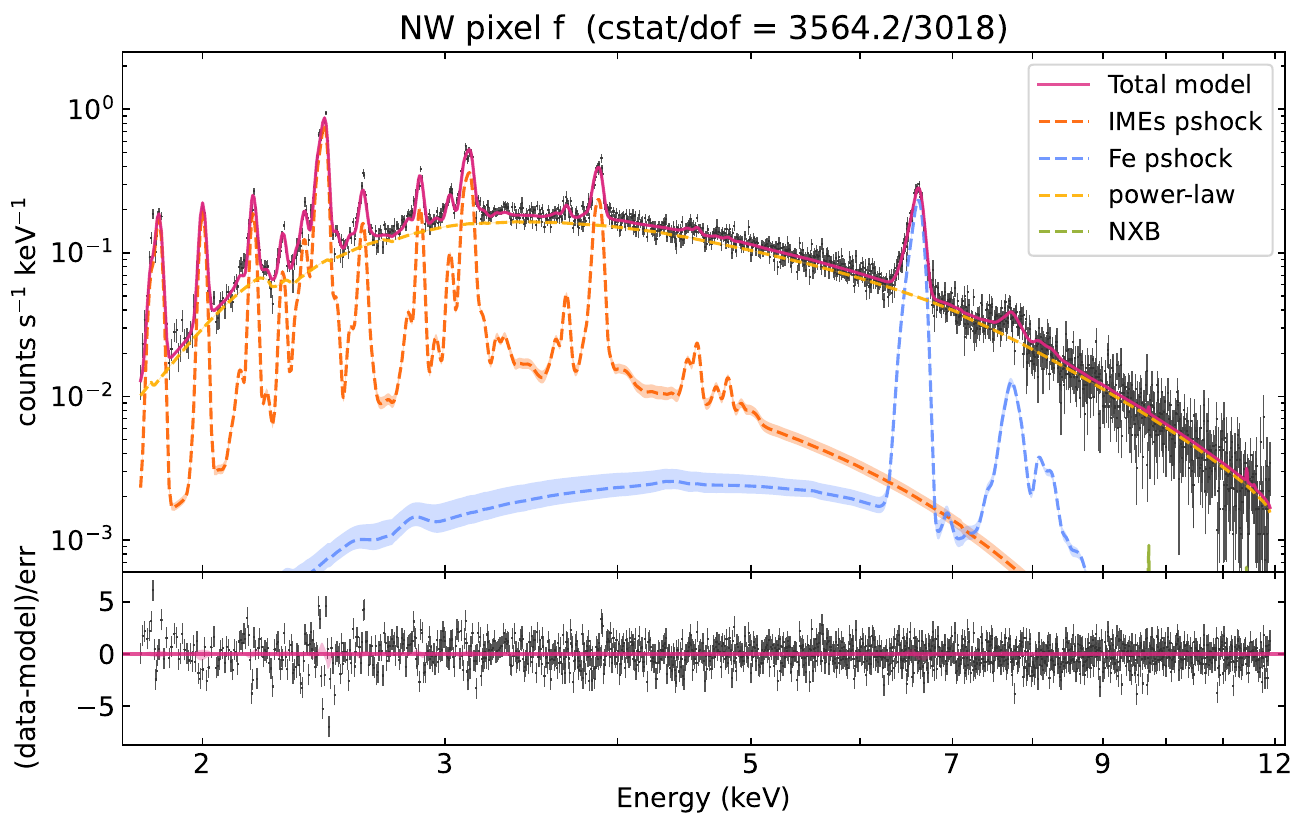}
  \hspace{2mm}
  \includegraphics[width=0.49\textwidth]{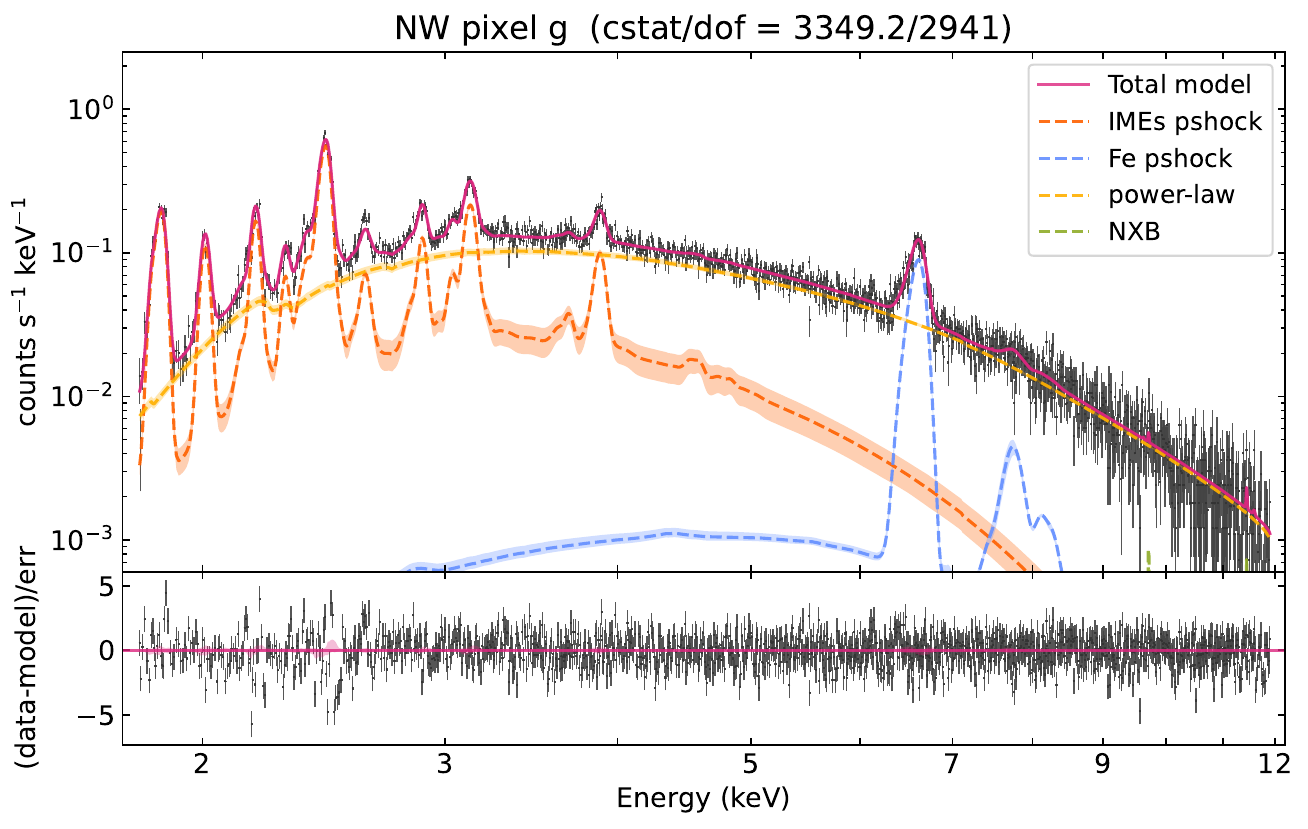}
}
\vspace{1mm}
\centerline{
  \includegraphics[width=0.49\textwidth]{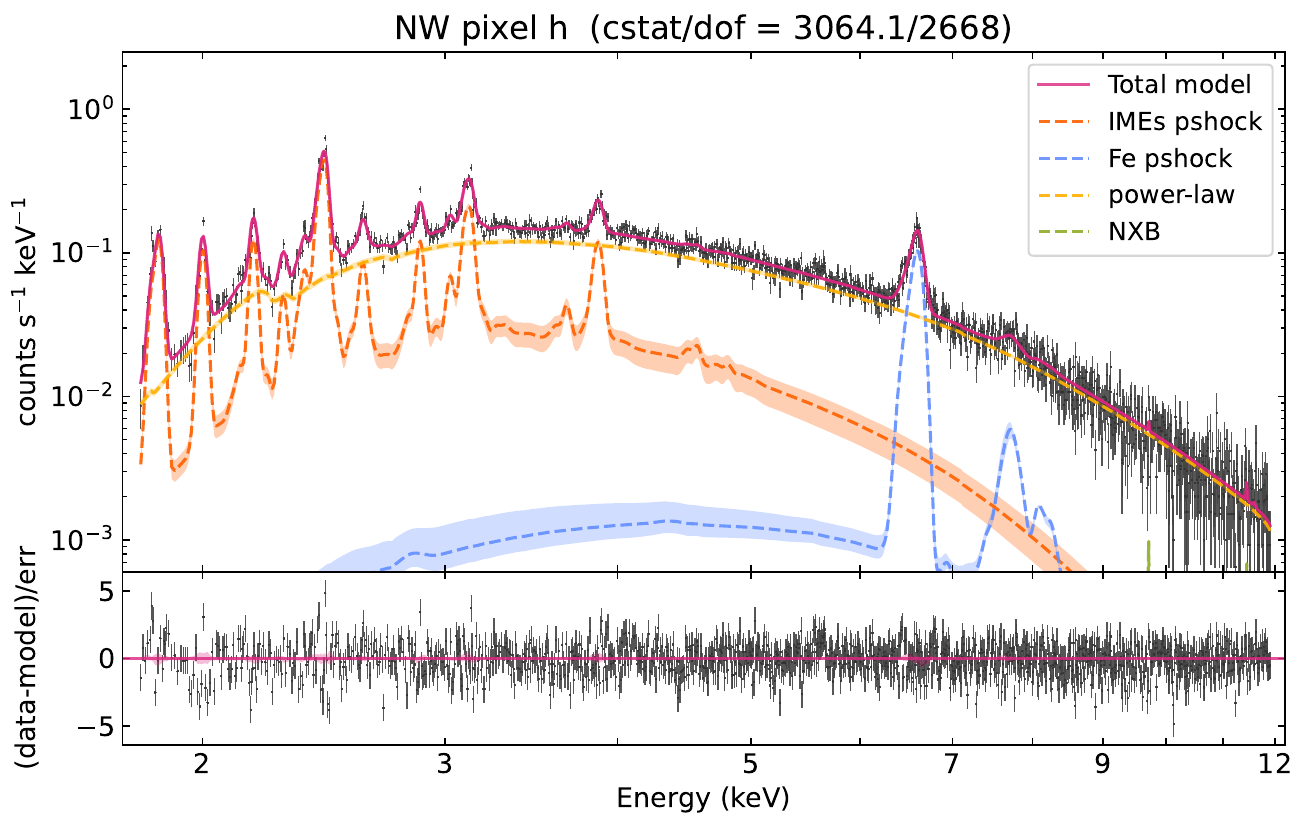}
  \hspace{2mm}
  \includegraphics[width=0.49\textwidth]{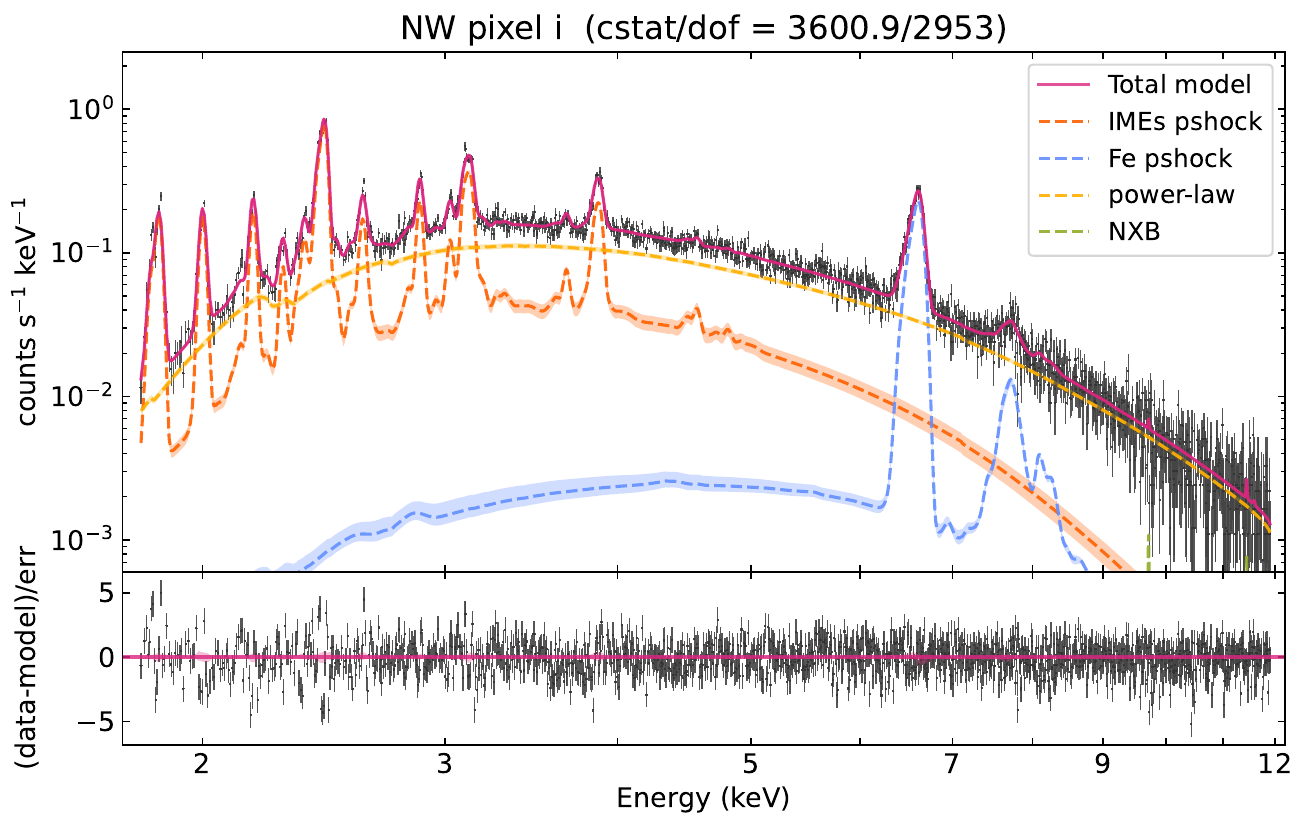}
}
\caption{Same as for Figure~\ref{fig:SE_spectra} but for the super-pixels in the NW pointing.
}
\label{fig:NW_spectra}
\end{figure*}

We fitted the \xrism/Resolve spectra from super-pixel regions with a pure-metal plasma model using {\em UltraSPEX}. 
Figures~\ref{fig:SE_spectra} and \ref{fig:NW_spectra} show the resulting fitted spectra along with the model components --- IMEs \texttt{pshock}, IGEs \texttt{pshock}, power-law, and NXB --- for all super-pixels in the SE and NW pointing, respectively, except for super-pixel {\em e} spectra, which are shown in Figure~\ref{fig:pixel_e_spectra}. 
The corresponding parameter values of the fitted model (median values of the posterior distribution) are listed in Tables \ref{tab:SE_median_par} and \ref{tab:NW_median_par} for the SE and NW pointing, respectively. 

\begin{deluxetable}{l c c c c c c c c c c}
    \tablecaption{Fitted parameter values obtained via {\em UltraSPEX} for the SE super-pixels.
    \label{tab:SE_median_par}}
    
    \tablehead{  & \colhead{Prior} &
    \colhead{SE pixel {\it a}} & \colhead{SE pixel {\it b}} & \colhead{SE pixel {\it c}} & \colhead{SE pixel {\it d}} & \colhead{SE pixel {\it e}} & \colhead{SE pixel {\it f}} & \colhead{SE pixel {\it g}} & \colhead{SE pixel {\it h}} & \colhead{SE pixel {\it i}} 
    }
    
    \startdata
    $N_{\rm H}\ (\times 10^{22}\ {\rm cm}^{-2} )$ & $-$ & 1.01 & 1.12 & 1.56 & 1.00 & 1.02 & 1.25 & 1.31 & 1.22 & 1.49 \\
    \hline
    IME-pshock & & & & & & & & & \\
    \hline
    $n_{\rm e}t\ (\times 10^{11}\ {\rm cm}^{-3}s $) & $\mathcal{LU}(0.5,10)$ & \makebox[0pt][r]{$3.00$}\makebox[0pt][l]{${}^{+0.43}_{-0.37}$} & \makebox[0pt][r]{$1.47$}\makebox[0pt][l]{${}^{+0.14}_{-0.12}$} & \makebox[0pt][r]{$1.05$}\makebox[0pt][l]{${}^{+0.14}_{-0.11}$} & \makebox[0pt][r]{$2.53$}\makebox[0pt][l]{${}^{+0.36}_{-0.35}$} & \makebox[0pt][r]{$2.23$}\makebox[0pt][l]{${}^{+0.14}_{-0.12}$} & \makebox[0pt][r]{$1.29$}\makebox[0pt][l]{${}^{+0.09}_{-0.09}$} & \makebox[0pt][r]{$2.24$}\makebox[0pt][l]{${}^{+0.48}_{-0.41}$} & \makebox[0pt][r]{$2.59$}\makebox[0pt][l]{${}^{+0.27}_{-0.25}$} & \makebox[0pt][r]{$1.84$}\makebox[0pt][l]{${}^{+0.17}_{-0.15}$} \\
    $kT_{\rm e}$ (keV) & $\mathcal{U}(1,5)$ & \makebox[0pt][r]{$1.70$}\makebox[0pt][l]{${}^{+0.10}_{-0.09}$} & \makebox[0pt][r]{$1.74$}\makebox[0pt][l]{${}^{+0.09}_{-0.09}$} & \makebox[0pt][r]{$1.79$}\makebox[0pt][l]{${}^{+0.14}_{-0.13}$} & \makebox[0pt][r]{$1.93$}\makebox[0pt][l]{${}^{+0.16}_{-0.13}$} & \makebox[0pt][r]{$1.76$}\makebox[0pt][l]{${}^{+0.05}_{-0.05}$} & \makebox[0pt][r]{$1.82$}\makebox[0pt][l]{${}^{+0.08}_{-0.07}$} & \makebox[0pt][r]{$1.77$}\makebox[0pt][l]{${}^{+0.20}_{-0.16}$} & \makebox[0pt][r]{$1.74$}\makebox[0pt][l]{${}^{+0.09}_{-0.08}$} & \makebox[0pt][r]{$1.60$}\makebox[0pt][l]{${}^{+0.07}_{-0.07}$} \\ 
    O$^*$/Si & $\mathcal{LU}(0.01,10)$ & \makebox[0pt][r]{$0.88$}\makebox[0pt][l]{${}^{+0.32}_{-0.31}$} & \makebox[0pt][r]{$1.10$}\makebox[0pt][l]{${}^{+0.12}_{-0.13}$} & \makebox[0pt][r]{$1.14$}\makebox[0pt][l]{${}^{+0.25}_{-0.24}$} & \makebox[0pt][r]{$0.06$}\makebox[0pt][l]{${}^{+0.15}_{-0.04}$} & \makebox[0pt][r]{$0.85$}\makebox[0pt][l]{${}^{+0.13}_{-0.14}$} & \makebox[0pt][r]{$1.10$}\makebox[0pt][l]{${}^{+0.14}_{-0.14}$} & \makebox[0pt][r]{$0.93$}\makebox[0pt][l]{${}^{+0.62}_{-0.60}$} & \makebox[0pt][r]{$0.62$}\makebox[0pt][l]{${}^{+0.22}_{-0.20}$} & \makebox[0pt][r]{$1.25$}\makebox[0pt][l]{${}^{+0.19}_{-0.18}$} \\ 
    S/Si & $\mathcal{U}(0.7,2.0)$ & \makebox[0pt][r]{$1.11$}\makebox[0pt][l]{${}^{+0.03}_{-0.03}$} & \makebox[0pt][r]{$1.03$}\makebox[0pt][l]{${}^{+0.02}_{-0.02}$} & \makebox[0pt][r]{$0.88$}\makebox[0pt][l]{${}^{+0.02}_{-0.02}$} & \makebox[0pt][r]{$1.02$}\makebox[0pt][l]{${}^{+0.03}_{-0.03}$} & \makebox[0pt][r]{$1.12$}\makebox[0pt][l]{${}^{+0.02}_{-0.02}$} & \makebox[0pt][r]{$1.02$}\makebox[0pt][l]{${}^{+0.02}_{-0.02}$} & \makebox[0pt][r]{$1.01$}\makebox[0pt][l]{${}^{+0.05}_{-0.04}$} & \makebox[0pt][r]{$1.06$}\makebox[0pt][l]{${}^{+0.02}_{-0.02}$} & \makebox[0pt][r]{$1.00$}\makebox[0pt][l]{${}^{+0.02}_{-0.02}$} \\ 
    Ar/Si & $\mathcal{U}(0.7,2.0)$ & \makebox[0pt][r]{$1.06$}\makebox[0pt][l]{${}^{+0.05}_{-0.04}$} & \makebox[0pt][r]{$0.92$}\makebox[0pt][l]{${}^{+0.02}_{-0.03}$} & \makebox[0pt][r]{$0.80$}\makebox[0pt][l]{${}^{+0.03}_{-0.03}$} & \makebox[0pt][r]{$0.96$}\makebox[0pt][l]{${}^{+0.04}_{-0.04}$} & \makebox[0pt][r]{$0.99$}\makebox[0pt][l]{${}^{+0.02}_{-0.02}$} & \makebox[0pt][r]{$0.87$}\makebox[0pt][l]{${}^{+0.02}_{-0.02}$} & \makebox[0pt][r]{$0.84$}\makebox[0pt][l]{${}^{+0.07}_{-0.06}$} & \makebox[0pt][r]{$0.97$}\makebox[0pt][l]{${}^{+0.04}_{-0.03}$} & \makebox[0pt][r]{$0.82$}\makebox[0pt][l]{${}^{+0.03}_{-0.03}$} \\ 
    Ca/Si & $\mathcal{U}(0.7,2.0)$ & \makebox[0pt][r]{$1.17$}\makebox[0pt][l]{${}^{+0.06}_{-0.06}$} & \makebox[0pt][r]{$1.02$}\makebox[0pt][l]{${}^{+0.03}_{-0.04}$} & \makebox[0pt][r]{$0.85$}\makebox[0pt][l]{${}^{+0.05}_{-0.05}$} & \makebox[0pt][r]{$0.98$}\makebox[0pt][l]{${}^{+0.05}_{-0.05}$} & \makebox[0pt][r]{$1.11$}\makebox[0pt][l]{${}^{+0.03}_{-0.03}$} & \makebox[0pt][r]{$0.89$}\makebox[0pt][l]{${}^{+0.03}_{-0.03}$} & \makebox[0pt][r]{$0.92$}\makebox[0pt][l]{${}^{+0.10}_{-0.09}$} & \makebox[0pt][r]{$0.94$}\makebox[0pt][l]{${}^{+0.04}_{-0.04}$} & \makebox[0pt][r]{$0.81$}\makebox[0pt][l]{${}^{+0.04}_{-0.03}$} \\ 
    Redshift (\kms) & $\mathcal{U}(-4500,4500)$ & \makebox[0pt][r]{$-1064$}\makebox[0pt][l]{${}^{+26}_{-3}$} & \makebox[0pt][r]{$-811$}\makebox[0pt][l]{${}^{+13}_{-4}$} & \makebox[0pt][r]{$-4$}\makebox[0pt][l]{${}^{+60}_{-35}$} & \makebox[0pt][r]{$-1125$}\makebox[0pt][l]{${}^{+5}_{-17}$} & \makebox[0pt][r]{$-979$}\makebox[0pt][l]{${}^{+1}_{-1}$} & \makebox[0pt][r]{$-1246$}\makebox[0pt][l]{${}^{+3}_{-9}$} & \makebox[0pt][r]{$-947$}\makebox[0pt][l]{${}^{+34}_{-80}$} & \makebox[0pt][r]{$-906$}\makebox[0pt][l]{${}^{+4}_{-27}$} & \makebox[0pt][r]{$-1107$}\makebox[0pt][l]{${}^{+54}_{-8}$} \\ 
    $\sigma_v$ (\kms) & $\mathcal{U}(500,3500)$ & \makebox[0pt][r]{$1538$}\makebox[0pt][l]{${}^{+23}_{-23}$} & \makebox[0pt][r]{$1513$}\makebox[0pt][l]{${}^{+20}_{-15}$} & \makebox[0pt][r]{$2200$}\makebox[0pt][l]{${}^{+47}_{-41}$} & \makebox[0pt][r]{$1207$}\makebox[0pt][l]{${}^{+18}_{-19}$} & \makebox[0pt][r]{$1284$}\makebox[0pt][l]{${}^{+9}_{-9}$} & \makebox[0pt][r]{$1437$}\makebox[0pt][l]{${}^{+12}_{-12}$} & \makebox[0pt][r]{$1318$}\makebox[0pt][l]{${}^{+40}_{-37}$} & \makebox[0pt][r]{$1167$}\makebox[0pt][l]{${}^{+16}_{-16}$} & \makebox[0pt][r]{$1285$}\makebox[0pt][l]{${}^{+15}_{-15}$} \\ 
    norm ($\times 10^{53}\ {\rm cm}^{-3} $)  & $\mathcal{LU}(0.01,50)$  & \makebox[0pt][r]{$1.37$}\makebox[0pt][l]{${}^{+0.08}_{-0.07}$} & \makebox[0pt][r]{$3.49$}\makebox[0pt][l]{${}^{+0.16}_{-0.14}$} & \makebox[0pt][r]{$2.67$}\makebox[0pt][l]{${}^{+0.18}_{-0.17}$} & \makebox[0pt][r]{$1.11$}\makebox[0pt][l]{${}^{+0.07}_{-0.07}$} & \makebox[0pt][r]{$4.45$}\makebox[0pt][l]{${}^{+0.13}_{-0.12}$} & \makebox[0pt][r]{$4.21$}\makebox[0pt][l]{${}^{+0.15}_{-0.15}$} & \makebox[0pt][r]{$0.56$}\makebox[0pt][l]{${}^{+0.06}_{-0.05}$} & \makebox[0pt][r]{$1.77$}\makebox[0pt][l]{${}^{+0.08}_{-0.08}$} & \makebox[0pt][r]{$2.95$}\makebox[0pt][l]{${}^{+0.13}_{-0.12}$} \\ 
    \hline
    IGE-pshock & & & & & & & & & \\
    \hline
    $n_{\rm e}t\ (\times 10^{11}\ {\rm cm}^{-3}s $) & $\mathcal{LU}(0.5,10)$ & \makebox[0pt][r]{$2.30$}\makebox[0pt][l]{${}^{+0.42}_{-0.30}$} & \makebox[0pt][r]{$1.00$}\makebox[0pt][l]{${}^{+0.07}_{-0.06}$} & \makebox[0pt][r]{$0.87$}\makebox[0pt][l]{${}^{+0.16}_{-0.12}$} & \makebox[0pt][r]{$2.94$}\makebox[0pt][l]{${}^{+0.32}_{-0.29}$} & \makebox[0pt][r]{$1.53$}\makebox[0pt][l]{${}^{+0.08}_{-0.08}$} & \makebox[0pt][r]{$0.98$}\makebox[0pt][l]{${}^{+0.08}_{-0.06}$} & \makebox[0pt][r]{$2.17$}\makebox[0pt][l]{${}^{+0.33}_{-0.27}$} & \makebox[0pt][r]{$2.20$}\makebox[0pt][l]{${}^{+0.24}_{-0.18}$} & \makebox[0pt][r]{$1.05$}\makebox[0pt][l]{${}^{+0.14}_{-0.11}$} \\ 
    $kT_{\rm e}$ (keV) & $\mathcal{U}(1,10)$ & \makebox[0pt][r]{$2.73$}\makebox[0pt][l]{${}^{+0.62}_{-0.49}$} & \makebox[0pt][r]{$7.89$}\makebox[0pt][l]{${}^{+1.38}_{-1.90}$} & \makebox[0pt][r]{$6.79$}\makebox[0pt][l]{${}^{+1.97}_{-2.26}$} & \makebox[0pt][r]{$2.42$}\makebox[0pt][l]{${}^{+0.22}_{-0.16}$} & \makebox[0pt][r]{$3.52$}\makebox[0pt][l]{${}^{+0.29}_{-0.24}$} & \makebox[0pt][r]{$8.32$}\makebox[0pt][l]{${}^{+1.15}_{-1.76}$} & \makebox[0pt][r]{$3.25$}\makebox[0pt][l]{${}^{+0.61}_{-0.51}$} & \makebox[0pt][r]{$3.01$}\makebox[0pt][l]{${}^{+0.34}_{-0.29}$} & \makebox[0pt][r]{$4.23$}\makebox[0pt][l]{${}^{+0.94}_{-0.80}$} \\ 
    Ni/Fe & $\mathcal{U}(0.05,2.0)$ & \makebox[0pt][r]{$1.31$}\makebox[0pt][l]{${}^{+0.25}_{-0.26}$} & \makebox[0pt][r]{$0.54$}\makebox[0pt][l]{${}^{+0.16}_{-0.16}$} & \makebox[0pt][r]{$0.29$}\makebox[0pt][l]{${}^{+0.23}_{-0.16}$} & \makebox[0pt][r]{$0.77$}\makebox[0pt][l]{${}^{+0.17}_{-0.16}$} & \makebox[0pt][r]{$0.64$}\makebox[0pt][l]{${}^{+0.09}_{-0.09}$} & \makebox[0pt][r]{$0.42$}\makebox[0pt][l]{${}^{+0.17}_{-0.15}$} & \makebox[0pt][r]{$1.13$}\makebox[0pt][l]{${}^{+0.27}_{-0.24}$} & \makebox[0pt][r]{$0.71$}\makebox[0pt][l]{${}^{+0.17}_{-0.16}$} & \makebox[0pt][r]{$0.18$}\makebox[0pt][l]{${}^{+0.16}_{-0.10}$} \\ 
    Redshift (\kms) & $\mathcal{U}(-4500,4500)$ & \makebox[0pt][r]{$-1230$}\makebox[0pt][l]{${}^{+305}_{-364}$} & \makebox[0pt][r]{$-1259$}\makebox[0pt][l]{${}^{+174}_{-164}$} & \makebox[0pt][r]{$338$}\makebox[0pt][l]{${}^{+464}_{-427}$} & \makebox[0pt][r]{$-1154$}\makebox[0pt][l]{${}^{+160}_{-173}$} & \makebox[0pt][r]{$-1574$}\makebox[0pt][l]{${}^{+89}_{-91}$} & \makebox[0pt][r]{$-719$}\makebox[0pt][l]{${}^{+208}_{-193}$} & \makebox[0pt][r]{$-1018$}\makebox[0pt][l]{${}^{+173}_{-174}$} & \makebox[0pt][r]{$-974$}\makebox[0pt][l]{${}^{+120}_{-114}$} & \makebox[0pt][r]{$-1363$}\makebox[0pt][l]{${}^{+268}_{-288}$} \\ 
    $\sigma_v$ (\kms) & $\mathcal{U}(500,4500)$ & \makebox[0pt][r]{$1523$}\makebox[0pt][l]{${}^{+142}_{-186}$} & \makebox[0pt][r]{$1708$}\makebox[0pt][l]{${}^{+88}_{-79}$} & \makebox[0pt][r]{$3387$}\makebox[0pt][l]{${}^{+155}_{-167}$} & \makebox[0pt][r]{$1430$}\makebox[0pt][l]{${}^{+78}_{-92}$} & \makebox[0pt][r]{$1421$}\makebox[0pt][l]{${}^{+46}_{-45}$} & \makebox[0pt][r]{$2128$}\makebox[0pt][l]{${}^{+92}_{-92}$} & \makebox[0pt][r]{$1294$}\makebox[0pt][l]{${}^{+99}_{-98}$} & \makebox[0pt][r]{$1333$}\makebox[0pt][l]{${}^{+65}_{-70}$} & \makebox[0pt][r]{$1595$}\makebox[0pt][l]{${}^{+129}_{-133}$} \\ 
    norm ($\times 10^{53}\ {\rm cm}^{-3} $)  & $\mathcal{LU}(0.01,50)$  & \makebox[0pt][r]{$0.66$}\makebox[0pt][l]{${}^{+0.34}_{-0.21}$} & \makebox[0pt][r]{$0.32$}\makebox[0pt][l]{${}^{+0.09}_{-0.04}$} & \makebox[0pt][r]{$0.25$}\makebox[0pt][l]{${}^{+0.15}_{-0.05}$} & \makebox[0pt][r]{$1.38$}\makebox[0pt][l]{${}^{+0.23}_{-0.22}$} & \makebox[0pt][r]{$2.11$}\makebox[0pt][l]{${}^{+0.24}_{-0.23}$} & \makebox[0pt][r]{$0.35$}\makebox[0pt][l]{${}^{+0.08}_{-0.03}$} & \makebox[0pt][r]{$0.38$}\makebox[0pt][l]{${}^{+0.12}_{-0.08}$} & \makebox[0pt][r]{$0.92$}\makebox[0pt][l]{${}^{+0.18}_{-0.15}$} & \makebox[0pt][r]{$0.42$}\makebox[0pt][l]{${}^{+0.15}_{-0.09}$} \\ 
    \hline
    power-law & & & & & & & & & \\
    \hline
    $\Gamma$ & $\mathcal{U}(2.7,3.5)$ & \makebox[0pt][r]{$3.08$}\makebox[0pt][l]{${}^{+0.07}_{-0.08}$} & \makebox[0pt][r]{$3.34$}\makebox[0pt][l]{${}^{+0.05}_{-0.05}$} & \makebox[0pt][r]{$3.24$}\makebox[0pt][l]{${}^{+0.04}_{-0.04}$} & \makebox[0pt][r]{$3.16$}\makebox[0pt][l]{${}^{+0.05}_{-0.06}$} & \makebox[0pt][r]{$3.43$}\makebox[0pt][l]{${}^{+0.04}_{-0.04}$} & \makebox[0pt][r]{$3.31$}\makebox[0pt][l]{${}^{+0.03}_{-0.03}$} & \makebox[0pt][r]{$2.95$}\makebox[0pt][l]{${}^{+0.08}_{-0.10}$} & \makebox[0pt][r]{$3.25$}\makebox[0pt][l]{${}^{+0.04}_{-0.04}$} & \makebox[0pt][r]{$3.27$}\makebox[0pt][l]{${}^{+0.05}_{-0.06}$} \\ 
    norm $(\times 10^{44})$ & $\mathcal{U}(0.01,4.00)$ & \makebox[0pt][r]{$0.15$}\makebox[0pt][l]{${}^{+0.02}_{-0.03}$} & \makebox[0pt][r]{$0.50$}\makebox[0pt][l]{${}^{+0.04}_{-0.04}$} & \makebox[0pt][r]{$0.64$}\makebox[0pt][l]{${}^{+0.04}_{-0.06}$} & \makebox[0pt][r]{$0.21$}\makebox[0pt][l]{${}^{+0.02}_{-0.02}$} & \makebox[0pt][r]{$0.60$}\makebox[0pt][l]{${}^{+0.05}_{-0.05}$} & \makebox[0pt][r]{$0.79$}\makebox[0pt][l]{${}^{+0.05}_{-0.05}$} & \makebox[0pt][r]{$0.08$}\makebox[0pt][l]{${}^{+0.02}_{-0.02}$} & \makebox[0pt][r]{$0.30$}\makebox[0pt][l]{${}^{+0.03}_{-0.03}$} & \makebox[0pt][r]{$0.44$}\makebox[0pt][l]{${}^{+0.05}_{-0.06}$} \\
    (ph s$^{-1}$ keV$^{-1}$ at 1.0 keV) & & & & & & & & & \\
    \hline
    Fit statistics & & & & & & & & & \\
    \hline
    C-statistic & & 3133.86 & 3916.85 & 3369.40 & 3005.10 & 3789.51 & 3675.47 & 2671.17 & 2807.73 & 3067.07 \\
    Degrees of freedom & & 2581 & 2741 & 2877 & 2602 & 2854 & 2918 & 2493 & 2539 & 2723
    \enddata
    
    \tablecomments{The prior limits used for the model parameter fitting via {\em UltraSPEX} are given in the second column along with the corresponding prior distribution --- logarithmic uniform $(\mathcal{LU})$ and uniform $(\mathcal{U})$. \\
    Normalization (emission measure) is calculated considering the distance as 3.4 kpc.
    }
    \end{deluxetable}

    \begin{deluxetable}{l c c c c c c c c c c}
    \tablecaption{Fitted parameter values obtained via {\em UltraSPEX} for the NW super-pixels.
    \label{tab:NW_median_par}}
    
    \tablehead{  & \colhead{Prior} &
    \colhead{NW pixel {\it a}} & \colhead{NW pixel {\it b}} & \colhead{NW pixel {\it c}} & \colhead{NW pixel {\it d}} & \colhead{NW pixel {\it e}} & \colhead{NW pixel {\it f}} & \colhead{NW pixel {\it g}} & \colhead{NW pixel {\it h}} & \colhead{NW pixel {\it i}} 
    }
    
    \startdata
    $N_{\rm H}\ (\times 10^{22}\ {\rm cm}^{-2} )$ & $-$ & 1.20 & 1.30 & 1.59 & 1.45 & 1.45 & 2.30 & 1.71 & 1.61 & 1.74 \\
    \hline
    IME-pshock & & & & & & & & & \\
    \hline
    $n_{\rm e}t\ (\times 10^{11}\ {\rm cm}^{-3}s $)  & $\mathcal{LU}(0.5,10)$  & \makebox[0pt][r]{$1.46$}\makebox[0pt][l]{${}^{+0.10}_{-0.10}$} & \makebox[0pt][r]{$1.51$}\makebox[0pt][l]{${}^{+0.10}_{-0.10}$} & \makebox[0pt][r]{$1.49$}\makebox[0pt][l]{${}^{+0.27}_{-0.21}$} & \makebox[0pt][r]{$1.66$}\makebox[0pt][l]{${}^{+0.13}_{-0.12}$} & \makebox[0pt][r]{$1.55$}\makebox[0pt][l]{${}^{+0.13}_{-0.12}$} & \makebox[0pt][r]{$3.44$}\makebox[0pt][l]{${}^{+0.45}_{-0.41}$} & \makebox[0pt][r]{$2.36$}\makebox[0pt][l]{${}^{+0.31}_{-0.30}$} & \makebox[0pt][r]{$2.27$}\makebox[0pt][l]{${}^{+0.38}_{-0.31}$} & \makebox[0pt][r]{$2.06$}\makebox[0pt][l]{${}^{+0.22}_{-0.20}$} \\ 
    $kT_{\rm e}$ (keV)  & $\mathcal{U}(1,5)$  & \makebox[0pt][r]{$1.89$}\makebox[0pt][l]{${}^{+0.09}_{-0.07}$} & \makebox[0pt][r]{$2.03$}\makebox[0pt][l]{${}^{+0.10}_{-0.08}$} & \makebox[0pt][r]{$2.07$}\makebox[0pt][l]{${}^{+0.25}_{-0.21}$} & \makebox[0pt][r]{$1.61$}\makebox[0pt][l]{${}^{+0.07}_{-0.06}$} & \makebox[0pt][r]{$2.01$}\makebox[0pt][l]{${}^{+0.11}_{-0.10}$} & \makebox[0pt][r]{$1.43$}\makebox[0pt][l]{${}^{+0.07}_{-0.07}$} & \makebox[0pt][r]{$1.30$}\makebox[0pt][l]{${}^{+0.08}_{-0.06}$} & \makebox[0pt][r]{$1.56$}\makebox[0pt][l]{${}^{+0.11}_{-0.10}$} & \makebox[0pt][r]{$1.77$}\makebox[0pt][l]{${}^{+0.10}_{-0.09}$} \\ 
    O$^*$/Si  & $\mathcal{LU}(0.01,10)$  & \makebox[0pt][r]{$0.90$}\makebox[0pt][l]{${}^{+0.14}_{-0.13}$} & \makebox[0pt][r]{$0.76$}\makebox[0pt][l]{${}^{+0.18}_{-0.17}$} & \makebox[0pt][r]{$0.54$}\makebox[0pt][l]{${}^{+0.44}_{-0.36}$} & \makebox[0pt][r]{$1.10$}\makebox[0pt][l]{${}^{+0.20}_{-0.21}$} & \makebox[0pt][r]{$2.24$}\makebox[0pt][l]{${}^{+0.24}_{-0.25}$} & \makebox[0pt][r]{$0.09$}\makebox[0pt][l]{${}^{+0.10}_{-0.06}$} & \makebox[0pt][r]{$1.09$}\makebox[0pt][l]{${}^{+0.37}_{-0.42}$} & \makebox[0pt][r]{$2.03$}\makebox[0pt][l]{${}^{+0.46}_{-0.48}$} & \makebox[0pt][r]{$2.15$}\makebox[0pt][l]{${}^{+0.31}_{-0.34}$} \\ 
    S/Si  & $\mathcal{U}(0.7,2.0)$  & \makebox[0pt][r]{$1.06$}\makebox[0pt][l]{${}^{+0.02}_{-0.02}$} & \makebox[0pt][r]{$1.12$}\makebox[0pt][l]{${}^{+0.02}_{-0.02}$} & \makebox[0pt][r]{$1.09$}\makebox[0pt][l]{${}^{+0.04}_{-0.04}$} & \makebox[0pt][r]{$1.09$}\makebox[0pt][l]{${}^{+0.02}_{-0.02}$} & \makebox[0pt][r]{$1.11$}\makebox[0pt][l]{${}^{+0.02}_{-0.02}$} & \makebox[0pt][r]{$1.02$}\makebox[0pt][l]{${}^{+0.03}_{-0.02}$} & \makebox[0pt][r]{$0.90$}\makebox[0pt][l]{${}^{+0.02}_{-0.02}$} & \makebox[0pt][r]{$0.99$}\makebox[0pt][l]{${}^{+0.03}_{-0.03}$} & \makebox[0pt][r]{$1.12$}\makebox[0pt][l]{${}^{+0.03}_{-0.03}$} \\ 
    Ar/Si  & $\mathcal{U}(0.7,2.0)$  & \makebox[0pt][r]{$0.94$}\makebox[0pt][l]{${}^{+0.02}_{-0.02}$} & \makebox[0pt][r]{$0.96$}\makebox[0pt][l]{${}^{+0.03}_{-0.03}$} & \makebox[0pt][r]{$0.93$}\makebox[0pt][l]{${}^{+0.06}_{-0.05}$} & \makebox[0pt][r]{$1.01$}\makebox[0pt][l]{${}^{+0.03}_{-0.03}$} & \makebox[0pt][r]{$1.02$}\makebox[0pt][l]{${}^{+0.03}_{-0.03}$} & \makebox[0pt][r]{$1.04$}\makebox[0pt][l]{${}^{+0.04}_{-0.04}$} & \makebox[0pt][r]{$0.82$}\makebox[0pt][l]{${}^{+0.04}_{-0.04}$} & \makebox[0pt][r]{$1.03$}\makebox[0pt][l]{${}^{+0.05}_{-0.05}$} & \makebox[0pt][r]{$1.12$}\makebox[0pt][l]{${}^{+0.04}_{-0.04}$} \\ 
    Ca/Si  & $\mathcal{U}(0.7,2.0)$  & \makebox[0pt][r]{$1.15$}\makebox[0pt][l]{${}^{+0.04}_{-0.04}$} & \makebox[0pt][r]{$1.09$}\makebox[0pt][l]{${}^{+0.04}_{-0.04}$} & \makebox[0pt][r]{$0.99$}\makebox[0pt][l]{${}^{+0.08}_{-0.07}$} & \makebox[0pt][r]{$1.18$}\makebox[0pt][l]{${}^{+0.05}_{-0.05}$} & \makebox[0pt][r]{$1.26$}\makebox[0pt][l]{${}^{+0.05}_{-0.05}$} & \makebox[0pt][r]{$1.34$}\makebox[0pt][l]{${}^{+0.06}_{-0.06}$} & \makebox[0pt][r]{$0.83$}\makebox[0pt][l]{${}^{+0.06}_{-0.05}$} & \makebox[0pt][r]{$1.13$}\makebox[0pt][l]{${}^{+0.09}_{-0.08}$} & \makebox[0pt][r]{$1.31$}\makebox[0pt][l]{${}^{+0.06}_{-0.06}$} \\ 
    Redshift (\kms)  & $\mathcal{U}(-4500,4500)$  & \makebox[0pt][r]{$1360$}\makebox[0pt][l]{${}^{+7}_{-9}$} & \makebox[0pt][r]{$1421$}\makebox[0pt][l]{${}^{+30}_{-13}$} & \makebox[0pt][r]{$1181$}\makebox[0pt][l]{${}^{+28}_{-34}$} & \makebox[0pt][r]{$2029$}\makebox[0pt][l]{${}^{+11}_{-33}$} & \makebox[0pt][r]{$1992$}\makebox[0pt][l]{${}^{+12}_{-11}$} & \makebox[0pt][r]{$868$}\makebox[0pt][l]{${}^{+15}_{-20}$} & \makebox[0pt][r]{$-346$}\makebox[0pt][l]{${}^{+13}_{-73}$} & \makebox[0pt][r]{$779$}\makebox[0pt][l]{${}^{+41}_{-27}$} & \makebox[0pt][r]{$893$}\makebox[0pt][l]{${}^{+35}_{-4}$} \\ 
    $\sigma_v$ (\kms)  & $\mathcal{U}(500,3500)$  & \makebox[0pt][r]{$1950$}\makebox[0pt][l]{${}^{+18}_{-19}$} & \makebox[0pt][r]{$1664$}\makebox[0pt][l]{${}^{+15}_{-15}$} & \makebox[0pt][r]{$1492$}\makebox[0pt][l]{${}^{+32}_{-30}$} & \makebox[0pt][r]{$2190$}\makebox[0pt][l]{${}^{+24}_{-23}$} & \makebox[0pt][r]{$1941$}\makebox[0pt][l]{${}^{+24}_{-24}$} & \makebox[0pt][r]{$1482$}\makebox[0pt][l]{${}^{+29}_{-29}$} & \makebox[0pt][r]{$2093$}\makebox[0pt][l]{${}^{+34}_{-37}$} & \makebox[0pt][r]{$1757$}\makebox[0pt][l]{${}^{+47}_{-45}$} & \makebox[0pt][r]{$1543$}\makebox[0pt][l]{${}^{+24}_{-26}$} \\
    norm ($\times 10^{53}\ {\rm cm}^{-3} $)  & $\mathcal{LU}(0.01,50)$  & \makebox[0pt][r]{$4.48$}\makebox[0pt][l]{${}^{+0.15}_{-0.16}$} & \makebox[0pt][r]{$4.26$}\makebox[0pt][l]{${}^{+0.14}_{-0.15}$} & \makebox[0pt][r]{$1.66$}\makebox[0pt][l]{${}^{+0.14}_{-0.12}$} & \makebox[0pt][r]{$4.64$}\makebox[0pt][l]{${}^{+0.18}_{-0.18}$} & \makebox[0pt][r]{$3.20$}\makebox[0pt][l]{${}^{+0.13}_{-0.14}$} & \makebox[0pt][r]{$4.87$}\makebox[0pt][l]{${}^{+0.27}_{-0.27}$} & \makebox[0pt][r]{$4.22$}\makebox[0pt][l]{${}^{+0.26}_{-0.28}$} & \makebox[0pt][r]{$2.34$}\makebox[0pt][l]{${}^{+0.16}_{-0.15}$} & \makebox[0pt][r]{$3.15$}\makebox[0pt][l]{${}^{+0.15}_{-0.16}$} \\ 
    \hline
    IGE-pshock & & & & & & & & & \\
    \hline
    $n_{\rm e}t\ (\times 10^{11}\ {\rm cm}^{-3}s $)  & $\mathcal{LU}(0.5,10)$  & \makebox[0pt][r]{$0.84$}\makebox[0pt][l]{${}^{+0.05}_{-0.04}$} & \makebox[0pt][r]{$1.17$}\makebox[0pt][l]{${}^{+0.08}_{-0.07}$} & \makebox[0pt][r]{$1.21$}\makebox[0pt][l]{${}^{+0.28}_{-0.18}$} & \makebox[0pt][r]{$0.80$}\makebox[0pt][l]{${}^{+0.05}_{-0.05}$} & \makebox[0pt][r]{$1.03$}\makebox[0pt][l]{${}^{+0.07}_{-0.06}$} & \makebox[0pt][r]{$1.03$}\makebox[0pt][l]{${}^{+0.14}_{-0.09}$} & \makebox[0pt][r]{$0.94$}\makebox[0pt][l]{${}^{+0.14}_{-0.11}$} & \makebox[0pt][r]{$1.04$}\makebox[0pt][l]{${}^{+0.18}_{-0.13}$} & \makebox[0pt][r]{$0.85$}\makebox[0pt][l]{${}^{+0.06}_{-0.05}$} \\ 
    $kT_{\rm e}$ (keV)  & $\mathcal{U}(1,10)$  & \makebox[0pt][r]{$7.20$}\makebox[0pt][l]{${}^{+1.52}_{-1.26}$} & \makebox[0pt][r]{$4.40$}\makebox[0pt][l]{${}^{+0.78}_{-0.54}$} & \makebox[0pt][r]{$3.89$}\makebox[0pt][l]{${}^{+1.50}_{-0.96}$} & \makebox[0pt][r]{$7.68$}\makebox[0pt][l]{${}^{+1.39}_{-1.60}$} & \makebox[0pt][r]{$7.46$}\makebox[0pt][l]{${}^{+1.61}_{-1.55}$} & \makebox[0pt][r]{$7.16$}\makebox[0pt][l]{${}^{+1.96}_{-2.35}$} & \makebox[0pt][r]{$8.41$}\makebox[0pt][l]{${}^{+1.13}_{-1.59}$} & \makebox[0pt][r]{$6.14$}\makebox[0pt][l]{${}^{+2.17}_{-2.06}$} & \makebox[0pt][r]{$7.70$}\makebox[0pt][l]{${}^{+1.48}_{-1.62}$} \\ 
    Ni/Fe  & $\mathcal{U}(0.05,2.0)$  & \makebox[0pt][r]{$0.51$}\makebox[0pt][l]{${}^{+0.10}_{-0.10}$} & \makebox[0pt][r]{$0.35$}\makebox[0pt][l]{${}^{+0.11}_{-0.11}$} & \makebox[0pt][r]{$0.75$}\makebox[0pt][l]{${}^{+0.29}_{-0.25}$} & \makebox[0pt][r]{$0.54$}\makebox[0pt][l]{${}^{+0.14}_{-0.13}$} & \makebox[0pt][r]{$0.24$}\makebox[0pt][l]{${}^{+0.13}_{-0.11}$} & \makebox[0pt][r]{$0.73$}\makebox[0pt][l]{${}^{+0.22}_{-0.21}$} & \makebox[0pt][r]{$0.49$}\makebox[0pt][l]{${}^{+0.32}_{-0.26}$} & \makebox[0pt][r]{$0.81$}\makebox[0pt][l]{${}^{+0.35}_{-0.32}$} & \makebox[0pt][r]{$0.94$}\makebox[0pt][l]{${}^{+0.18}_{-0.18}$} \\ 
    Redshift (\kms)  & $\mathcal{U}(-4500,4500)$  & \makebox[0pt][r]{$2025$}\makebox[0pt][l]{${}^{+145}_{-158}$} & \makebox[0pt][r]{$2165$}\makebox[0pt][l]{${}^{+131}_{-134}$} & \makebox[0pt][r]{$1386$}\makebox[0pt][l]{${}^{+384}_{-372}$} & \makebox[0pt][r]{$2375$}\makebox[0pt][l]{${}^{+199}_{-198}$} & \makebox[0pt][r]{$2367$}\makebox[0pt][l]{${}^{+167}_{-176}$} & \makebox[0pt][r]{$983$}\makebox[0pt][l]{${}^{+225}_{-228}$} & \makebox[0pt][r]{$413$}\makebox[0pt][l]{${}^{+447}_{-433}$} & \makebox[0pt][r]{$1335$}\makebox[0pt][l]{${}^{+367}_{-375}$} & \makebox[0pt][r]{$639$}\makebox[0pt][l]{${}^{+209}_{-213}$} \\ 
    $\sigma_v$ (\kms)  & $\mathcal{U}(500,4500)$  & \makebox[0pt][r]{$1921$}\makebox[0pt][l]{${}^{+66}_{-66}$} & \makebox[0pt][r]{$1622$}\makebox[0pt][l]{${}^{+61}_{-62}$} & \makebox[0pt][r]{$1794$}\makebox[0pt][l]{${}^{+159}_{-160}$} & \makebox[0pt][r]{$2592$}\makebox[0pt][l]{${}^{+85}_{-79}$} & \makebox[0pt][r]{$2064$}\makebox[0pt][l]{${}^{+74}_{-73}$} & \makebox[0pt][r]{$1793$}\makebox[0pt][l]{${}^{+101}_{-101}$} & \makebox[0pt][r]{$2611$}\makebox[0pt][l]{${}^{+179}_{-178}$} & \makebox[0pt][r]{$2005$}\makebox[0pt][l]{${}^{+184}_{-178}$} & \makebox[0pt][r]{$1744$}\makebox[0pt][l]{${}^{+103}_{-102}$} \\ 
    norm ($\times 10^{53}\ {\rm cm}^{-3} $)  & $\mathcal{LU}(0.01,50)$  & \makebox[0pt][r]{$0.82$}\makebox[0pt][l]{${}^{+0.16}_{-0.12}$} & \makebox[0pt][r]{$1.46$}\makebox[0pt][l]{${}^{+0.29}_{-0.26}$} & \makebox[0pt][r]{$0.58$}\makebox[0pt][l]{${}^{+0.33}_{-0.20}$} & \makebox[0pt][r]{$0.67$}\makebox[0pt][l]{${}^{+0.16}_{-0.08}$} & \makebox[0pt][r]{$0.67$}\makebox[0pt][l]{${}^{+0.16}_{-0.09}$} & \makebox[0pt][r]{$0.54$}\makebox[0pt][l]{${}^{+0.26}_{-0.09}$} & \makebox[0pt][r]{$0.23$}\makebox[0pt][l]{${}^{+0.04}_{-0.02}$} & \makebox[0pt][r]{$0.30$}\makebox[0pt][l]{${}^{+0.18}_{-0.07}$} & \makebox[0pt][r]{$0.60$}\makebox[0pt][l]{${}^{+0.14}_{-0.08}$} \\
    \hline
    power-law & & & & & & & & & \\
    \hline
    $\Gamma$  & $\mathcal{U}(2.7,3.5)$  & \makebox[0pt][r]{$3.13$}\makebox[0pt][l]{${}^{+0.04}_{-0.04}$} & \makebox[0pt][r]{$3.12$}\makebox[0pt][l]{${}^{+0.02}_{-0.02}$} & \makebox[0pt][r]{$3.19$}\makebox[0pt][l]{${}^{+0.03}_{-0.02}$} & \makebox[0pt][r]{$3.15$}\makebox[0pt][l]{${}^{+0.03}_{-0.04}$} & \makebox[0pt][r]{$3.16$}\makebox[0pt][l]{${}^{+0.03}_{-0.03}$} & \makebox[0pt][r]{$3.24$}\makebox[0pt][l]{${}^{+0.01}_{-0.01}$} & \makebox[0pt][r]{$3.13$}\makebox[0pt][l]{${}^{+0.05}_{-0.05}$} & \makebox[0pt][r]{$3.13$}\makebox[0pt][l]{${}^{+0.03}_{-0.03}$} & \makebox[0pt][r]{$3.11$}\makebox[0pt][l]{${}^{+0.03}_{-0.03}$} \\ 
    norm $(\times 10^{44})$  & $\mathcal{U}(0.01,4.00)$  & \makebox[0pt][r]{$0.51$}\makebox[0pt][l]{${}^{+0.05}_{-0.05}$} & \makebox[0pt][r]{$0.85$}\makebox[0pt][l]{${}^{+0.05}_{-0.05}$} & \makebox[0pt][r]{$0.81$}\makebox[0pt][l]{${}^{+0.04}_{-0.04}$} & \makebox[0pt][r]{$0.85$}\makebox[0pt][l]{${}^{+0.08}_{-0.08}$} & \makebox[0pt][r]{$1.21$}\makebox[0pt][l]{${}^{+0.07}_{-0.07}$} & \makebox[0pt][r]{$2.10$}\makebox[0pt][l]{${}^{+0.04}_{-0.05}$} & \makebox[0pt][r]{$1.06$}\makebox[0pt][l]{${}^{+0.13}_{-0.12}$} & \makebox[0pt][r]{$1.25$}\makebox[0pt][l]{${}^{+0.09}_{-0.09}$} & \makebox[0pt][r]{$1.17$}\makebox[0pt][l]{${}^{+0.08}_{-0.08}$} \\ 
    (ph s$^{-1}$ keV$^{-1}$ at 1.0 keV) & & & & & & & & & \\
    \hline
    Fit statistics & & & & & & & & & \\
    \hline
    C-statistic & & 3638.05 & 3402.30 & 3056.23 & 3495.56 & 3419.30 & 3564.18 & 3349.22 & 3064.05 & 3600.87 \\
    Degrees of freedom & & 2811 & 2868 & 2769 & 2828 & 2955 & 3018 & 2941 & 2668 & 2953
    \enddata
    
    \end{deluxetable}

\section{Examples of posterior distribution}\label{app:corner_plots}

\begin{figure*}
\centerline{
  \includegraphics[width=\textwidth]{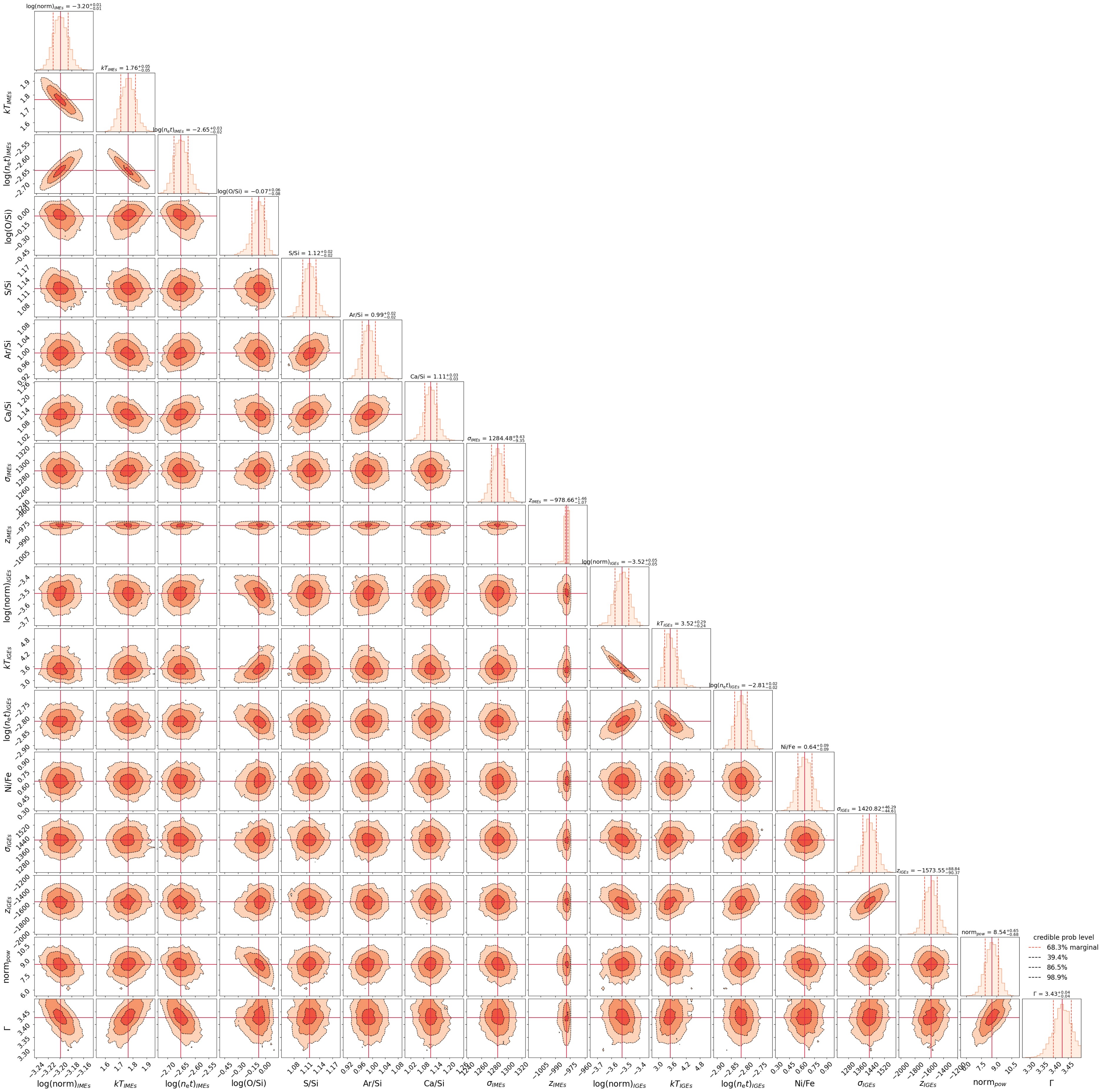}
  \vspace{2mm}
}
\caption{Corner plot of SE super-pixel {\em e} with all 17 parameters fitted with {\em UltraSPEX}.
The parameters fitted with log-uniform priors are shown in log space.
The top plot of each column shows the individual parameter histograms (marginal distribution), with the 1$\sigma$ (68.3\%) uncertainties shown as dashed orange lines and the medians marked as solid red lines. 
The contours represent (from darker to lighter orange) 39.4\%, 86.5\%, and 98.9\% confidence levels which correspond to 1$\sigma$, 2$\sigma$, and 3$\sigma$ significance levels, respectively, for a 2D Gaussian distribution.
For the IMEs and IGEs \texttt{pshock} components, the normalizations are given in units of $10^{58}$ cm$^{-3}$, temperatures are given in keV, the ionization timescales ($n_{\mathrm{e}}t$) are in $10^{14}$ cm$^{-3}$s, redshift and broadening are in \kms. For the power-law component, the normalization is in units of $10^{44}$ ph s$^{-1}$ keV$^{-1}$.
}
\label{fig:corner_SE_pix_e}
\end{figure*}

\begin{figure*}
\centerline{
  \includegraphics[width=\textwidth]{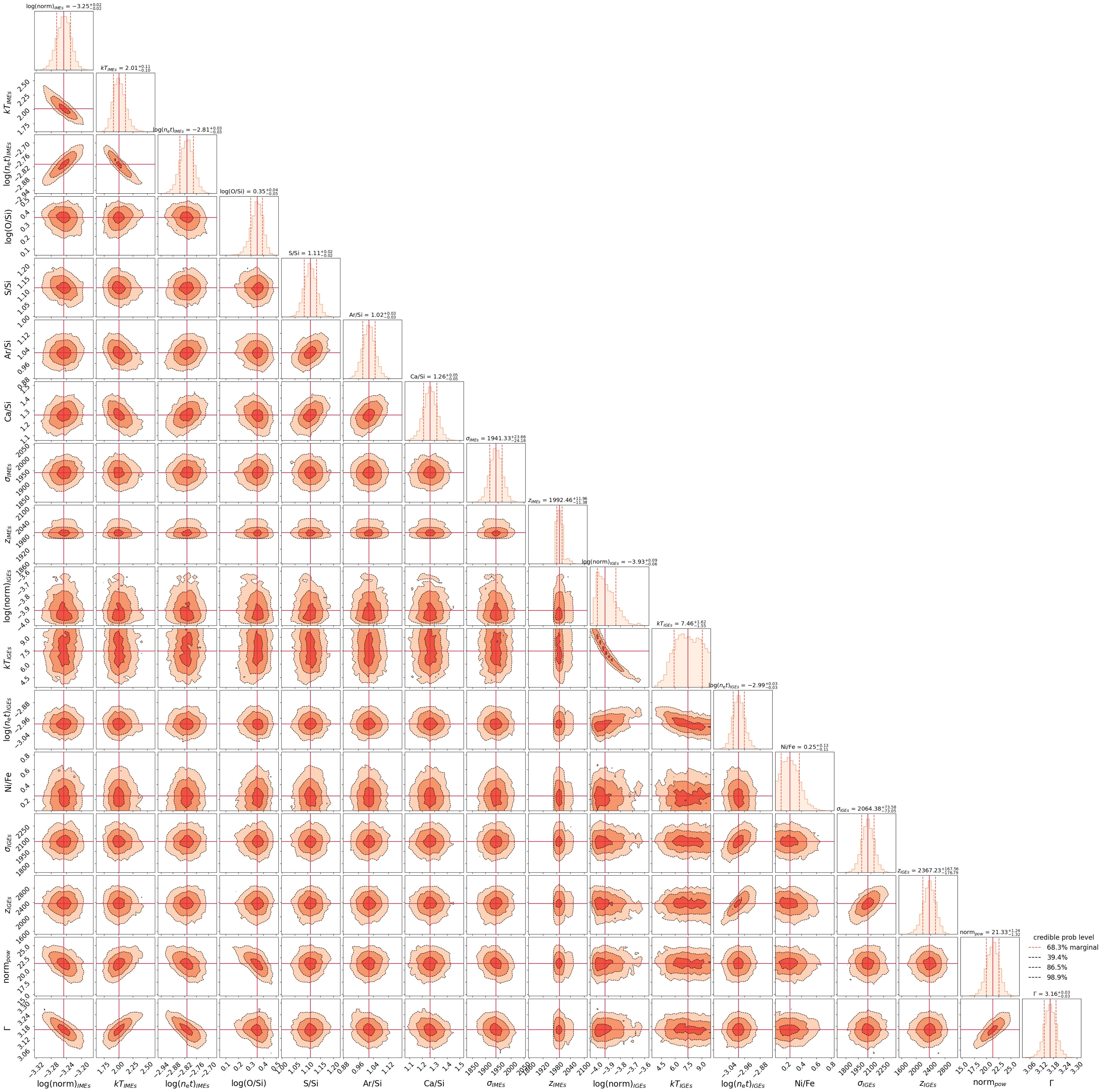}
  \vspace{2mm}
}
\caption{Same as for Figure~\ref{fig:corner_SE_pix_e} but for NW super-pixel {\em e}.
The complete corner plots for all super-pixels are available on Zenodo.
}
\label{fig:corner_NW_pix_e}
\end{figure*}

\begin{figure*}
\centerline{
  \includegraphics[width=0.45\textwidth]{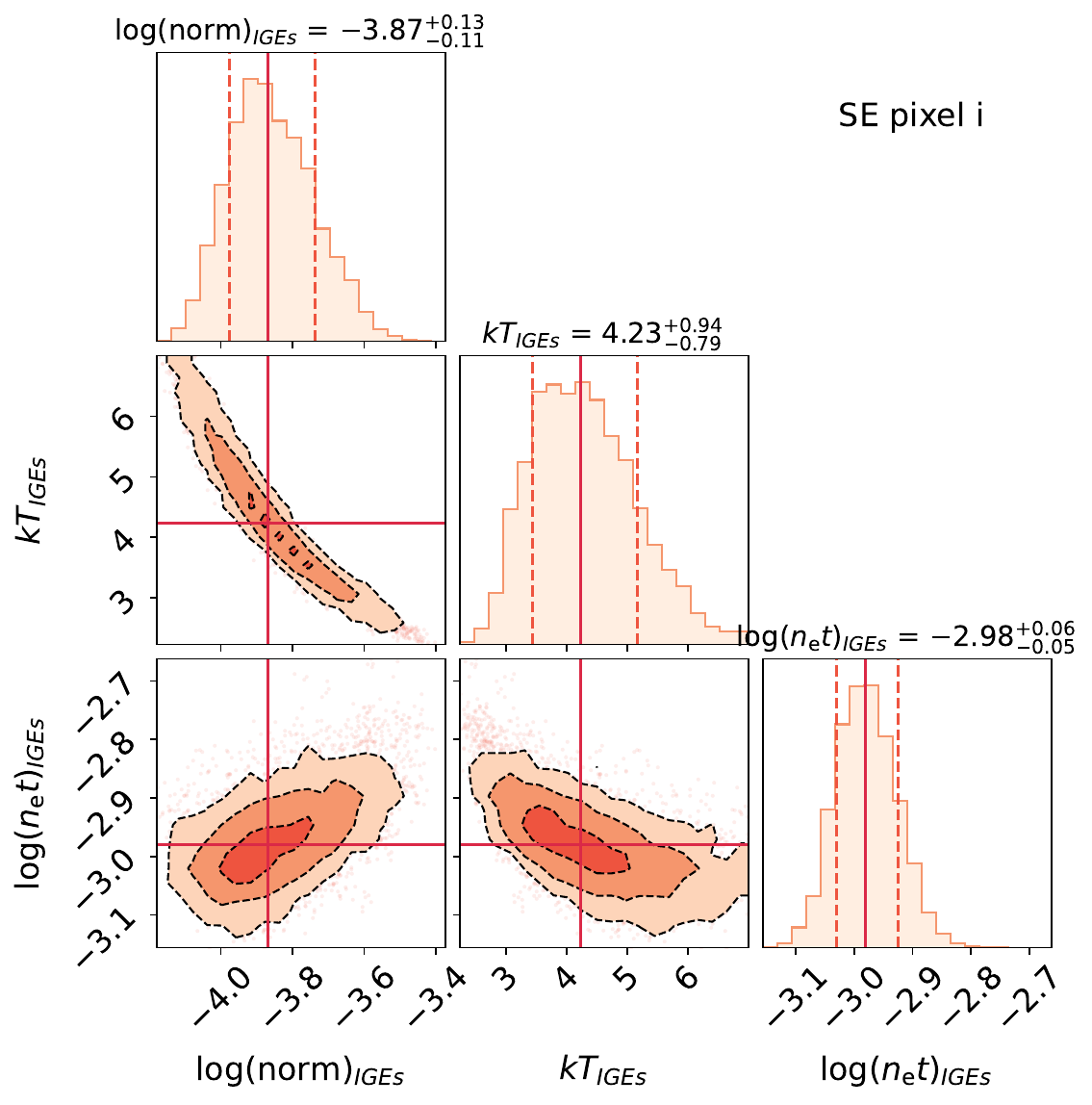}
  \hspace{5mm}
  \includegraphics[width=0.45\textwidth]{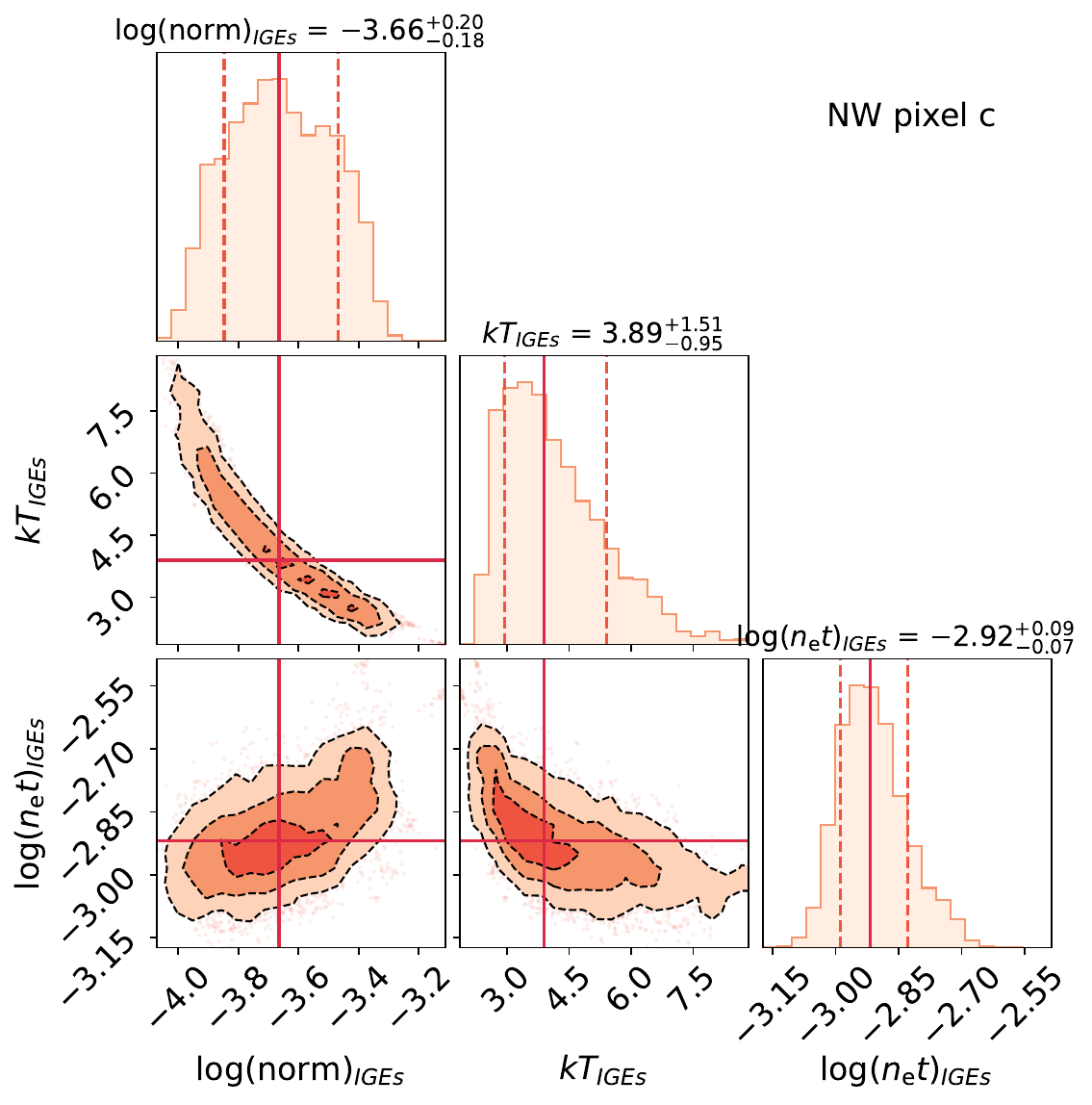}
}
\caption{Similar to corner plot in Figure~\ref{fig:corner_SE_pix_e} but for SE super-pixel {\em i} (left) and NW super-pixel {\em c} (right) with a few selected parameters (normalization, electron temperature, and ionization timescale) of the Fe-group \texttt{pshock} component. 
This shows the presence of multiple isolated islands of local minima in the parameter space.
}
\label{fig:multi_minima_corner_plots}
\end{figure*}

\begin{figure*}
\centerline{
  \includegraphics[width=\textwidth]{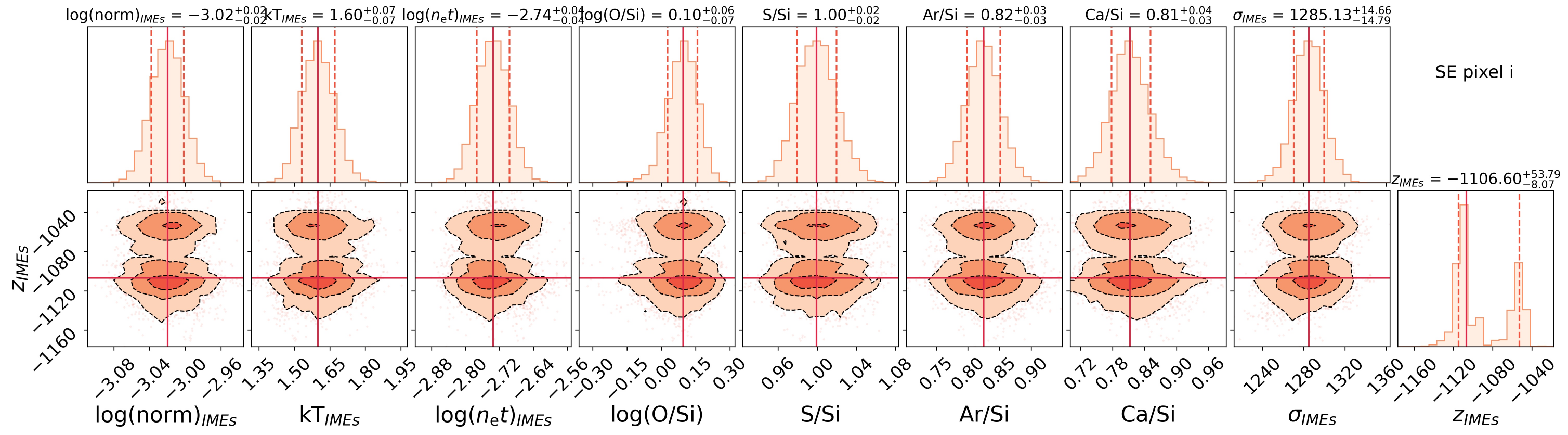}
}
\vspace{2mm}
\centerline{
  \includegraphics[width=\textwidth]{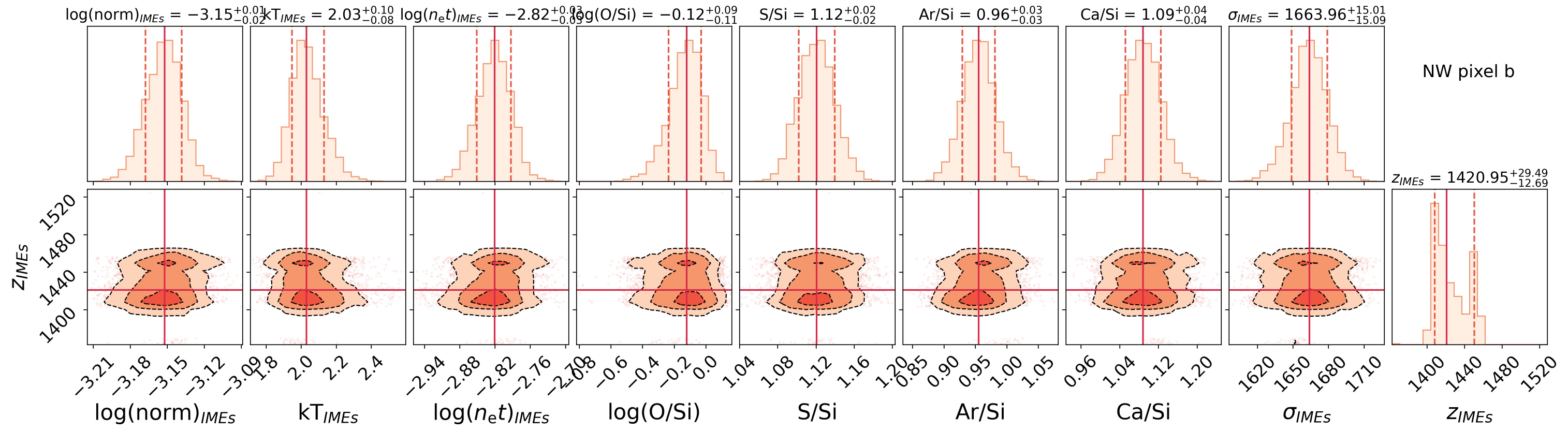}
}
\caption{Posterior probability distribution between all fitted parameters of the IMEs \texttt{pshock} component and the corresponding redshift. Corner plots of SE super-pixel {\em i} (top) and NW super-pixel {\em b} (bottom) are shown. The red solid line marks the median value of the posterior distribution and the dashed lines indicate the 1$\sigma$ uncertainties. The dark to light contours represent 39.4\%, 86.5\%, and 98.9\% confidence levels which correspond to 1$\sigma$, 2$\sigma$, and 3$\sigma$ significance levels, respectively, for a 2D Gaussian distribution. These corner plots highlight the asymmetric uncertainties of redshift along with the presence of multiple local minima. The full corner plots with all parameters fitted with {\em UltraSPEX} are available on Zenodo.
}
\label{fig:multimodal_z_posteriors}
\end{figure*}

The Bayesian-based Nested Sampling fits produce a posterior distribution for each parameter in the model and joint distributions between each pair of parameters, which are important diagnostics to access the quality of model fitting. They also provide crucial insights into the parameter terrain along with potential correlations and degeneracies.
The corner plots obtained from the model fits using {\em UltraSPEX} are shown in Figure~\ref{fig:corner_SE_pix_e} and Figure~\ref{fig:corner_NW_pix_e} for the SE super-pixel {\em e} and the NW super-pixel {\em e}, respectively. The corner plots for all other super-pixels are provided in the Zenodo repository. 
The median parameter values are highlighted with solid red lines. The 1$\sigma$ parameter uncertainties are marked in the marginal posterior distribution for each parameter, and the contours indicating the 1$\sigma$, 2$\sigma$, and 3$\sigma$ uncertainties are shown for the joint distribution between each pair of parameters.

The well-known correlations between normalization, $kT_\mathrm{e}$, and $n_{\mathrm{e}}t$ are clearly observed in the corner plots for both the IMEs and the IGEs \texttt{pshock} components.
This intrinsic physical degeneracy between $kT_\mathrm{e}$ and $n_{\mathrm{e}}t$ for an ionizing plasma in an NEI state is due to both an increase in temperature or $n_{\mathrm{e}}t$ shifting the plasma towards higher ionization states.
The parameter space with multiple ``banana-shaped" posterior distribution can be challenging for standard MCMC fitting routines and highlight the need for NS algorithms. This is further complicated by the presence of multiple local minima in the parameter terrain, as shown in Figure~\ref{fig:multi_minima_corner_plots}. It presents multiple 1$\sigma$ contours (marked in dark orange) in the distribution between the normalization and $n_{\mathrm{e}}t$ corresponding to the IGEs \texttt{pshock} component for the SE super-pixel {\em i} and NW super-pixel {\em c}. Investigating the model spectra at these different local minima, we find that the profile of the Fe K complex does not vary significantly; however, there is a different contribution of the corresponding thermal continuum, with the C-statistic being marginally better for models with lower thermal continuum. This is in line with our discussion in Section~\ref{subsec:net_vs_kT}, regarding the tendency of the IGEs component to fit models with higher $kT_\mathrm{e}$.

In the corner plots with all model parameters, we observe correlations between power-law index and normalization vs IMEs \texttt{pshock} parameters -- normalization, $kT_\mathrm{e}$, $n_{\mathrm{e}}t$, and O$^*$ abundance. These parameters jointly are responsible for the majority of the continuum emission in our fitted models. The degeneracies further underscore the importance of the Bayesian approach to fitting and the access to high-energy X-ray emission to constrain pure-metal ejecta models.
We find a positive correlation between redshift and $n_{\mathrm{e}}t$ and also between redshift and broadening for the Fe-group \texttt{pshock} component. The $kT_\mathrm{e}$ and $n_{\mathrm{e}}t$ are known to impact the line centroids by representing plasmas in different ionization states, and thus are expected to be degenerate with redshift (particularly for instruments with low spectral resolution). A similar effect was presented in detail in the Appendix of \cite{godinaud23} for the Si line at \chandra\ spectral resolution. Our fitted model posterior distributions only show this correlation for the Fe-group component.

In some cases, we note a multi-modal posterior distribution for the IMEs \texttt{pshock} redshift. In Figure~\ref{fig:multimodal_z_posteriors}, we present the joint distribution of the redshift with all other parameters in the IMEs \texttt{pshock} component for the SE super-pixel {\em i} and the NW super-pixel {\em b}. We observe multiple 1$\sigma$ contours in all parameter spaces; however, there is no correlation with any parameter, making this effect tricky to explain. A consequence of the multi-modal distribution is the highly asymmetric uncertainties for the redshift, which was also reported by \cite{plucinsky25} where they fit the \xrism\ data of Cas~A and scanned the parameter space using the traditional \texttt{`error'} command. 
The velocity shifts between the peaks (different 1$\sigma$ contours) are on the order of 50 \kms\ (which is within the \xrism/Resolve systematic uncertainties of $\sim$1 eV at the energy range of the IMEs) and therefore do not affect our results.
A similar multi-modal redshift distribution was also present in Bayesian-based spectral fits with \texttt{XSPEC} using \chandra\ data of Cas~A \citep{vink24a} and Tycho \citep{godinaud25}.
Therefore, it is an interesting feature of the parameter space that has been observed with different spectral fitting software, different X-ray observatories, and different targets, and which needs to be further investigated.

\section{Additional maps}\label{app:extra_maps}
\begin{figure*}
\centerline{
  \includegraphics[width=0.33\textwidth]{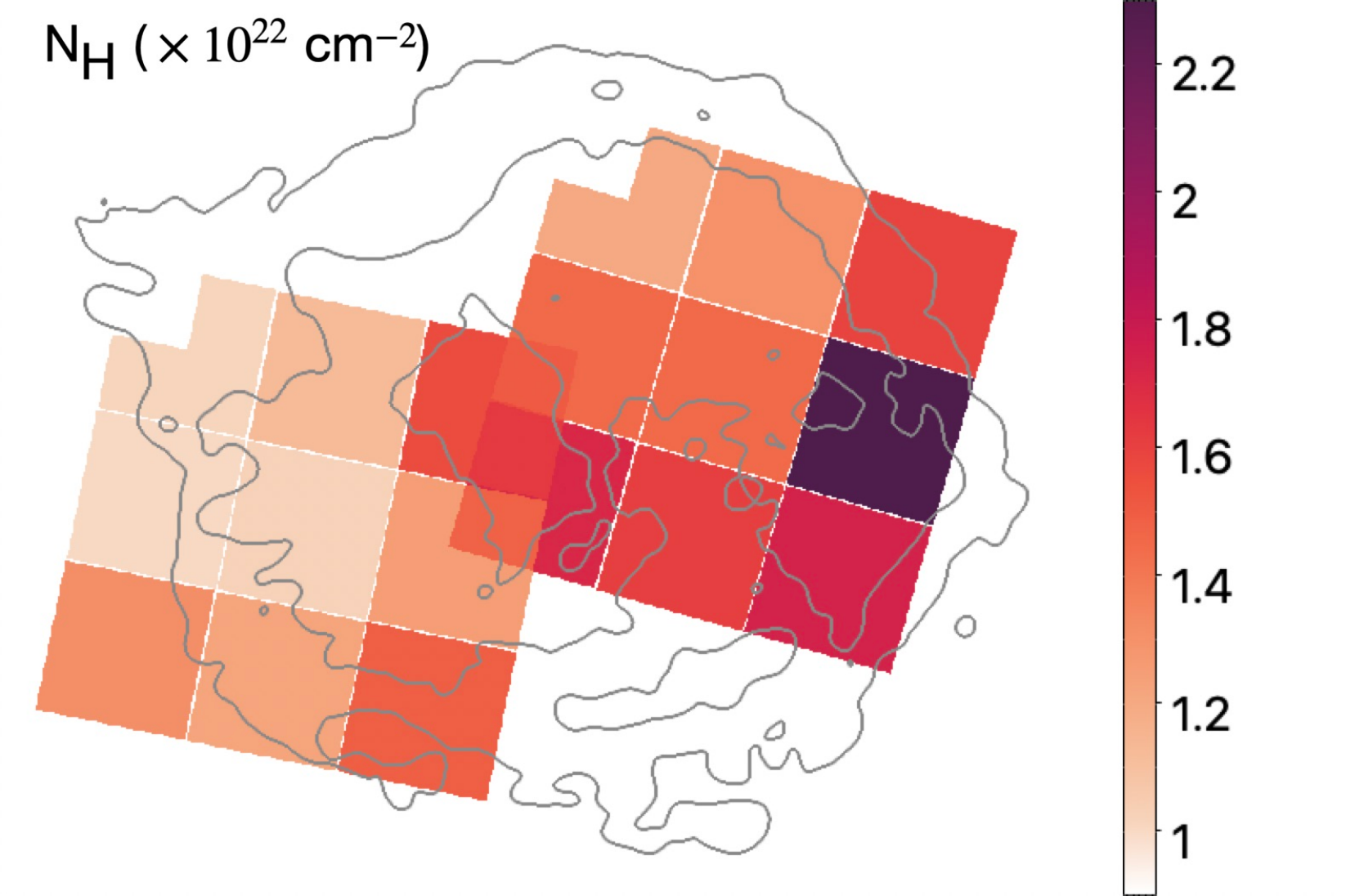}
   \hspace{2mm}
  \includegraphics[width=0.33\textwidth]{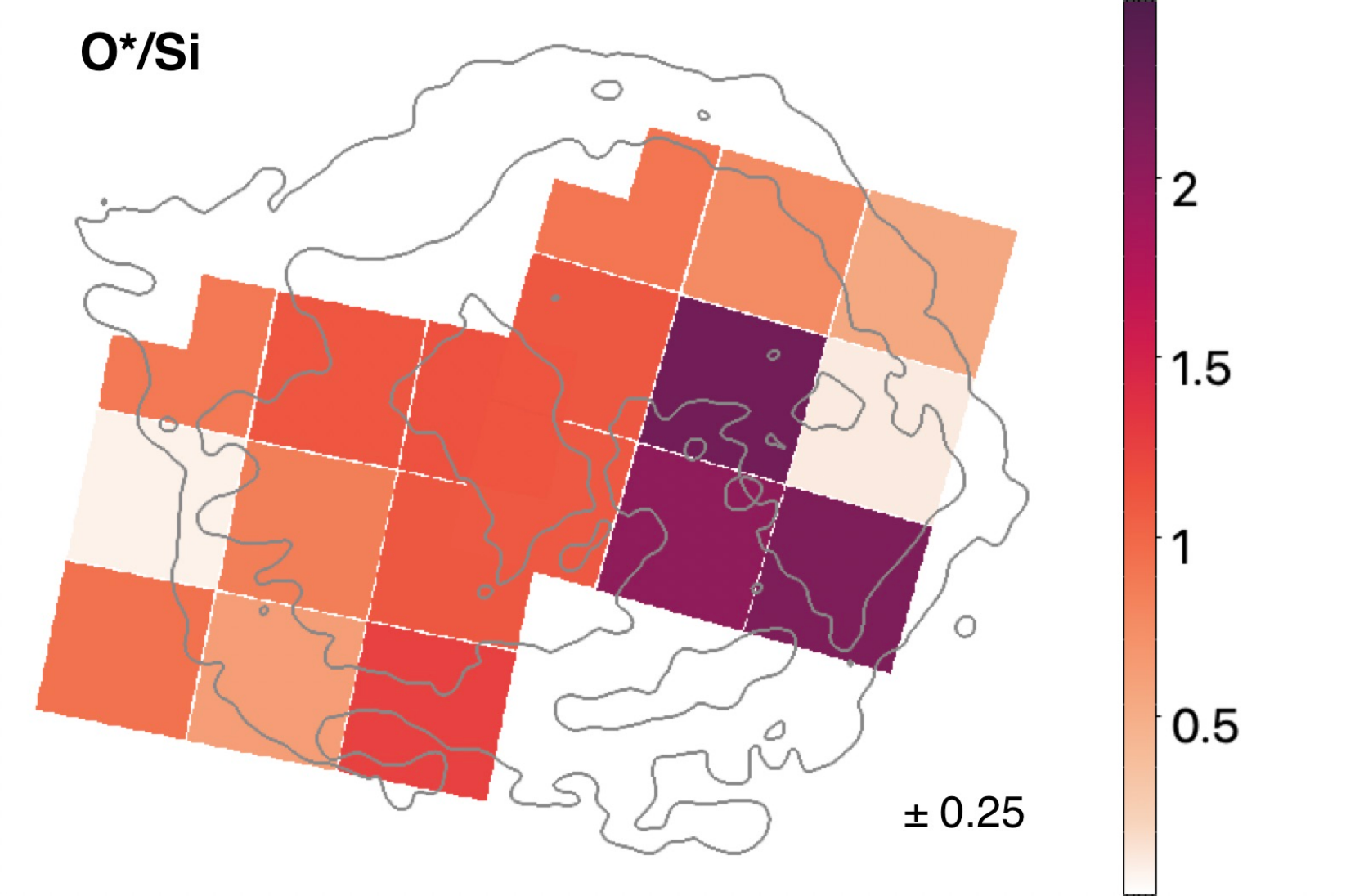}
   \hspace{2mm}
  \includegraphics[width=0.33\textwidth]{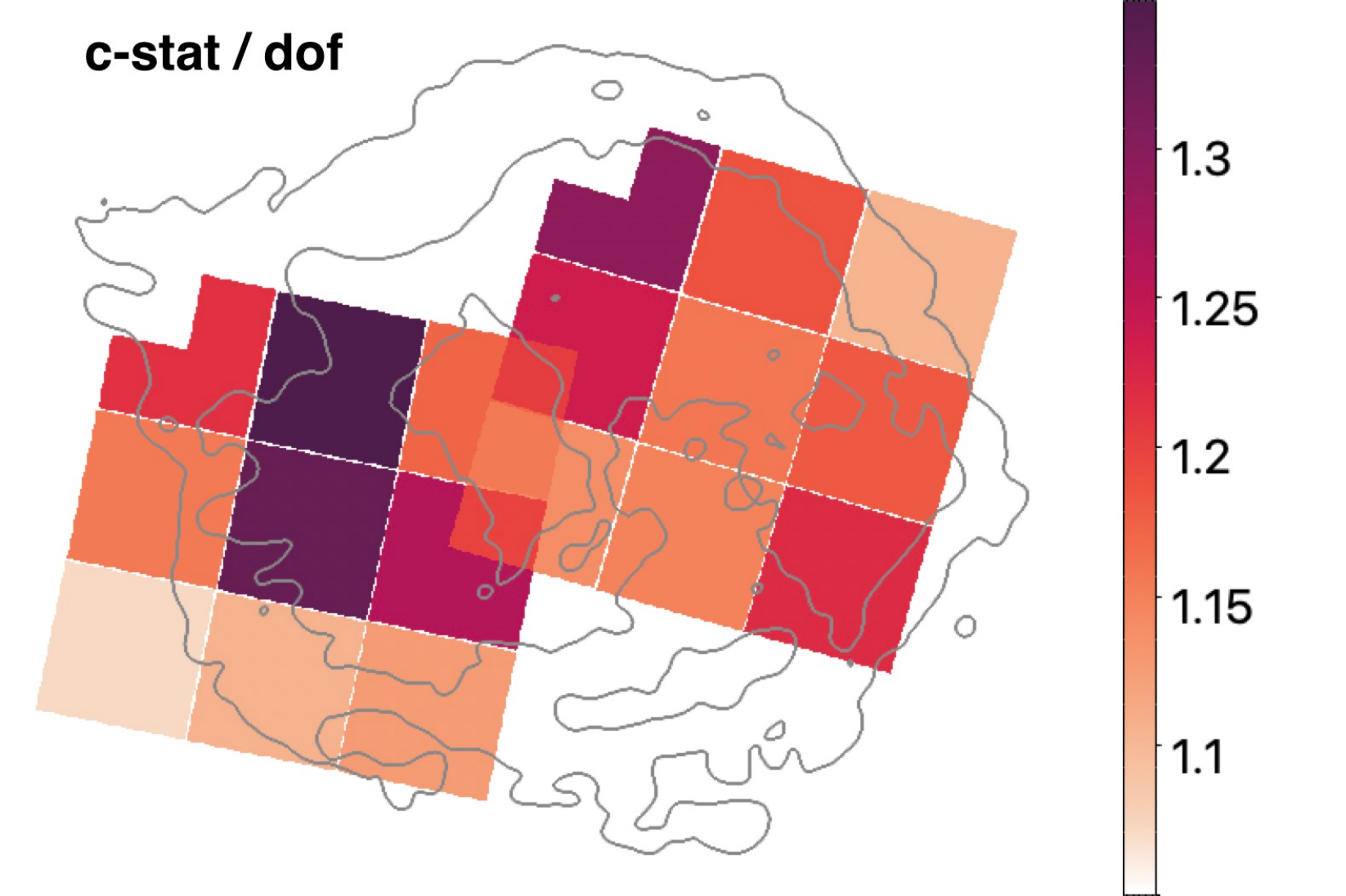}
}
\caption{
Left: Absorption map ({$N_\text{H}$}) used for spectral fitting.
Center: Map of fitted abundance ratio of O$^*$ to Si compared to the solar ratio (O/Si)$_{\odot}$. The average one sigma error is indicated on the bottom right.
Right: Map of C-statistic over degrees of freedom.
}
\label{fig:extra_maps}
\end{figure*}

The neutral hydrogen column density ({$N_\text{H}$}) is known to vary over different regions of Cas~A, however with the \xrism\ Gate Valve closed, the column density cannot be well constrained with Resolve alone.
For the absorption component in our model fits, we take the values of {$N_\text{H}$} as the average value in the corresponding super-pixel region from the absorption map in \citet[Appendix Figure~6]{vink22b}, which was produced by fitting \chandra/ACIS-S data in the 0.6--7 keV band with a model comprising of three pure-metal {\it vnei} models, a power-law component, and a {\it tbabs} Galactic absorption model. A map of the {$N_\text{H}$} values used in our analysis is shown in Figure \ref{fig:extra_maps} (left).

The fitted abundance ratio map of (O$^*$/Si)/(O/Si)$_{\odot}$ is shown in Figure \ref{fig:extra_maps} (center). 
There are no spectral lines of O in the fitted energy range of 1.8--11.9 keV; instead the abundance of O is set as a free parameter to model the combined thermal continuum emission from all elements lighter than Si (particularly O, Ne, Mg, He, H), which we refer to as O$^*$. So the abundance ratio O$^*$/Si is not representative of the true abundance of O.
The prior on the O$^*$/Si ratio during fitting was between 0.01 and 10 times (O/Si)$_{\odot}$. Despite no spectral lines, the O$^*$/Si ratio is well constrained with most regions within $\pm$0.5 times solar ratio.

The quality of the spectral fits is represented with a map of the C-statistic over the degrees of freedom (dof) in Figure \ref{fig:extra_maps} (right).

\begin{figure*}
\centerline{
  \includegraphics[width=0.33\textwidth]{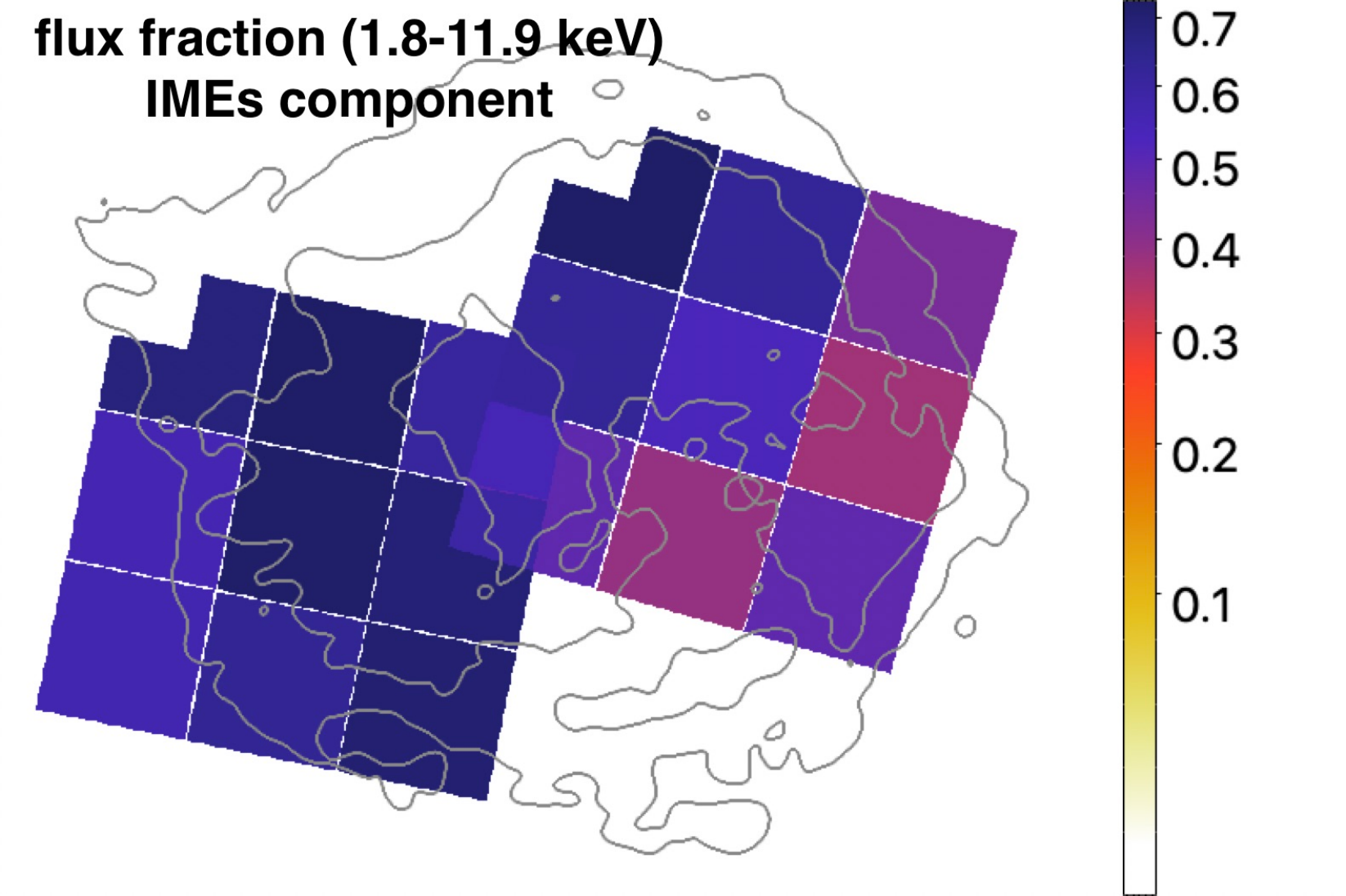}
   \hspace{2mm}
  \includegraphics[width=0.33\textwidth]{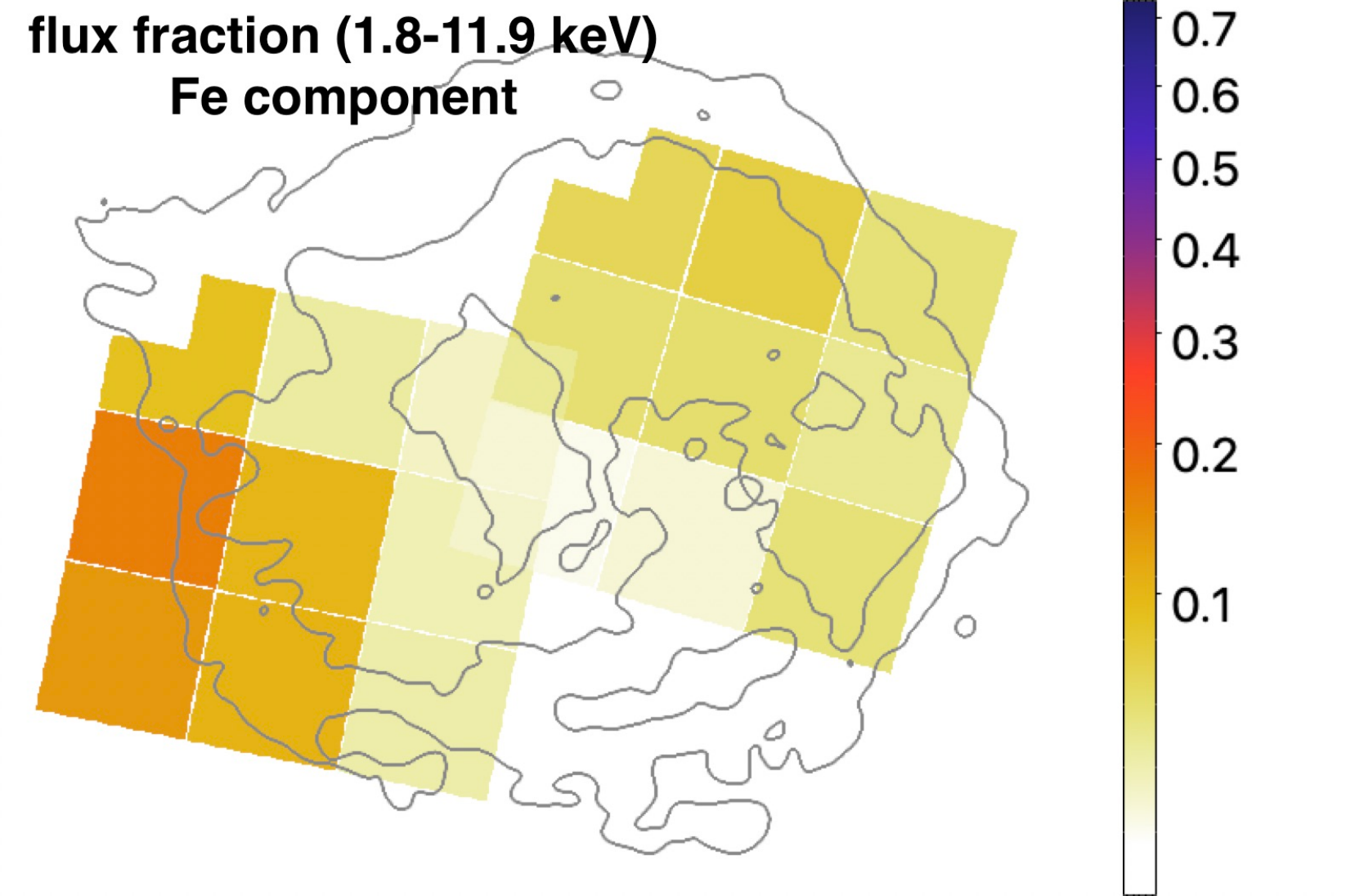}
   \hspace{2mm}
  \includegraphics[width=0.33\textwidth]{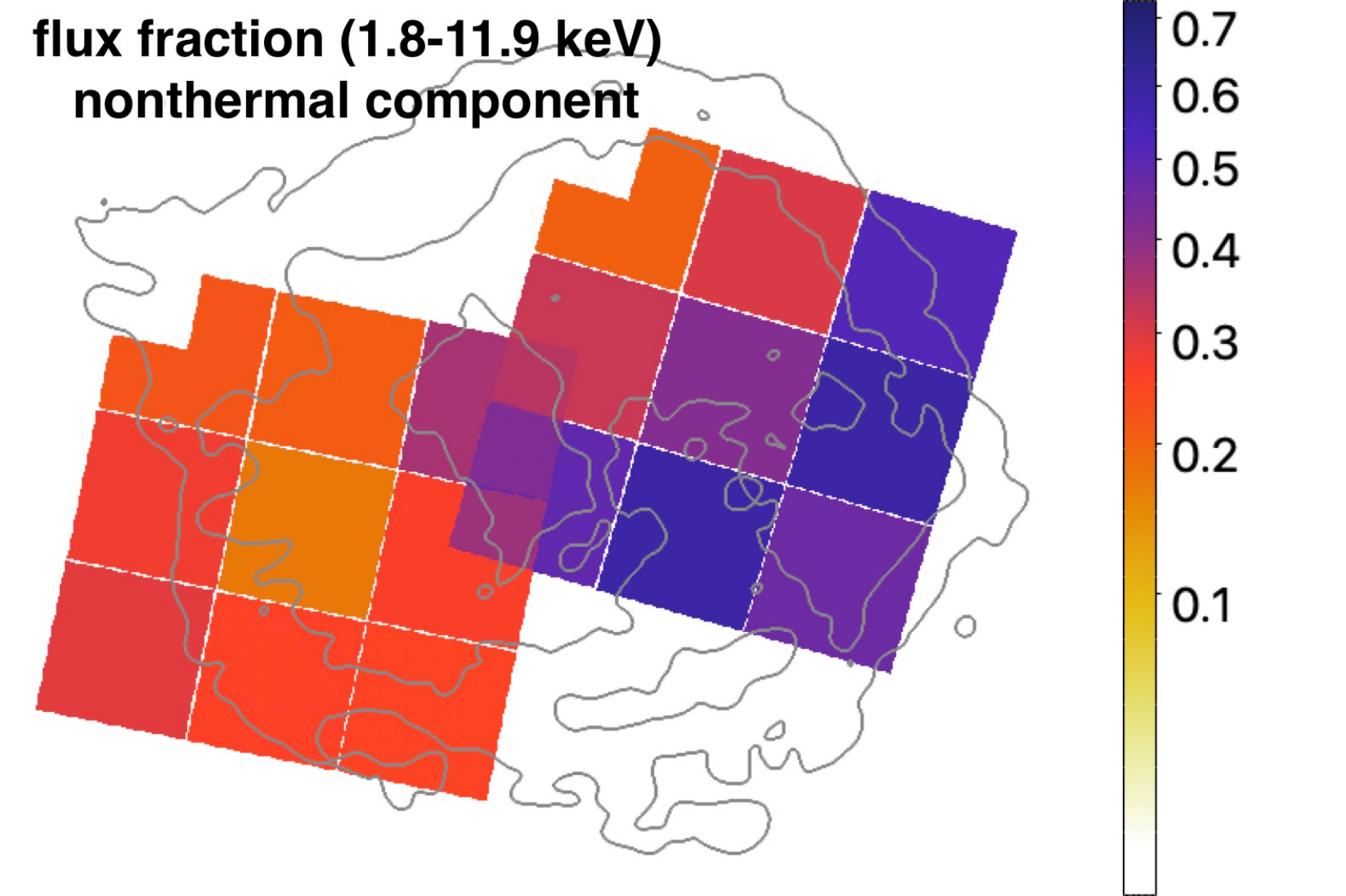}
}
\caption{
Maps of relative flux contribution from a model component with respect to the total flux in the 1.8--11.9 keV energy range. Left to right panels show the flux fraction map of IMEs component, IGEs component and the power-law component, respectively.
}
\label{fig:flux_fraction_maps}
\end{figure*}

Figure \ref{fig:flux_fraction_maps} shows the flux contribution of each fitted model component (IMEs \texttt{pshock}, IGEs \texttt{pshock}, and power-law) with respect to the total flux in the energy range 1.8--11.9 keV.
The IMEs and IGEs component flux fractions include the contribution of the different elements to the thermal continuum emission (Bremsstrahlung, two-photon emission, and radiative recombination continuum).
We detect highest synchrotron contribution in the western and central regions of the remnant, in line with other studies \citep{helder08,grefenstette15}.

\section{Higher velocities for Ca}\label{app:ca_velocities}
\begin{figure*}
\centerline{
  \includegraphics[width=0.34\textwidth]{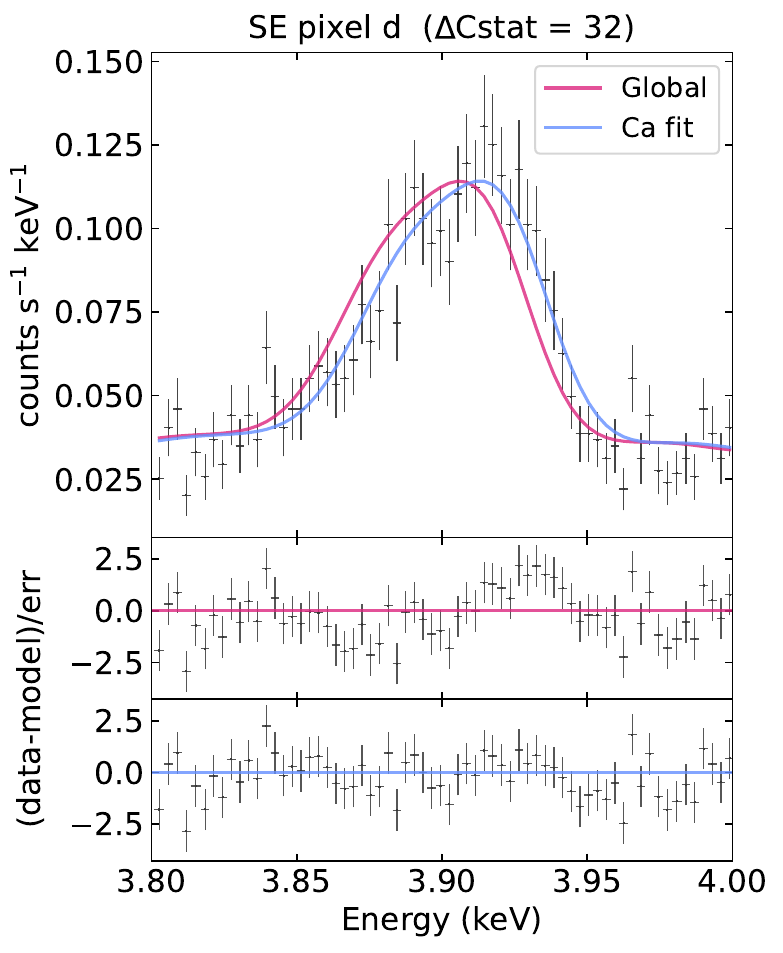}
   \hspace{2mm}
  \includegraphics[width=0.32\textwidth]{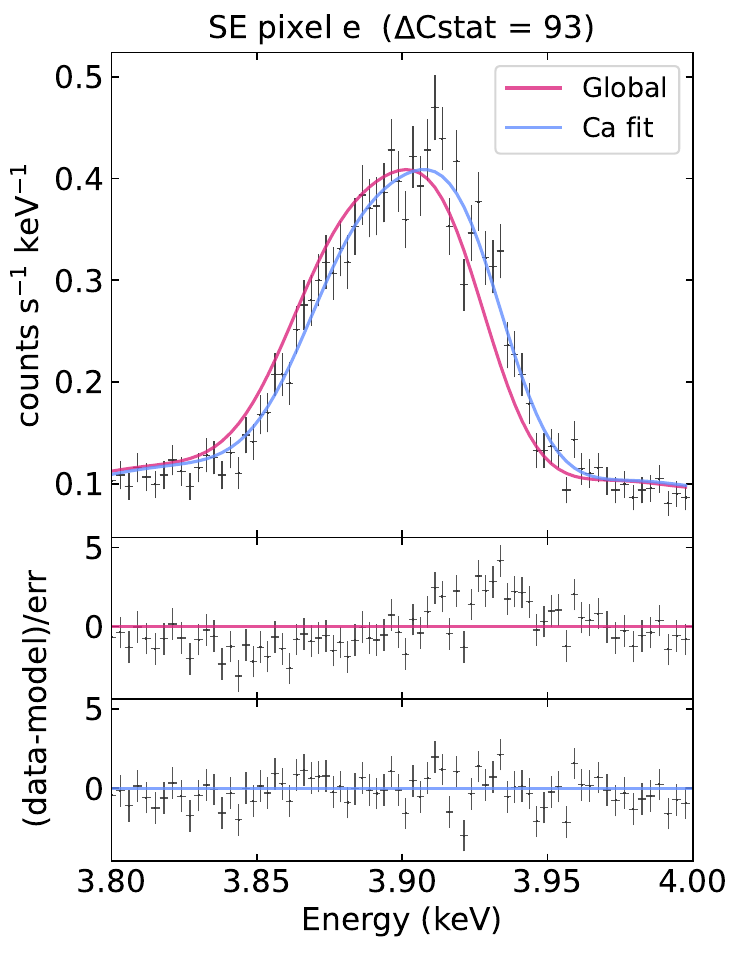}
   \hspace{2mm}
  \includegraphics[width=0.33\textwidth]{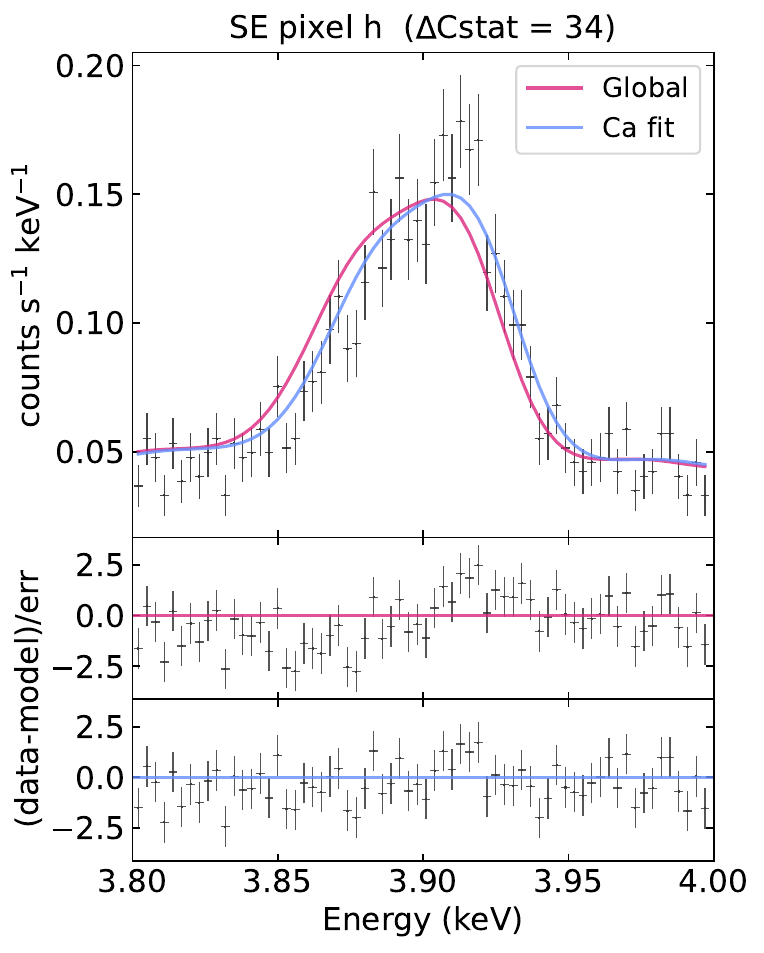}
}
\caption{
The \xrism/Resolve spectra near the Ca He$\alpha$ lines (3.8--4 keV) are presented for SE pixel {\em d} (left), SE pixel {\em e} (center), and SE pixel {\em h} (right). The global fitted spectra (fit to 1.8--11.9 keV band) is shown in red and the corresponding residuals are shown in the middle panel of each plot. The spectral fits to model the Doppler velocity of the Ca He$\alpha$ lines are shown in blue and the corresponding residuals are presented in the bottom panels. The fits were performed to optimally binned spectra. Unlike Fig.~\ref{fig:pixel_e_spectra}, no further binning was performed for display purposes.
}
\label{fig:ca_velocities}
\end{figure*}

In our spectral modeling, a single \texttt{pshock} component is attributed to model all the IMEs (Si, S, Ar, and Ca). Thus, the fitted component represents the overall plasma properties of the IMEs. 
Upon investigating the fitted spectra (1.8--11.9 keV), we observed excess residuals near Ca He$\alpha$ for some regions in the SE --- pixel {\em d}, pixel {\em e}, and pixel {\em h}. Figure~\ref{fig:ca_velocities} shows the zoomed in \xrism/Resolve spectra near Ca He$\alpha$ for these super-pixels. The sine-wave-shaped residuals (seen in the middle panels) clearly indicate a velocity offset for Ca compared to the fitted overall IMEs redshift. We perform spectral fitting in the 3.8--4 keV range while keeping all parameters fixed to the global best-fit values except for the redshift of the IMEs. Higher blueshift velocities are measured for all these regions, and the fitted values are presented in Table~\ref{tab:ca_velocities}. The measured velocities are faster than both IMEs and IGEs, except SE super-pixel {\em e} where it is slower than the Fe-group but still closer to it than the IMEs.

The shift in line profile can also be due to different plasma properties for Ca with respect to IMEs.
To investigate this, we also performed spectral fits with additional free parameters -- $n_{\mathrm{e}}t$, $kT_\mathrm{e}$, and $\sigma$ corresponding to the IMEs \texttt{pshock} component. The resulting velocity measurements were lower (within 1$\sigma$--3$\sigma$) than those in Table~\ref{tab:ca_velocities}, but still faster than the corresponding IMEs velocities. The fit also tended to prefer a slightly narrower broadening and higher $n_{\mathrm{e}}t$. However, only incremental improvement was measured in the $C$-statistic, indicating that the residuals for Ca are mainly caused by kinematic differences compared to the IMEs.
We also investigated the SE pixel {\em g} as it is in the same general region as the other three super-pixels. However, no statistically significant difference in velocity was measured.

\begin{table*}
\caption{The velocity measurements of Ca He$\alpha$ lines by fitting our model in 3.8--4 keV range with all parameters fixed except redshift corresponding to the IMEs \texttt{pshock} component. The new fitted velocity and the difference compared to the initial overall IMEs velocity is given in units of \kms\ in Column 2 and 3, respectively. The improvement in C-statistic compared to the model fitted to the global spectra (1.8--11.9 keV) is given in Column 4.
\label{tab:ca_velocities}}
\centering
\begin{tabular}{l r@{\,}l r@{\,}l c}
\noalign{\smallskip}\hline\noalign{\smallskip}
Region	&		\multicolumn{2}{c}{Ca $V_{z}$ (km/s)} 						&		\multicolumn{2}{c}{$\Delta V_{\text{Ca-IME}}$ (km/s)}						&	$\Delta$ Cstat	\\ 
\noalign{\smallskip}\hline\noalign{\smallskip}	
SE pixel $d$	&	-1656	&	$^{+	103	}_{	-88	}$	&	-531	&	$^{+	103	}_{	-90	}$	&	32	\\	
SE pixel $e$	&	-1445	&	$^{+	45	}_{	-32	}$	&	-466	&	$^{+	45	}_{	-32	}$	&	93	\\	
SE pixel $h$	&	-1355	&	$^{+	39	}_{	-79	}$	&	-449	&	$^{+	39	}_{	-83	}$	&	34	\\ 
\noalign{\smallskip}\hline\noalign{\smallskip}	
\end{tabular}
\end{table*}


\section{The ionization time scale and the reverse shock history}\label{app:ionization_age}

The ionization parameter, $n_{\rm e}t$, provides a measure of how long ago the X-ray emitting plasma was heated by the shock. The age of Cas A is about 350~yr, whereas if we assume a shocked ejecta mass of $M=2$~\msun being uniformly distributed with no clumping, we
expect an average electron density of 
\begin{equation}\label{eq:ne}
\begin{split}
n_{\rm e}&\approx 4 \frac{3M_{\rm ej}}{4\pi R_{\rm cd}^3} \times \frac{\bar{Z}}{\bar{A} m_{\rm p}} \\
&\approx 7.3\left(\frac{M_{\rm ej}}{2~M_\odot}\right)\left(\frac{R_{\rm cd}}{2~{\rm pc}}\right)^{-3} \left(\frac{\bar{Z}}{8}\right)\left(\frac{\bar{A}}{16}\right)^{-1}~{\rm cm^{-3}},
\end{split}
\end{equation}
with $M_{\rm ej}$ the shocked ejecta mass, $R_{\rm cd}$ the radius
of the contact discontinuity, $\bar{Z}$ the average ion charge,
and $\bar{A} m_{\rm p}$ the average ion mass. The factor 4 arises from the shock compression. We picked here approximately values for Cas~A.
A crude expectation for the ionization age would, therefore, be
$n_{\rm e}t\approx 8\times 10^{10}~{\rm cm^{-3}s}$. However, that
does not take into account the fact that in the past the radius was much smaller, and the density was likely higher, but also that the age of
Cas A provides an upper limit to, $\Delta t$, the time available for ionization.

In order to provide a better understanding of what the measured distribution of $n_{\rm e}t$ tells us about the past evolution
of Cas~A, we resort to the analytic shock model described by
\citet{micelotta16}, which is an extension of the shock models
by \citet{truelove99}, but applied to SNRs developing in a wind-density profile \citep[see also][]{laming03,hwang12,tang17}. This model has
as parameters the total ejecta mass ($M_{\rm ej}$), the outer density profile index ($n, \rho_{\rm SN}(r) \propto r^{-n}$), the progenitor's wind mass loss parameter ($\dot{M}$), and the explosion energy ($E_{\rm sn}$). We coded this model and tested a variety of parameters against the
measured shock proper motions \citep{vink22a}, and forward- and reverse-shock locations \citep{arias18} of Cas~A.

\begin{figure*}
\centerline{
    \includegraphics[trim=20 0 30 20,clip=true,width=0.48\textwidth]{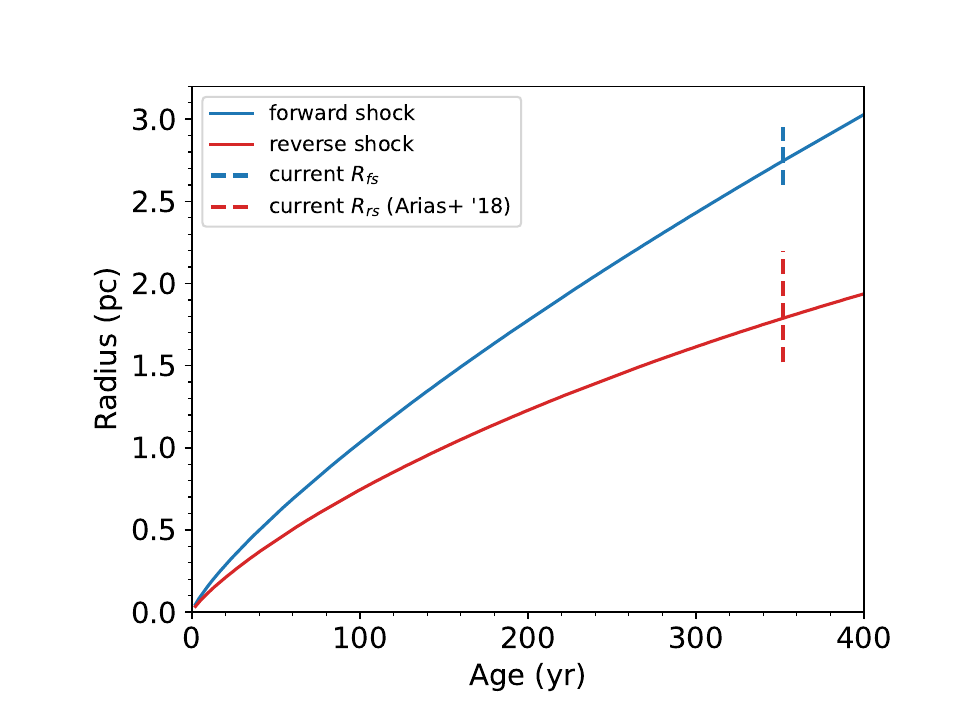}
    \includegraphics[trim=20 0 30 20,clip=true,width=0.48\textwidth]{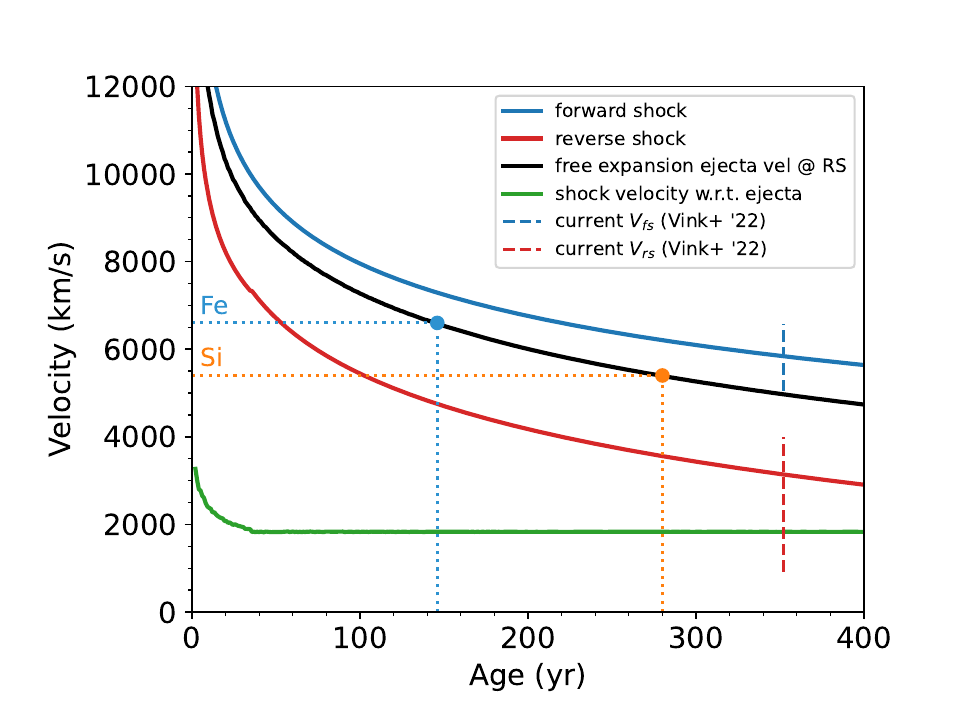}
    }   
\caption{Forward- and reverse- shock evolution according to 
the \citet{micelotta16} model with $M_{\rm ej}=3.5$~\msun, 
$E_{\rm sn}=3.6\times 10^{51}$~erg, $\dot{M}=4.2\times 10^{-4}$~\msun\ yr$^{-1}$ (assuming $v_{\rm w}=10~$~\kms), and $n=7$.
Left: the shock radii, with vertical dotted lines indicating the observed radii. 
Right: shock velocities. Besides the shock progressions
$v_{\rm fs,obs}=dR_{\rm fs}/dt$ (blue solid line) and $v_{\rm rs,obs}=dR_{\rm rs}/dt$ (red solid line), we also show the velocity of the freely expanding ejecta before entering the reverse shock
($v_{\rm ej}=R_{\rm rs}/t$, black line), and the 
ejecta velocity in the frame of the reverse shock, i.e. the velocity with which the stellar ejecta is shock heated ($V_{\rm S} = |v_{\rm ej,rs}|=|v_{\rm ej}-dR_{\rm rs}/dt| \approx 1800$~\kms, green line).
The blue dot indicates the age at which $v_{\rm ej}\approx 6600$~\kms\ is shocked, indicative of the outer Fe-rich knots in the SE.
Similarly, the orange dot is for $v_{\rm ej}\approx 5400$~\kms, the approximate velocity of the Si-rich shell in the SE.
\label{fig:truelove}}
\end{figure*}

\begin{figure*}
\centerline{
    \includegraphics[trim=20 0 30 20,clip=true,width=0.48\textwidth]{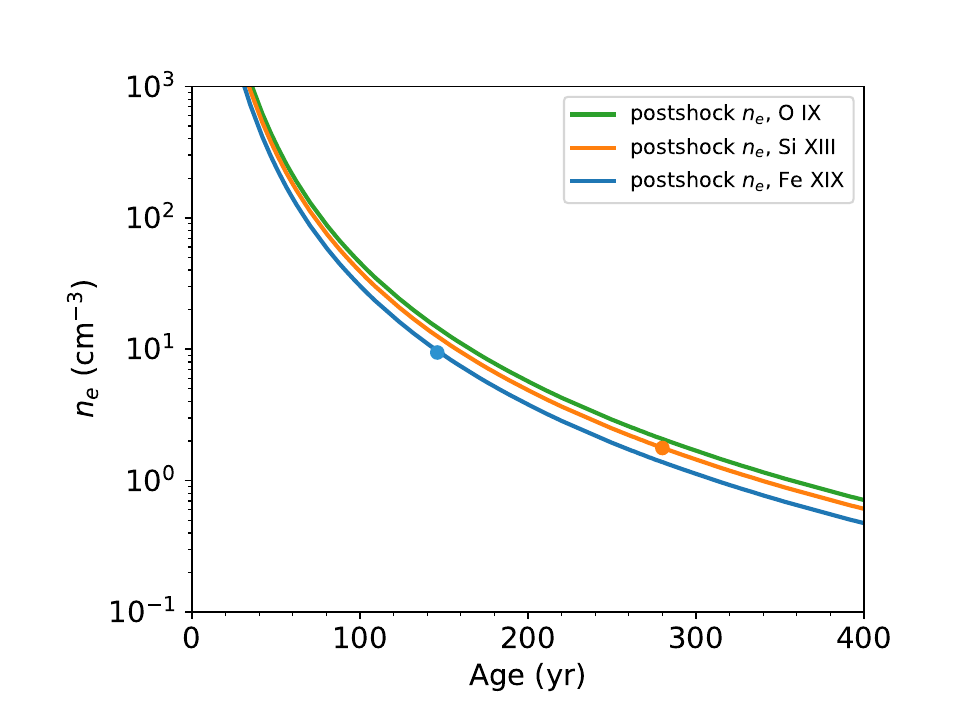}
    \includegraphics[trim=20 0 30 20,clip=true,width=0.48\textwidth]{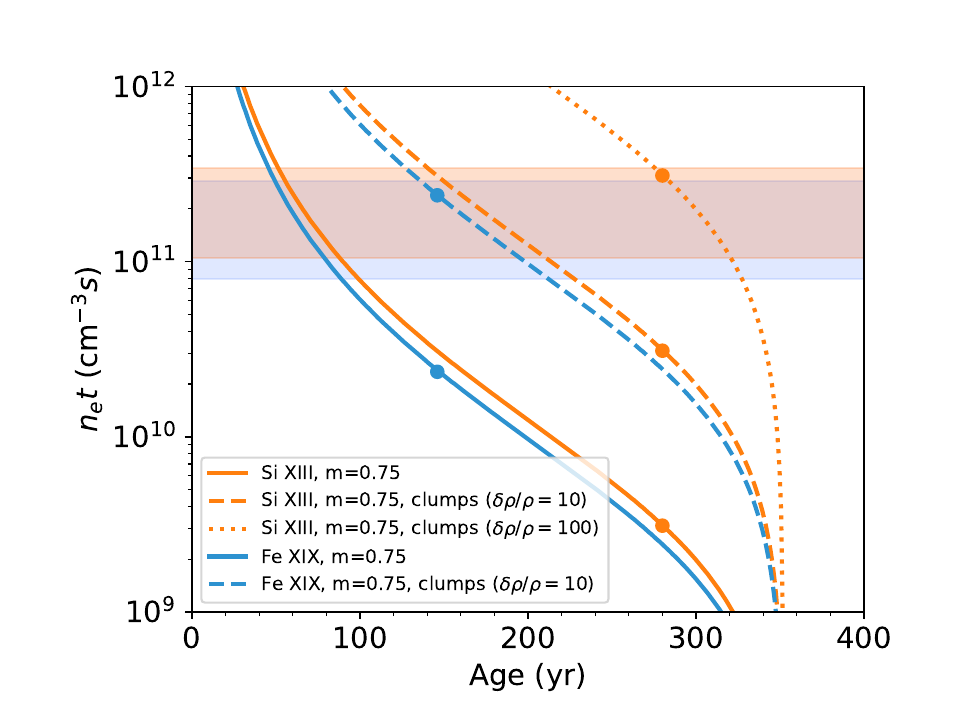}
    }
    \caption{\label{fig:truelove_ne_t}
    Left: estimated postshock electron density at the reverse shock. The blue dot and orange dot mark the ejecta shocked at an age of 145 yr (Fe ejecta) and 280 yr (Si ejecta), respectively.
    Right: the expected current $n_{\rm e}t$ for plasma shocked at 
    a certain age and overdensity. The measured range of $n_{\rm e}t$ values for IMEs and IGEs is shown in an orange- and blue- shaded band, respectively.
    This shows that an overdensity of factor $\sim10$ (for IGEs) and up to $\sim100$ (for IMEs) are required to explain observed $n_{\rm e}t$ values.
    }
\end{figure*}

There are several parameter combinations
that can reproduce the shock characteristics of Cas~A, but these parameters tend to cluster around $M_{\rm ej}=3.5$~\msun, 
$E_{\rm sn}=3.6\times 10^{51}$~erg, $\dot{M}=4.\times 10^{-4}$~\msun\ yr$^{-1}$, and $n=7$, in general agreement with earlier estimates
\citep{vink96,willingale03,laming03,hwang12}. 
We show the forward and reverse shock radii and velocities in Fig.~\ref{fig:truelove}.
However, a clear caveat is that 
observational evidence indicates that some observable characteristics of Cas~A are inconsistent with the analytical models. In particular,
in the analytical models \citep[and also numerical models][]{orlando16} the reverse shocks moves outward, whereas there
is clear evidence that at least in the western part of Cas~A it is at a standstill or even moving back to the center \citep{sato18,vink22a,fesen25}.  However, for the SE part of Cas~A the model appears to describe the general characteristics of Cas~A.

Having a semi-analytical model now allows us to calculate the electron density at the moment the ejecta gets shocked, using Eq.~\ref{eq:ne} but
with the unshocked ejecta mass $M_{\rm ej,unsh}$ instead of the total ejecta mass ($M_{\rm ej}$) and with the reverse shock radius, $R_{\rm rs}$,
rather than $R_{\rm cd}$. Note that for $t\gtrsim 38$~yr the density profile is flat ($\rho_{ej,unsh}(r)=$constant) in the \citet{truelove99,micelotta16} models.

Having the ejecta density at the moment of shock, we can calculate
$n_{\rm e}t$ by assuming that the density scales with expansion:
\begin{equation}
n_{\rm e}(r) =  n_{\rm e,0} \left(\frac{r}{R_{\rm rs,0}}\right)^{-3}=
n_{\rm e,0} \left(\frac{t}{t_{\rm 0}}\right)^{-3m}
=n_{\rm e}(t),
\end{equation}
with $n_{\rm e,0}$ the postshock electron-density at $t_0$, which is the time
at which the ejecta element entered the reverse shock. We used here
a homologous expansion of the remnant: $r \propto t^m$, with 
$m\approx 0.75$ the expansion rate of Cas~A \citep[e.g.][]{vink22a}.
(See \citealt{laming03} for a more elaborate model.)
The ionization age of a given element is then
\begin{equation}\label{eq:net}
n_{\rm e}t=\int_{t_0}^{t} n_{\rm e}(t)dt =\frac{n_{\rm e,0}t_0}{3m-1}\left[
1 - \left(\frac{t}{t_0}\right)^{-3m+1}
\right].
\end{equation}
We see that for $t\gg t_0$, $n_{\rm e}t$ depends on the time and density at which the element got shocked and not on the age of the SNR.
We can use this equation together with the semi-analytical model to calculate $n_{\rm e}t$ as a function of $t_0$, the age at which the plasma
was shock heated. 

We show the $n_{\rm e}$ and $n_{\rm e}t$ as a function of age (or better $t_0$) in Fig.~\ref{fig:truelove_ne_t}.
A caveat here is that in pure-metal plasmas ionization 
progresses, which leads to a small overestimate of the true $n_{\rm e}t$.
We also show in the Fig.~\ref{fig:truelove} and \ref{fig:truelove_ne_t} the approximate ages at which the Fe-knots
and main Si-shell entered the reverse shock, based on the projected 
average velocities $r/t$ of knots in the SE, to which we added $\delta V= \frac{1}{4}V_{\rm rs}\approx 500$~\kms,
taking into account that the ejecta slow down after having been shocked.
This gives free expansion ejecta velocities $v_{\rm ej} \approx 6600$~\kms\ and $\approx 5300$~\kms\ for Fe and Si, respectively, using projected radius of $1\farcm7$ and $2\farcm2$ \citep{milisavljevic13, vink22a}.
We see that at the shock we expected Fe to have been shocked
when Cas~A was $\approx 145$~yr old, 
with $n_{\rm e,0}\approx 9.4~{\rm cm^{-3}}$. Si was according to this
model shocked much later around $t_0\approx280$~yr, and $n_{\rm e,0}\approx 1.8~{\rm cm^{-3}}$ (Fig.~\ref{fig:truelove_ne_t} left panel). 
Calculating $n_{\rm e}t$ using Eq.~\ref{eq:net} now
reveals that it greatly underestimates the measured $n_{\rm e}t$ values.

This could imply two things: the ejecta were entering the reverse shock
much earlier than indicated by the semi-analytic model. For example,
the Fe knots may have been shocked when Cas~A was just $\sim $50 yr old.
This would likely indicate a much slower reverse shock motion ($dR_{\rm rs}/dt$) than the model suggests, perhaps due to an early interaction
with a dense circumstellar shell.

However, it is more likely that the ejecta are highly clumped. The
emission of the ejecta scales with $n_{\rm e}^2$, so the overall emission from the ejecta is biased towards the denser regions.
In Fig.~\ref{fig:truelove_ne_t} we also plot the $n_{\rm e}t$ values
for overdensity factors of 10 and 100, and see that for Fe an overdensity
$\sim 10$ would be sufficient, whereas for Si overdensities of $\sim$100 should be considered. Such overdensities are also consistent with
the density estimates by \citet{lazendic06} based on \chandra\ grating
measurements of Si-bright knots, indicating $n_{\rm e}\approx 20$--$200~{\rm cm^{-3}}$. However, by its nature, this study was biased towards the ejecta with more fragmentation.

Note that it is unlikely that the total ejecta mass is dominated by the clumps.
The ejecta mass is proportional to $M\propto n_{\rm e}V$.
If we define a clumping density ratio of $C = n_{\rm c}/n_{\rm diff}$, where $n_{\rm c}$ and $n_{\rm diff}$ are the electron densities in the clumped and diffused ejecta material, respectively.
We take a clumping filling fraction as $f =V_{\rm c}/V$. The condition that the ejecta mass ($M = n_{\rm diff}V_{\rm diff} + n_{\rm c}V_{\rm c}$) is dominated by the diffuse ejecta suggests that $f< 1/(C+1)$.
However, the high $n_{\rm e}t$ suggests that the emission measure ($EM = n_{\rm diff}^2(1-f)V + n_{\rm c}^2 fV$), which scales as $n_{\rm e}^2V$, is dominated by the clumps. This suggests that $f>1/(C^2+1)$.
Thus, for $C\approx 10$ (IGEs), this suggests plausible values for the clumping filling factor of $1\% \lesssim f \lesssim 10\%$. For the IMEs ($C\sim 100$),
this could be as small as $0.01\% \lesssim f \lesssim 1\%$.

Therefore, the high $n_{\rm e}t$ values for Cas~A probably testify to the
clumpy nature of its ejecta. It does, however, imply that there is more diffuse plasma with lower densities and lower $n_{\rm e}t$ values, which
still contribute to the overall spectrum. This provides another justification for using the \texttt{pshock} model, but with the range in ionization
parameters driven by the density spectrum as well as by a range
in plasma ages.

\section{Thermal Continuum}\label{app:thermal_continuum}

\begin{figure*}
\centerline{
  \includegraphics[width=0.5\textwidth]{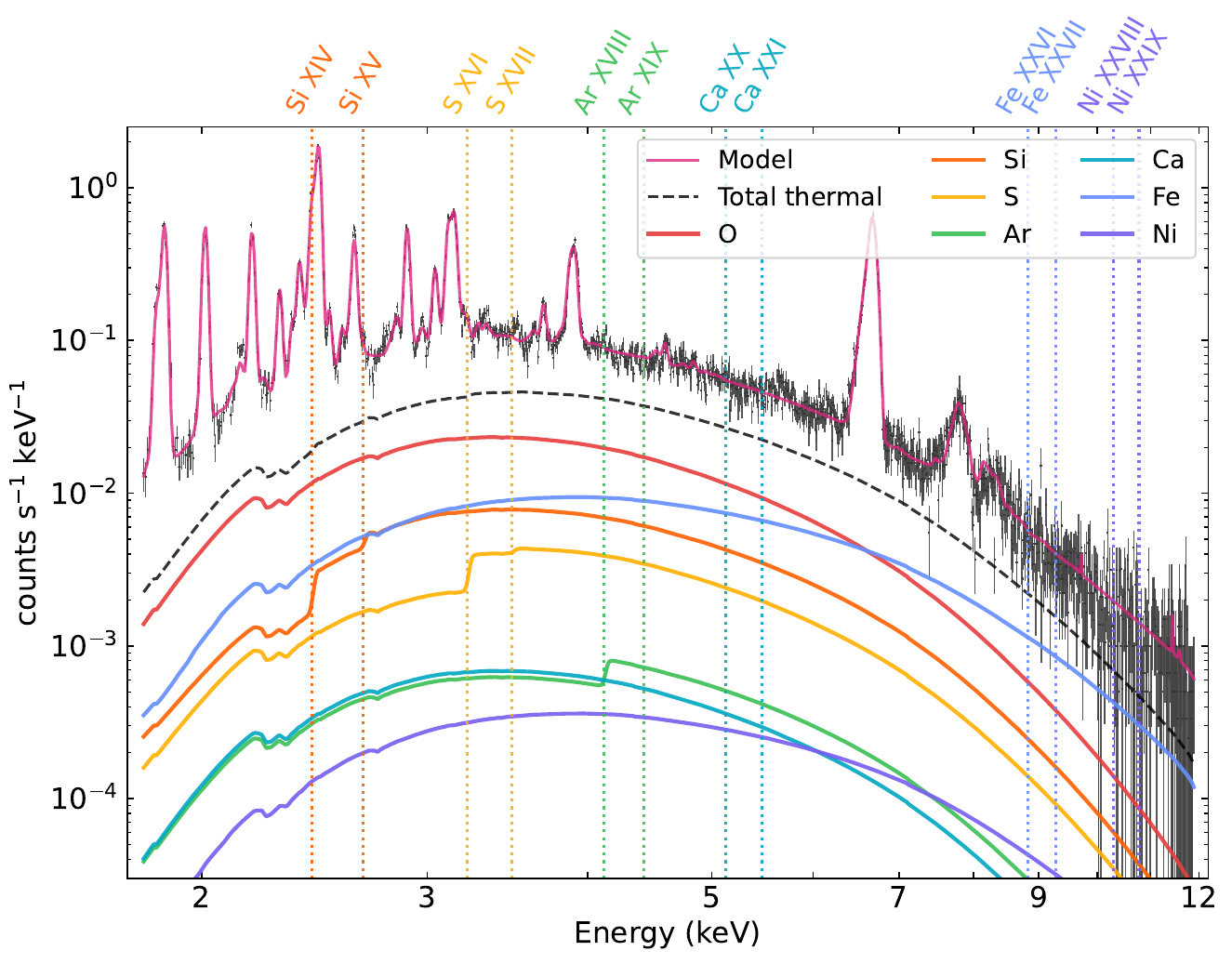}
  \includegraphics[width=0.5\textwidth]{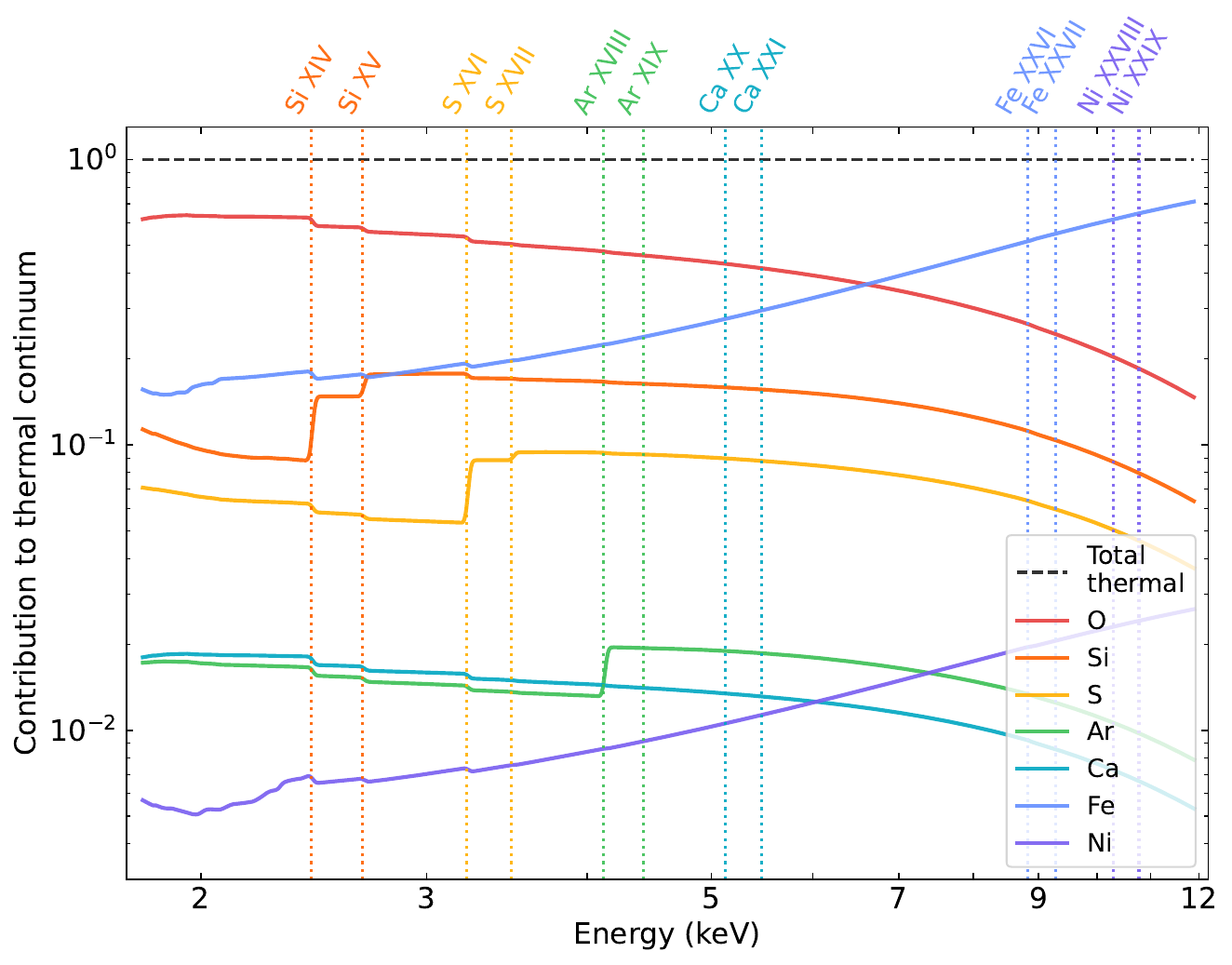}
}
\caption{
The thermal continuum --- bremsstrahlung and RRC --- for each element in our fitted model for super-pixel {\em e} is shown. The vertical dotted lines mark the binding energy of the ions, where the RRC edge of K-shell recombination is expected. Left: The contribution of the thermal continuum with respect to the observed spectrum. Right: The relative continuum contribution from each element in our model to the total thermal continuum.
}
\label{fig:thermal_continuum_plots}
\end{figure*}

The SE super-pixel {\em e} has the highest relative contribution of the thermal component to the total flux. Furthermore, it is among the regions with small Doppler broadening. Thus, it is the best region to investigate the thermal continuum --- thermal bremsstrahlung and radiative recombination continuum (RRC). 
Fig~\ref{fig:thermal_continuum_plots} shows the element-wise thermal continuum contribution to the total fitted model. 
The RRC edge, smoothed with Doppler broadening, is visible at the ionization energy of the corresponding He-like and H-like ions. Despite the sharp rise visible in the thermal continuum component for each element, the RRC only accounts for a small fraction of the total flux.
The continuum is dominated by non-thermal synchrotron emission and the thermal bremsstrahlung of oxygen, which in our model accounts for the thermal continuum contribution of all elements lighter than Si.

For the IMEs, all the L-shell edges are below the \xrism\ bandpass. Additionally, from the spectra and the fitted plasma parameters, we find a low concentration of the H-like ion of Ca, and fully ionized ions of all IMEs, thus their corresponding RRC features will also be weak. This leaves only the K-shell edges of Si XIV and S XVI at $\sim$2.44 keV and $\sim$3.22 keV, respectively. 
The edge energy of Si XIV is close to the forbidden (z) line of S He$\alpha$, which makes it blend with the He$\alpha$ triplet. 
For the S XVI edge, an excess near 3.2 keV was clearly identified during the early analysis of \xrism\ data and was suspected to be a potential sign of RRC or charge exchange. However, with the latest atomic databases (SPEX 3.08.02), it is explained by emission from high {\em n} transitions of the Rydberg series of Ar \citep{plucinsky25}. The Ar XVIII edge is relatively isolated from other line emission, but it only contributes $\sim1\%$ of the total thermal continuum. 

The Fe XXVI (recombining into Fe XXV) and Fe XXVII (recombining into Fe XXVI) ions have an RRC edge at $\sim$8.83 keV and $\sim$9.28 keV from K-shell recombination, respectively, and at $\sim$2.13 keV and $\sim$2.30 keV from L-shell recombination, respectively. However, from the fitted plasma properties, we find that the Fe ionisation states are primarily between Fe XXII and Fe XXV, while the ion concentrations of Fe XXVI and Fe XXVII are low.
For the lower ionization states of Fe, there is no K-shell edge, since the K-shell is already filled up and the L-shell edges are below 2 keV where the \xrism/Resolve effective area is severely affected by the Gate Valve closed. Similarly, for Ni, the ionisation states are too low to produce K-shell RRCs.


\bibliography{snrs,xray,supernovae,extras,methods_and_bayesian}
\bibliographystyle{aasjournalv7}



\end{document}